\documentclass[11pt]{article}
\usepackage[T1]{fontenc}
\usepackage{lmodern,amsmath,amssymb,amsthm,mathtools}
\usepackage[nopatch=footnote]{microtype}
\usepackage[margin=1in]{geometry}
\usepackage{booktabs,array,tabularx,enumitem,natbib,needspace}
\usepackage[colorlinks=true,linkcolor=blue,citecolor=blue,urlcolor=blue]
{hyperref}
\usepackage{setspace}
\usepackage{float,graphicx}
\floatstyle{boxed}
\newfloat{procedure}{H}{lop}
\floatname{procedure}{Procedure}
\floatstyle{plain}
\newtheorem{theorem}{Theorem}
\newtheorem{proposition}{Proposition}
\newtheorem{lemma}{Lemma}
\newtheorem{corollary}{Corollary}
\theoremstyle{definition}
\newtheorem{assumption}{Assumption}
\newtheorem{definition}{Definition}
\theoremstyle{remark}

\newcommand{\E}{\mathbb E}
\newcommand{\PP}{\mathbb P}
\newcommand{\R}{\mathbb R}
\newcommand{\Var}{\operatorname{Var}}
\newcommand{\Cov}{\operatorname{Cov}}
\newcommand{\ind}{\mathbf 1}
\newcommand{\ip}[2]{\langle #1,#2\rangle}

\newcommand{\cl}{\overline}
\newcommand{\spanof}{\operatorname{span}}
\newcommand{\tr}{\operatorname{tr}}
\newcommand{\op}{\mathrm{op}}
\newcommand{\eff}{\mathrm{eff}}
\newcommand{\sym}{\operatorname{sym}}
\newcommand{\Pn}{\mathbb{P}_n}
\newcommand{\Un}{\mathbb{U}_n}
\newcommand{\dto}{\Rightarrow}

\newcommand{\cK}{\mathcal K}
\newcommand{\cH}{\mathcal H}
\newcommand{\cT}{\mathcal T}
\newcommand{\cN}{\mathcal N}
\newcommand{\cA}{\mathcal A}
\newcommand{\cF}{\mathcal F}
\newcommand{\cD}{\mathcal D}

\numberwithin{equation}{section}
\title{The Statistical Cost of Second-Order Orthogonality\thanks{Research supported by MICIN/AEI/10.13039/501100011033, grant
CEX2021-001181-M, Comunidad de Madrid, grants EPUC3M11 (V PRICIT) and
H2019/HUM-5891, and grant PID2025-170712NB-I00 (MCI/AEI/FEDER, UE).}}
\author{Juan Carlos Escanciano\thanks{Universidad Carlos III de Madrid.}
\and Jo\"el R. Terschuur\thanks{Technical University of Munich.}}
\date{November 2026}
\begin{document}
\maketitle
\begin{abstract}
Debiased machine learning removes the first-order effect of nuisance
estimation, but the remaining quadratic bias can invalidate inference
when nuisance functions are hard to learn. Second-order orthogonal
moments remove this bias at the cost of additional variance. We
characterize this cost as the minimum variance of a correction that
offsets the quadratic sensitivity of the moment to nuisance errors,
and show that exact correction is possible only when this cost is finite.
Because the cost varies across moments, the first-order efficient
moment need not minimize total variance. We therefore choose the
moment and its correction jointly, and derive the resulting minimum
variance, a second-order analogue of the semiparametric efficiency
bound. The optimal moment may trade first-order efficiency for a lower
correction cost. In a binary choice logit model, it uses information beyond
the conditional likelihood only in well-populated covariate cells; in
difference-in-differences with continuous covariates, exact correction
forgoes the efficiency gains from exclusion restrictions. In the
partially linear model with continuous controls, the cost is infinite
for every moment, so that only part of the bias can be removed. For
such cases, we propose a data-driven adaptive rule that chooses how much bias
to remove within a bound on variance inflation, and we show that
inference remains valid after selection. In simulations for the
partially linear model, the coverage of nominal 95\% intervals rises
from 32--80\% to 94--97\% in all designs but a stress test.
\end{abstract}\noindent\textbf{Keywords:}
Second-order Neyman orthogonality;
automatic debiasing;
semiparametric efficiency;
Riesz representation;
U-statistics.
\newpage
%%%%%%%%%%%%%%%%%%%%%%%%%%%%%%%%%%%%%%%%%%
\section{Introduction}
\label{sec:intro}
Econometric inference often concerns a parameter of interest
whose estimation depends on other unknown quantities, such as
conditional means or treatment probabilities. Estimation errors
in these nuisance quantities can distort inference. First-order Neyman orthogonality removes the linear effect of
nuisance-estimation errors, but the remaining quadratic bias can still
be large relative to sampling variation. A pairwise correction, an
average over pairs of observations, can remove this quadratic bias \citep{Robins2008,Robins2017},
but it also adds variance. The relevant choice is therefore not simply
whether to debias. It is which estimating equation to debias, how much
bias to remove, and at what variance cost. These choices become
particularly important with infinite-dimensional nuisance parameters,
for which removing quadratic bias in every direction may be impossible
at finite variance. This paper studies these choices. Second-order
orthogonality has a statistical cost, determined by the model and by
the chosen moment. We characterize this cost, show how to minimize it,
and show how to estimate and control it in practice.

We first ask what it costs to correct a given first-order orthogonal
moment. The correction must reproduce the curvature of the moment, its
quadratic sensitivity to nuisance-estimation errors, and the variance
it adds is the price of doing so. We characterize the smallest such
variance, which we call the correction cost, and show that an exact
second-order correction with finite variance exists if and only if
this cost is finite. The cost depends on the moment's curvature relative to the nuisance's information: it
is large when the moment is sensitive to nuisance directions that the
data reveal poorly, and it can be infinite when curvature is spread
over infinitely many directions, even if each of them is well
informed. For moments depending pointwise on a nonparametric regression,
we obtain a sharp dichotomy: with continuously distributed
covariates and an otherwise unrestricted law, the correction
cost is zero if the first-order orthogonal moment has no
quadratic nuisance bias, and infinite otherwise.

The partially linear model (PLM), in which an outcome depends linearly
on a treatment and nonparametrically on controls, is the workhorse
example. Debiased machine learning (DML) \citep{DML2018} relates the
residuals from the regressions of the outcome and the treatment on the
controls. Its bias is the expected product of the treatment-regression
error and the implied error in the structural regression on controls. A correction along a set of functions of the controls
removes the part of this product that lies in their span, and its
variance grows with the number of functions. When that number is
proportional to the sample size, the correction changes first-order
inference, and correcting every direction of a nonparametric
regression requires infinite variance. In this case, exact second-order
orthogonality is out of reach, and the practical question becomes how
much of the bias to remove. The obstruction comes from the infinite-dimensional nuisance
parameters: with finite-dimensional nuisance
parameters, higher-order orthogonal moments are available
\citep{BJN2026,CST2026}.

The correction cost depends on the first-order moment, so we ask
whether choosing the moment together with its correction improves the
trade-off: a moment with larger first-order variance can have smaller
curvature, and hence smaller total variance. Extending the classical
theory of optimal moments \citep{Hansen1982,Chamberlain1987}, we
characterize the minimum variance over all first-order orthogonal
moments and their corrections by a second-order information, and show
that the first-order efficient moment need not attain it. The excess
over the first-order efficiency bound is the minimum combined cost of
departing from the efficient moment and correcting the curvature of
the resulting moment. Whether moment choice helps depends on the
sensitivity of the alternative moments to the nuisance parameters.
When some relevant moments are less sensitive to them, the optimal
moment trades first-order efficiency for a lower correction cost. When
the costly sensitivity is shared by all relevant moments, the choice
of moment cannot reduce it, and the question is only how much bias to
remove. The PLM, defined by a conditional mean restriction, is of the second
kind: every relevant moment is sensitive to the product of the errors
in the outcome and treatment regressions, and none can be corrected
exactly with continuous controls. For this reason,
Section~\ref{sec:approx} takes the usual DML moment as given in the
PLM and chooses only how much of its bias to correct.

The infinite-dimensional problem is the limit of finite ones. Requiring second-order orthogonality in increasingly many nuisance directions, while keeping full first-order orthogonality, gives a nondecreasing sequence of minimum variances. A finite-variance correction of the full problem exists if and only if this sequence is bounded. As the sample grows, the weight on the correction variance vanishes, and the second-order information approaches the first-order information exactly when the efficient moment is a mean-square limit of moments with finite correction cost. This can happen even when the efficient moment itself
has no finite-variance correction. With finitely many restrictions,
the problem reduces to a finite quadratic program, and allowing some
curvature to remain traces a frontier between variance and remaining
bias.

Two economic models illustrate when the choice of moment lowers the
correction cost and when the cost is shared by all moments. In a binary-choice
model, two outcomes share a nonparametric baseline index and differ by
a common shift in log odds, as when two consumers face prices that
differ by a fixed amount. The conditional-likelihood moment
\citep{Rasch1960,Andersen1970,Chamberlain1980} is insensitive to the baseline index at every
order, as it does not depend on it, while the efficient moment uses additional information at the
price of sensitivity. We solve the joint problem over the full class
of first-order moments. The optimum combines the conditional-likelihood
moment with a variance-reducing term whose weight grows with the size
of the covariate cell: the additional first-order information is used
where cells are well populated, and only the conditional likelihood
remains on continuous components of the covariates. With a discrete covariate
taking infinitely many values, the optimum recovers the full
first-order information even when the efficient moment cannot be exactly corrected.

In difference-in-differences (DiD), restrictions on which
covariates affect treatment assignment and untreated outcome
growth improve first-order precision, but the moments that
exploit them are more sensitive to the nuisance parameters
than the standard doubly robust moment \citep{SZ2020}. With continuously distributed
covariates, exact correction is incompatible with any efficiency gain
from these exclusions: only the doubly robust moment can be corrected
exactly, and it can be corrected if and only if the joint law of the
assignment and outcome covariates has a square-integrable density
relative to the product of its marginals. When the two share a
continuous covariate, exact correction is impossible. Along directions
of a shared covariate, the sensitivity is the same for every moment:
the variance cost of correcting these directions grows linearly with
their number, whatever the moment. The cost of second-order
orthogonality is then a feature of the statistical problem itself. In this DiD model with covariate exclusions, approximate
correction attains the first-order efficiency bound under
explicit smoothness and rate conditions. The construction
consistently estimates the efficient moment and corrects
the three nuisance-error products generated by its estimation.

Exact correction is thus often impossible or too costly, and inference
must rely on approximate corrections chosen in light of their cost. We
construct them automatically for moments that depend on regressions,
with the PLM as the leading case, extending \citet{CNS2022} from first-
to second-order orthogonality. A correction along a set of covariate
functions removes the bias in their span, and which set works best
depends on how the regression errors are distributed, which is
unknown. Ridge suits errors spread thinly over many directions;
projection on selected functions suits errors concentrated in a few.
Our rule, ADA, adapts to this structure: within a budget for variance
inflation, it chooses the correction that captures the largest share
of the regression errors, estimated by products of residuals from
different observations. Procedure~\ref{proc:plm} in Section~\ref{sec:auto-selection} collects
the steps for the PLM, from the first-step fits to the confidence
interval.

With the cross-fitting of \citet{ET2026}, in which each pair of
observations uses nuisance estimates trained without both, the
selected correction satisfies an oracle inequality relative to the
best candidate within the budget. When the correction variance is
negligible, inference after selection is asymptotically equivalent to inference with
known nuisance functions. More generally, the correction can
contribute to first-order variance. This case is technically more
involved, and we analyze it in the PLM as an illustration. There, the
standard error includes this contribution, and inference after
selection remains valid under a stability condition on the
correction, which ridge corrections satisfy.

The simulations quantify these trade-offs. In nonparametric and sparse
designs for the PLM, first-order DML intervals cover the parameter in
17--80\% of the replications. ADA restores coverage to 94--97\% in all
designs but a stress test. ADA adapts as intended: it selects ridge when the
regression errors are spread over many frequencies and selected
projection in the sparse linear designs, in at least 93\% of the
replications in each case. Higher-order influence-function estimators
\citep{Robins2008,Robins2017,LMNR2026,Liu2026} and the triple-debiased
Lasso of \citet{CST2026} perform well in the sparse linear designs, for
which the latter is built, and undercover in the nonparametric ones.
Sharing the same fits between the first-order term and the correction
of each pair, as in \citet{ET2026}, gives the smallest mean squared
error in most designs across different cross-fitting methods.

Taken together, the results show that the benefits of second-order
orthogonality come at a variance cost that depends on the model, on
the moment, and on the directions corrected. The cost can be infinite,
it can require giving up valid economic restrictions, and it can be
the same for every relevant moment for a given target. It can also be estimated and
controlled, and second-order corrections chosen in light of their cost
deliver reliable inference where first-order methods do not.

\subsection{Relation to existing results}

First-order orthogonalization goes back to \citet{Neyman1959};
subsequent developments include doubly robust estimation, targeted learning, and debiased
machine learning
\citep{ScharfsteinEtAl1999,BangRobins2005,vanderLaanRobins2003,Newey1994,LR2022,DML2018}.
\citet{AE2026} distinguish the existence of an orthogonal moment from
its relevance for the target. We study these questions under
second-order restrictions and quantify their variance cost.

Higher-order influence-function theory supplies the higher-order tangent space representation
and approximation framework
\citep{HasminskiiIbragimov1979,Robins2008,vdV2014,Robins2017}. This
literature recognizes that exact second-order corrections can
fail and that approximation trades bias for variance. We give a necessary and sufficient criterion for
finite-variance correction, a decomposition of the minimum oracle
variance over the joint choice of moment and correction, and
conditions for recovering first-order efficient information. In the
applications, we identify the entire exactly correctable moment class
and show how the covariate distribution restricts the information it
can retain. Feasible higher-order estimation with estimated projection
and Gram operators is studied by \citet{LMNR2026}, \citet{Liu2026},
and \citet{ZLZ2026}.

In econometrics, \citet{BJN2026} have recently constructed
higher-order orthogonal moments in finite-dimensional moment-condition
models, allowing the nuisance dimension to grow with the sample size;
related constructions include  \citet{CST2026}, and
\citet{BJW2026}. \citet{MSZ2018} study the limitations of higher-order orthogonality with infinite-dimensional parameters in the PLM, based on one-observation moments. We use U-statistics, as in \citet{ET2026}, but for second-order orthogonality rather than first, and optimize over the first-order moment
as well as its correction. 

Our automatic construction extends
Riesz-based debiasing \citep{CNS2022} to second order, and ADA complements the bias-based selection of nuisance learners of \citet{CT2024}.
Many-regressor and small-bandwidth asymptotics provide a related
variance perspective, in which degenerate U-statistic components
matter at first order as their complexity grows
\citep{CJN2018,KSS2020,CCJ2014,CJ2018}; here the size of these
components also depends on the first-order moment.

Section~\ref{sec:motivation} sets out the problem and illustrates it in the PLM.
Section~\ref{sec:theory} develops the exact cost theory and the joint
choice of moment and correction. Section~\ref{sec:approx} develops
approximate correction, its data-driven selection, and inference, and
evaluates them in the PLM. Section~\ref{sec:applications} studies
binary choice and DiD, and Section~\ref{sec:conclusion} concludes. The
appendices contain proofs and implementation details.

%%%%%%%%%%%%% SECTION 2 %%%%%%%%%%%
\section{The problem: correcting bias at a variance cost}
\label{sec:motivation}

Let $W_1,\ldots,W_n$ be independent and identically
distributed (iid) observations from $P_0$. The target is
$\theta_0\in\mathbb R$, and $\eta_0$ collects the nuisance
parameters, which may be infinite dimensional. Write
$\lambda=(\theta,\eta)$ and $\lambda_0=(\theta_0,\eta_0)$.
Write $\E=\E_0$ for expectation under $P_0$, using its product law
for independent observations. We also write
$\phi(W,\lambda)=\phi(W,\eta,\theta)$, and similarly for pairwise
functions, $k(W_i,W_j,\lambda)=k(W_i,W_j,\eta,\theta)$.
Unless a measure is specified, $\|f\|^2=\E[f(W)^2]$,
$\|h\|^2=\E[h(X)^2]$, and $\|k\|^2=\E[k(W_i,W_j)^2]$ for $i\ne j$.
For vector-valued functions the integrand is the squared Euclidean norm;
$\langle f,g\rangle=\E[f'g]$. We use $\|\cdot\|_2$ for Euclidean
norms, $\|\cdot\|_\infty$ for essential suprema, and
$\|\cdot\|_{\op}$ and $\|\cdot\|_F$ for operator and Frobenius norms. We focus on a scalar target for simplicity; orthogonalization
applies componentwise to moments identifying a vector target.

Starting from a first-order orthogonal estimating function
$\phi(W,\lambda)$, we introduce a symmetric pairwise
correction $k(W_i,W_j,\lambda)$ and form
\[
 \Psi(W_i,W_j,\lambda)
 =\frac{\phi(W_i,\lambda)+\phi(W_j,\lambda)}{2}
  -k(W_i,W_j,\lambda).
\]
For pairwise functions, subscripts $ij$ indicate evaluation
at observations $i$ and $j$. A subscript $0$ denotes evaluation at the true
law, for example $\phi_0(W)=\phi(W,\lambda_0)$ and
$k_0(W_i,W_j)=k_{ij}(\lambda_0)$.

A valid pairwise moment satisfies $\E[\Psi_0]=0$.
It is \emph{second-order Neyman-orthogonal} if
\begin{equation}
 \left.
 \frac{d^l}{dt^l}
 \E[\Psi(W_i,W_j,\eta_0+th,\theta_0)]
 \right|_{t=0}=0,
 \qquad l=1,2,\quad h\in\mathcal H,
 \label{eq:second-order-orthogonality}
\end{equation}
where $\mathcal H$ is a linear space of admissible
nuisance perturbations. The derivatives vary the nuisance
argument while keeping the target fixed.
First-order orthogonality imposes only the condition
for $l=1$; the additional condition removes quadratic
sensitivity to nuisance-estimation errors.

The corrected oracle estimating equation is
\begin{equation}
 \begin{gathered}
 \Un\Psi(\cdot,\cdot,\eta_0,\theta)
 =\Pn\phi(\cdot,\eta_0,\theta)-\Un k(\cdot,\cdot,\eta_0,\theta)=0,\\
 \Pn f:=\frac1n\sum_{i=1}^n f(W_i),
 \qquad
 \Un k:=\binom n2^{-1}\sum_{i<j}k(W_i,W_j).
 \end{gathered}
 \label{eq:estimating_eq}
\end{equation}
Suppose $\phi_0$ and $k_0$
are square-integrable, $\E[\phi_0]=0$, and the correction
is degenerate at the truth:
$\E[k_0(W_i,W_j)\mid W_i]=0$.
The two sample terms are then uncorrelated, giving
the exact oracle variance
\begin{equation}
 \Var(\Pn\phi_0-\Un k_0)
 =\frac{\E[\phi_0^2]}{n}
  +\frac{2\E[k_0^2]}{n(n-1)}.
 \label{eq:oracle_variance}
\end{equation}
This variance evaluates both terms at the true nuisance
values. For comparisons across moments,
Section~\ref{sec:theory} normalizes their sensitivity
to the target parameter. The correction adds variance,
and its contribution can become substantial as the
number of corrected nuisance directions grows.

\paragraph{Running example: the partially linear model (PLM).}
Let $W=(Y,D,X)$, where $Y$ is an outcome, $D$ a treatment
or policy variable, and $X$ observed controls. Consider
\begin{equation}
 Y=\theta_0D+c_0(X)+\varepsilon,
 \qquad \E[\varepsilon\mid D,X]=0.
 \label{eq:plm-model}
\end{equation}
The target $\theta_0$ measures the conditional mean response
to a unit change in $D$, holding $X$ fixed. 
Define $\gamma_0(X)=\E[D\mid X]$,
$\mu_0(X)=\E[Y\mid X]$, and
$c_0=\mu_0-\theta_0\gamma_0$ and $V=D-\gamma_0(X)$, with $\E[V^2]>0$.
For candidate regressions $\eta=(\mu,\gamma)$,
put $\widetilde D=D-\gamma(X)$ and $e(W,\mu,\gamma,\theta)=Y-\mu(X)-\theta\widetilde D$.
The usual DML moment is
$\phi(W,\lambda)=e(W,\mu,\gamma,\theta)\widetilde D$ \citep{Robinson1988}.

Consider a nuisance path $\eta_t=\eta_0+th$, where
$h=(h_\mu,h_\gamma)$, and write
$h_e=h_\mu-\theta_0h_\gamma$. Along this path,
$e(W,\mu_t,\gamma_t,\theta_0)=\varepsilon-th_e(X)$ and
$\widetilde D=V-th_\gamma(X)$, giving
\begin{equation}
 \E[\phi(W,\eta_t,\theta_0)]
 =t^2\E[h_e(X)h_\gamma(X)].
 \label{eq:plm-product-bias}
\end{equation}
The first derivative at zero vanishes, but the second
equals $2\E[h_eh_\gamma]$, which is generally nonzero.
Thus, the DML moment is first-order but not
second-order orthogonal.

To correct this quadratic term, choose a vector $p(X)$
of $d$ functions with nonsingular $G_0=\E[p(X)p(X)']$.
Define the population projection kernel $K(x,x')=p(x)'G_0^{-1}p(x')$ and the pairwise correction
\begin{equation}
 k(W_i,W_j,\lambda)
 =\frac{K(X_i,X_j)}{2}
 \left\{
 e(W_i,\mu,\gamma,\theta)\widetilde D_j
 +e(W_j,\mu,\gamma,\theta)\widetilde D_i
 \right\}.
 \label{eq:plm-projection-correction}
\end{equation}
Independence of the two observations gives
\[
 \E[k(W_i,W_j,\eta_t,\theta_0)]
 =t^2\E[p(X)h_e(X)]'G_0^{-1}\E[p(X)h_\gamma(X)].
\]
If either $h_e$ or $h_\gamma$ belongs to the span of $p$,
the correction cancels \eqref{eq:plm-product-bias} exactly
along this path. Thus, $\Psi(\cdot,\cdot,\lambda)$ is
second-order orthogonal on $\mathcal{H}$ whenever this
condition holds for every admissible direction.
For general nonparametric paths, its remaining bias is
the expected product of the error components outside
that span. With nonatomic covariates, eliminating this
bias in every nuisance direction requires a correction
with infinite variance, as we show below.

At the truth, the correction is degenerate and has
variance
\begin{equation}
 \Var(\Un k_0)
 =\frac{\E[K(X_i,X_j)^2\varepsilon_i^2V_j^2]}
        {n(n-1)}.
 \label{eq:plm-projection-cost}
\end{equation}
Since $\E[K(X_i,X_j)^2]=d$, this cost is of order
$d/n^2$ when $\E[\varepsilon^2\mid X]$ and
$\E[V^2\mid X]$ are bounded above and away from zero.
Increasing $d$ therefore corrects more nuisance
directions at a growing variance cost. When $d$
is proportional to $n$, the correction contributes
to first-order sampling variance. For example, suppose $c_0$ and $\gamma_0$ are H\"older-$s$
on $[0,1]^q$, where larger $s$ means greater smoothness
($s=1$ means Lipschitz continuity; $s=2$ means Lipschitz
first derivatives). Assume the density of $X$ is bounded
above and away from zero. Let the correction space be a
$d$-dimensional tensor-product B-spline space of order
exceeding $s$, on a quasi-uniform partition, containing
the first-step fits. The population projection then
leaves bias $O(d^{-2s/q})$, while, under the variance
bounds above, the additional oracle variance is of order
$d/n^2$. Balancing the squared-bias bound and correction
variance gives $d\asymp n^{2q/(4s+q)}$: sublinear in $n$
when $s>q/4$, proportional to $n$ when $s=q/4$, and
superlinear when $s<q/4$. The PLM thus illustrates the additional
difficulty beyond the finite-dimensional constructions.

Whether a different first-order moment can avoid this
obstruction motivates the general theory.
Section~\ref{sec:theory} develops the general conditions
for existence, minimum variance cost, and joint choice
of the moment and its correction. Readers interested
primarily in implementation can proceed directly to
Section~\ref{sec:approx}, which develops the automatic
procedure and its inference guarantees, continuing
with the PLM as a running example.
%%%%%%%%%%% SECTION 3 %%%%%%%%%%%%
\section{Existence, relevance, and minimum cost}
\label{sec:theory}

Let $P_0=P_{\theta_0,\eta_0}$ be the data distribution in the semiparametric model
$\{P_{\theta,\eta}:\theta\in\Theta,\eta\in\mathcal E\}$.
For any probability law $P$, let $\E_P$ denote expectation
under $P$ and define
\[
 L_2(P)=\{f:\E_P[f^2]<\infty\},
 \qquad
 L_2^0(P)=\{f\in L_2(P):\E_P[f]=0\}.
\]
Expectations under $P_\lambda$ or its product law are
denoted by $\E_\lambda$. For a linear subspace $\mathcal S$ of $L_2(P)$, the closure
$\overline{\mathcal S}$ is the set of limits in $L_2(P)$ of sequences
in $\mathcal S$, the orthocomplement is
$\mathcal S^\perp=\{f\in L_2(P):\E_P[fs]=0\text{ for all }s\in\mathcal S\}$,
and $\Pi_{\mathcal S}$ denotes the orthogonal projection onto
$\overline{\mathcal S}$. The same definitions apply in $L_2(P^2)$ to
functions of two independent observations.  Henceforth, $L_2\equiv L_2(P_0)$ or $L_2\equiv L_2(P_0^2)$, which will be clear from the context.

Assumption~\ref{ass:regularity} is maintained throughout
this section without being repeated in each result. Pathwise regularity is a standard condition in the semiparametric literature.

\begin{assumption}[Sampling and regularity]
\label{ass:regularity}
The observations $W_1,\ldots,W_n$ are iid with law $P_0=P_{\theta_0,\eta_0}$, and the
pathwise regularity conditions in
Appendix~\ref{app:regularity_paths} hold.
\end{assumption}

\subsection{Moment families and curvature matching}

Let $\mathcal H$ be a linear space of admissible nuisance
directions. Holding the target fixed at $\theta_0$, we assume the
following second-order likelihood-ratio expansion for
$\eta_0+th$, for $t$ near zero:
\[
\frac{dP_{\theta_0, \eta_0 + th}}{dP_0} = 1 + t s_{h} + \frac{1}{2} t^2u_{h,h} + o_{L_2(P_0)}(t^2),
\]
where, under the conditions in Appendix \ref{app:regularity_paths},
\[
 s_h=\left.\frac{\partial}{\partial t}
 \frac{dP_{\theta_0,\eta_0+th}}{dP_0}\right|_{t=0},
 \qquad
 u_{h,g}=\left.\frac{\partial^2}{\partial t\,\partial v}
 \frac{dP_{\theta_0,\eta_0+th+vg}}{dP_0}\right|_{t=v=0},
\]
and the maps $h\in\mathcal{H} \mapsto s_h$ and $(h,g)\in\mathcal{H}^2 \mapsto u_{h,g}$ are linear and bilinear scores in $L_2^0(P_0)$, respectively. 
We also denote the sensitivity of the statistical model to the target parameter by the score $\ell = d/ dt \,dP_{\theta_0 +t,\eta_0}/dP_0|_{t = 0} \in L_2^0(P_0)$. 

Define
\[
\mathcal{T}_\eta = \overline{\text{span}\{s_h: h \in \mathcal{H}\}}, \quad \mathcal{T} = \overline{\mathcal{T}_\eta + \text{span}\{\ell\}}, \quad \mathcal{N} = \mathcal{T}_\eta^\perp \cap L_2^0(P_0), \quad \mathcal{A} = \mathcal{T}^\perp \cap L_2^0(P_0).
\]

The restricted tangent space $\mathcal{T}_\eta$ includes all scores in the
statistical model that preserve the target at $\theta_0$. The space $\cN$ contains the first-order orthogonal moments. The direction space $\mathcal{H}$ can consist of bounded functions whose
scores generate $\mathcal{T}_\eta$, i.e. the likelihood ratio expansion need not
hold for every unrestricted $L_2$ direction.

Let the space of symmetric degenerate kernels be
\[
 \cK_\lambda=
 \left\{
 k\in L_2(P_\lambda^2):
 k(w_i,w_j)=k(w_j,w_i),\;
 \E_\lambda[k(W_i,W_j)\mid W_i]=0
 \right\},
\]
for $\lambda$ in an open neighborhood of $\lambda_0$, with the identities understood almost surely, and $\cK_{\lambda_0} = \cK$.

\begin{definition}[Admissible moment families]
\label{def:admissible}
A family $(\phi(\cdot,\lambda),k(\cdot,\cdot,\lambda))$ is admissible if, for all $\lambda$ in an open neighborhood of $\lambda_0$, $k(\cdot,\cdot,\lambda) \in \cK_\lambda$, $\phi(\cdot,\lambda) \in L_2(P_\lambda)$,  and for every $h \in \cH$, satisfy
\begin{equation}
\begin{aligned}
 &\text{(Local validity): }\E_\lambda[\phi(W,\lambda)]=0,\\
 &\text{(Local Neyman orthogonality): }
 \left.\frac{d}{dt}\E_{\theta_0,\eta}[\phi(W,\eta+th,\theta_0)]\right|_{t=0}=0.
\end{aligned}
\label{eq:uniformfirst}
\end{equation}
\end{definition}
Define $\mathcal M(\tilde\lambda;P_\lambda) = \E_\lambda[\phi(W,\tilde\lambda) - k(W_i,W_j,\tilde\lambda)]$, $\mathcal M_0(\lambda) = \E[\phi(W,\lambda) - k(W_i,W_j,\lambda)]$, and let
\begin{equation}
 B_\phi(h,g)=\tfrac12\E[\phi u_{h,g}],
 \qquad
 (\mathcal Rk)(h,g)
 =\E[k(W_i,W_j)s_h(W_i)s_g(W_j)].
 \label{eq:BR}
\end{equation}

\begin{lemma}[Curvature decomposition]\label{lem:pathwise}
For any admissible family, $\phi\in\cN$,
$-\partial_\theta\mathcal M_0(\theta_0,\eta_0)=\E[\phi\ell]$, and,
for every $h\in \cH$,
\[
\mathcal M_0(\theta_0,\eta_0+th)
=t^2[B_\phi(h,h)-(\mathcal Rk)(h,h)]+o(t^2),
\qquad t\to0.
\]
Thus, second-order orthogonality is equivalent to $\mathcal Rk=B_\phi$.
\end{lemma}
Since $B_\phi$ and $\mathcal Rk$ are symmetric bilinear forms,
their difference $A=B_\phi-\mathcal Rk$ satisfies the polarization identity
\[
  2A(h,g)=A(h+g,h+g)-A(h,h)-A(g,g),
  \qquad h,g\in \mathcal{H}.
\]
Thus, vanishing quadratic sensitivity along every scalar path
$\eta_0+th$ is equivalent to curvature matching in every pair of
directions. It also implies that, for any finite collection
$h_1,\ldots,h_d\in \mathcal{H}$,
\[
  \mathcal M_0\!\left(\theta_0,\eta_0+\sum_{j=1}^d t_jh_j\right)
  =o(\|\mathbf t\|_2^2),
  \qquad \mathbf t=(t_1,\ldots,t_d)\to0,
\]
under the maintained pathwise regularity conditions. Lemma \ref{lem:pathwise} separates the quadratic sensitivity of the first-order moment, $B_\phi$, from the curvature of the pairwise correction, $\mathcal{R}k$. 

A pair $(\phi,k)\in\cN\times\cK$ is \emph{compatible} if
$\mathcal Rk=B_\phi$. It is \emph{relevant} if
$\E[\phi\ell]\ne0$, and \emph{normalized} if
$\E[\phi\ell]=1$. A value $\phi$ is \emph{correctable} if
it belongs to a compatible pair. We denote the set of compatible pairs as
\[
\mathcal{C} = \{(\phi,k)\in\cN\times\cK: \mathcal Rk=B_\phi\}.
\]

By definition and by Lemma \ref{lem:pathwise}, admissible families $(\phi(\cdot,\lambda), k(\cdot,\cdot,\lambda))$ with $\mathcal Rk=B_\phi$ are such that $(\phi, k) \in \mathcal{C}$. Conversely, not every pair in $\mathcal{C}$ necessarily extends to an admissible family. The optimization problems below concern compatible pairs, while attainment of the optimum by an admissible family requires an admissible extension, as defined in Appendix \ref{app:extensions}.

\subsection{Existence and the cost of correcting a given moment}

Note that, from (\ref{eq:BR}),
\[
(\mathcal{R}k)(h,g) = \mathbb{E}\left[k(W_i,W_j) \frac{s_h(W_i)s_g(W_j) + s_h(W_j)s_g(W_i)}{2}\right],
\]
so defining
\[
Q_{h,g}(w_i,w_j) = \frac{s_h(w_i)s_g(w_j) + s_h(w_j)s_g(w_i)}{2}, \quad \mathcal{Q} = \overline{\text{span}\{Q_{h,g}: h,g \in \mathcal{H}\}}^{L_2(P_0^2)} \subseteq \mathcal{K},
\]
we can see that $(\mathcal{R}k)(h,g) = \mathbb{E}[k Q_{h,g}]$ is the inner product of $k$ and $Q_{h,g}$ in $L_2(P_0^2)$. Hence, if there exists a linear bounded functional $L_\phi: \mathcal{Q} \to \mathbb{R}$ satisfying
\[
L_\phi(Q_{h,g}) = B_\phi(h,g), \quad \forall h,g \in \mathcal{H},
\] 
the Riesz representation theorem would allow us to write $B_\phi$ as an inner product with some $k_\phi \in \mathcal{K}$. Then, we would have that $(\mathcal{R}k_\phi)(h,g) = B_\phi(h,g)$ for all $h,g \in \mathcal{H}$. Define $c_\phi$ as the infimum of all $C \geq 0$ such that
\begin{equation}
\label{eq:cost-criterion}
\biggl| \sum_{j=1}^J a_j B_\phi(h_j,g_j) \biggr| \leq C \biggl\| \sum_{j=1}^J a_j Q_{h_j,g_j} \biggr\|_{L_2(P_0^2)}, \quad \forall J \in \mathbb{N}, a_j \in \mathbb{R}, h_j, g_j \in \mathcal{H}.
\end{equation}
The infimum is $+\infty$ if no finite $C$ satisfies the above inequality. The norm of $Q_{h,g}$ represents the information the statistical model contains about the departures in directions $h$ and $g$. If the information about these directions is small but the curvature of $\phi$ in these directions is large, then $c_\phi$ has to be large. A finite variance correction will require curvature to be small relative to combinations of scores with little information.

Our first result is then an application of the Riesz Representation Theorem.

\begin{theorem}[Existence is finite correction cost]
\label{thm:cost}
For fixed $\phi\in\cN$, a kernel $k\in\cK$ satisfying
$\mathcal Rk=B_\phi$ exists if and only if $c_\phi<\infty$.
When it exists, the unique minimum-norm kernel $k_\phi$
belongs to $\mathcal Q$ and satisfies
\begin{equation}
 c_\phi^2
 =\E[k_\phi^2]
 =\min_{\substack{k\in\cK\\\mathcal Rk=B_\phi}}\E[k^2].
 \label{eq:mincost}
\end{equation}
If $c_\phi<\infty$ and $(\phi,k_\phi)$ admits an
admissible extension, as defined in
Appendix~\ref{app:extensions}, this extension is
second-order orthogonal at $\lambda_0$ and attains
\eqref{eq:mincost}. It is relevant if and only if
$\E[\phi\ell]\ne0$.
\end{theorem}

By \eqref{eq:oracle_variance}, the minimum-norm correction gives\begin{equation}
 \Var(\Pn\phi-\Un k_\phi)
 =\frac{\E[\phi^2]}n+\frac{2c_\phi^2}{n(n-1)}.
 \label{eq:oracle-cost}
\end{equation}
Hence
$2c_\phi^2/\{n(n-1)\}$ is the least additional variance. For a moment with
target sensitivity $J_\phi=\E[\phi\ell]\ne0$, divide
\eqref{eq:oracle-cost} by $J_\phi^2$.

Theorem \ref{thm:cost} gives a general characterization of the minimum cost. It turns the existence of a second-order correction into a variance, which says whether exact correction is possible and at what price, and it shows that the minimum kernel is built from products of nuisance scores alone. The criterion \eqref{eq:cost-criterion} is, however, abstract, since it involves all finite combinations of nuisance directions. To gain insight into how the cost depends on the curvature relative to the statistical information, suppose that there is a nuisance score operator $S: h\in \mathcal{H} \to s_h\in L_2(P_0)$ which admits a singular system. This means there exist orthonormal bases $\{e_j\}$ of $(\text{ker } S)^\perp$, the identifiable nuisance directions ($\text{ker } S = \{h \in \mathcal{H}: Sh = 0\}$), and $\{f_j\}$ of the nuisance tangent space, $\mathcal{T}_\eta$, such that
\begin{equation} 
\label{SVD}
 Se_j=\sqrt{I_{\eta,j}}f_j,
 \qquad I_{\eta,j}>0.
\end{equation}

Since the basis $\{f_j\}$ is orthonormal, $\|S e_j\|^2_{L_2(P_0)} = I_{\eta,j}$, so $I_{\eta,j}$ is the Fisher information in the $j$th nuisance direction. 

Let $b_{jl} = B_\phi(e_j,e_l)$. We formalize this heuristic discussion in the following proposition. 
\begin{proposition}[Information in nuisance directions]
\label{prop:spectral}
Suppose $\mathcal{H}$ is a Hilbert space and $S: \mathcal{H} \mapsto \mathcal{T}_\eta$ is a bounded linear score operator admitting a singular system $(e_j, f_j, I_{\eta,j})$ satisfying (\ref{SVD}). Suppose further that $B_\phi$ is a continuous bilinear form on $\mathcal{H}$. If $B_\phi(h,g)=0$ whenever either argument belongs to
$\ker S$, then
\begin{equation}
 c_\phi^2
 =\sum_{j,l}\frac{b_{jl}^2}{I_{\eta,j}I_{\eta,l}}.
 \label{eq:spectral}
\end{equation}
When the series is finite, the minimum kernel is
\[
 k_\phi(w_i,w_j)
 =\sum_{q,l}
 \frac{b_{ql}}{\sqrt{I_{\eta,q}I_{\eta,l}}}
 f_q(w_i)f_l(w_j),
\]
with convergence in $L_2(P_0^2)$.
If the series diverges, or $B_\phi(h,g) \neq 0$ for some $h$ or $g$ in $\ker S$, then no $L_2$ correction exists.
\end{proposition}
Proposition \ref{prop:spectral} gives a representation of the squared correction cost $c_\phi^2$ as the sum of curvature towards each pair of nuisance directions relative to the information in those directions. It also gives an interpretation of the minimum kernel as an orthogonal projection since $f_q(W_i)f_l(W_j)$ are mean zero and orthogonal, compatibility determines the following regression coefficients
\begin{align}
 \label{eq:spectral-regression-coefficients}
\frac{\Cov(k_\phi(W_i,W_j), f_q(W_i)f_l(W_j))}{\Var(f_q(W_i)f_l(W_j))} = \frac{b_{ql}}{\sqrt{I_{\eta,q}I_{\eta,l}}}.
\end{align}
Divergence of $c_\phi^2$ can arise either from curvature in directions with little information or from the sum of infinitely many directions with information bounded away from zero not converging.

\subsection{A conditional-score criterion}

Theorem~\ref{thm:cost} provides a general existence criterion.
We now consider a structure that arises naturally in conditional
moment models. Suppose
\begin{equation}
  \E[\rho(W,\eta_0,\theta_0)\mid X]=0,
  \qquad
  \left.\frac{d}{dt}
  \E[\rho(W,\eta_0+th,\theta_0)\mid X]\right|_{t=0}
  =G(X)h(X),
\label{CMR}
\end{equation}

and write $\rho_0=\rho(W,\eta_0,\theta_0)$ and
$\Omega(X)=\E[\rho_0\rho_0'\mid X]$. Along a nuisance path
$\eta_t=\eta_0+th$ that holds $\theta_0$ and the marginal law
of $X$ fixed, differentiation gives
$\E[\rho_0s_h\mid X]=-G(X)h(X)$. When $\Omega$ is invertible,
the minimum conditional variance solution is
\[
  s_h(W)=R(W)'h(X),
  \qquad R(W)=-G(X)'\Omega(X)^{-1}\rho_0(W).
\]
Under regularity ensuring that these scores belong to the
nuisance tangent space, this supplies a conditional-score block;
see \citet[Theorem~4.1]{ChenSantos2018} for tangent-space
regularity conditions. We maintain in addition the second-order
pathwise regularity of Assumption~\ref{ass:regularity}.
Motivated by this structure, suppose a block of nuisance scores
has the form
\begin{equation}
  s_h(W)=R(W)'h(X),\qquad
  \E[R\mid X]=0,\qquad
  \Sigma(X)=\E[RR'\mid X],
  \label{eq:cond_score_crit}
\end{equation}
where $R\in\R^r$ and $h$ ranges over bounded measurable
$\R^r$-valued functions representing directions in $\mathcal{H}$.
Suppose the eigenvalues of $\Sigma$ are bounded above and away
from zero, and
\begin{equation}
  B_\phi(h,g)=\E[h(X)'C_\phi(X)g(X)]
  \label{eq:conditional-score-curvature}
\end{equation}
for a symmetric matrix function $C_\phi$ with integrable entries.
This additional condition makes curvature pointwise: it depends
on the directions only through their values at the same $X$.
Let $\mathcal X_a=\{x:P_{0,X}\{x\}>0\}$,
$\pi_x=P_{0,X}\{x\}$, and $\|A\|_F^2=\tr(A'A)$.

\begin{proposition}[Support and conditional variances]
\label{prop:support}
Suppose \eqref{eq:cond_score_crit} and
\eqref{eq:conditional-score-curvature} hold. This score block has
an $L_2$ correction if and only if $C_\phi=0$ almost surely on
the nonatomic part of $P_{0,X}$ and
\begin{equation}
  \sum_{x\in\mathcal X_a}
  \|\Sigma(x)^{-1/2}C_\phi(x)\Sigma(x)^{-1/2}\|_F^2<\infty.
  \label{eq:atomiccost}
\end{equation}
The series is the minimum squared kernel norm for this block.
The minimizing kernel is
\[
  k_\phi(W_i,W_j)=
  \sum_{x\in\mathcal X_a}
  \frac{\ind\{X_i=X_j=x\}}{\pi_x}
  R(W_i)'\Sigma(x)^{-1}C_\phi(x)\Sigma(x)^{-1}R(W_j).
\]
\end{proposition}

Thus, with nonatomic covariates, any nonzero pointwise curvature
on this block rules out an exact $L_2$ correction. A correction
for the full model must also correct this block.

\paragraph{Example: PLM, cont.}
For the PLM \eqref{eq:plm-model}, first-order orthogonal moments
have the form
\[
  \phi=\varepsilon r(D,X),\qquad
  \E[r\mid X]=0,\qquad J_\phi=\E[rV];
\]
see, e.g., \citet[Theorem~4.1]{ChenSantos2018}, applied with
$\theta=\theta_0$ fixed.

The general discussion above applies with $\eta=(c,\gamma)$ and
$\rho=(\omega_0\{Y-\theta D-c(X)\},D-\gamma(X))'$
in \eqref{CMR}, where
$\omega_0=\{\E[\varepsilon^2\mid D,X]\}^{-1}$ is held fixed.
Writing $\tau_0=\E[\omega_0\mid X]$ and
$v_0=\E[V^2\mid X]$, the score block
$s_h=\omega_0\varepsilon h_c+(V/v_0)h_\gamma$ gives,
on $\{v_0>0\}$,
\[
 G=-\operatorname{diag}(\tau_0,1),\qquad
 \Omega=\operatorname{diag}(\tau_0,v_0),\qquad
 C_\phi=\frac{\E[rV\mid X]}{2v_0}
 \begin{pmatrix}0&1\\1&0\end{pmatrix}.
\]
Since $J_\phi=\E[rV]\ne0$, $C_\phi$ is nonzero on a set
of positive probability. The proof uses localization and
bounded scores to obtain the following conclusion from
Proposition~\ref{prop:support}.

\begin{corollary}[No exact correction in the PLM]
\label{cor:plm-impossible}
Under the maintained regularity conditions, suppose that,
at the true law, for constants $v_->0$ and $C_4<\infty$,
\[
 \E[\varepsilon^2\mid D,X]\ge v_-,
 \qquad
 \E[\varepsilon^4\mid D,X]\le C_4
 \quad\text{almost surely}.
\]
If $P_{0,X}$ is nonatomic, every relevant first-order
orthogonal moment in the PLM has infinite correction cost.
\end{corollary}

Corollary~\ref{cor:plm-impossible} illustrates one of two situations.
In the PLM, the correction cost is infinite for every relevant
moment, so choosing a different moment cannot avoid it. A related
obstruction arises along covariate directions shared by the two
nuisance functions in DiD, where the curvature is identical for
every normalized moment (Theorem~\ref{thm:shared}). In other models,
the cost varies across moments. In binary choice, the
conditional-likelihood moment has zero curvature, and hence zero
cost, while moments that use additional first-order information
have infinite cost on continuous components of the covariates
(Section~\ref{sec:binary}). When the cost varies, the first-order
moment should be chosen together with its correction, which we do
in the next subsection.

\subsection{Joint choice of the moment and its correction}

So far we have considered correcting a given first-order moment. However, there may be many first-order orthogonal moments for the same target parameter. First-order moments that are first-order efficient need not minimize the variance once the second-order correction is taken into account. We now consider the joint choice of the first-order moment and its second-order correction.

We first introduce some objects related to first-order efficiency results. The efficient score is $\ell_{\eff}=\ell-\Pi_{\cT_\eta}\ell$ and the first-order efficient information is $I_1=\E[\ell_{\eff}^2]$.
If $I_1=0$, every first-order orthogonal moment has zero target
sensitivity. Suppose henceforth that $I_1>0$, and define the first-order efficient influence function
$\phi_{\eff}=\ell_{\eff}/I_1$.
The full class of normalized first-order values is
\begin{equation}
 \cF=\{\phi\in\cN:\E[\phi\ell]=1\}
 =\phi_{\eff}+\cA.
 \label{eq:fullclass}
\end{equation}
This is the first-order class characterized, under the
corresponding attainability conditions, by \citet{AE2026}. The additional requirement here is finite correction cost. Every normalized admissible first-order moment is the efficient moment plus $a \in \mathcal{A}$. Adding $a \in \mathcal{A}$ increases first-order variance but may reduce the curvature to be corrected. The optimization problem we aim to solve is
\[
V_{2,n}^* = \inf_{\substack{(\phi,k)\in\mathcal{C},\\
        \E[\phi\ell]=1}} 
 \frac{\E[\phi^2]}n+\frac{2\E[k^2]}{n(n-1)}.
\]
To characterize the solution, equip $\cN\times\cK$ with the weighted norm $\|(\phi,k)\|_n^2 = \E[\phi^2]+\frac{2}{n-1}\E[k^2]$. The objective function above is $\|(\phi,k)\|_n^2/n$. Analogous to first-order efficiency theory, the efficient second-order variance is characterized by a projection, now onto the space of compatible pairs $\mathcal{C}$. Let $v_n$ be the projection of $(\ell_{\eff},0)$ onto $\mathcal{C}$, which is proven to be closed below, in the norm $\|\cdot\|_n$. Projection preserves target sensitivity:
$\E[\phi\ell]=\langle v_n,(\phi,k)\rangle_n$
for every $(\phi,k)\in\mathcal C$.
Cauchy--Schwarz therefore shows that multiples of $v_n$
maximize squared target sensitivity relative to variance,
giving
\begin{equation}
 I_{2,n}\equiv\|v_n\|_n^2
 =\sup_{(\phi,k)\in\mathcal{C}\setminus\{0\}}
 \frac{\{\E[\phi\ell]\}^2}{\E[\phi^2]+\frac{2}{n-1}\E[k^2]}.
 \label{eq:information2}
\end{equation}
Second order information is $I_{2,n}$ in the sense of the result below.

\begin{theorem}[Minimum oracle variance over the full class]
\label{thm:joint}
The space $\mathcal{C}$ is closed. A normalized compatible pair
exists if and only if $I_{2,n} >0$. When it exists, the minimum oracle variance is attained uniquely and equals
\begin{align}
 V_{2,n}^*
 &=\frac1{nI_{2,n}}\nonumber\\
 &=\frac1{nI_1}
 +\frac1n
 \min_{\substack{a\in\cA,\ k\in\cK\\
        \mathcal Rk=B_{\phi_{\eff}}+B_a}}
 \{\|a\|^2+\frac{2}{n-1}\|k\|^2\}
 \label{eq:joint-decomp}\\
 &=\min_{\phi\in\cF}
 \left\{
 \frac{\E[\phi^2]}n+
 \frac{2c_\phi^2}{n(n-1)}
 \right\}.
 \label{eq:profile}
\end{align}
Infinite $c_\phi$ excludes a value from the last minimization.
The minimizing pair is $v_n/I_{2,n}$.
If this pair admits an admissible extension, as defined
in Appendix~\ref{app:extensions}, that family is
second-order orthogonal at $\lambda_0$ and attains
$V_{2,n}^*$.
\end{theorem}

Theorem \ref{thm:joint} gives three characterizations of the minimum oracle variance in the joint problem. The first is a second-order analogue of the first-order efficient bound. The second is a decomposition into the first-order efficiency bound and the minimum combined cost of departing from the efficient first-order moment and correcting its nuisance curvature. The third profiles out the second-order correction and reduces the problem to choosing the first-order moment. Relative to \citet{Hansen1982} and
\citet{Chamberlain1987}, both the variance criterion and the
admissible class now account for second-order correction.

Theorem~\ref{thm:joint} imposes the second-order restrictions in every
pair of nuisance directions. We now impose them for finitely many
directions $h_1,\ldots,h_d\in\mathcal H$ and define
\[
 \mathcal C_d=\bigl\{(\phi,k)\in\cN\times\cK:
 (\mathcal Rk)(h_j,h_l)=B_\phi(h_j,h_l),\ 1\le j,l\le d\bigr\}.
\]
As $d$ increases, the restrictions accumulate and $\mathcal C_d$
shrinks; we choose the directions so that
$\bigcap_d\mathcal C_d=\mathcal C$. Let $V_{2,n,d}^*$ be the minimum
oracle variance over normalized pairs in $\mathcal C_d$, with minimizer
$(\phi_{n,d}^*,k_{n,d}^*)$, and set $V_{2,n,d}^*=\infty$ if no such
pair exists. Only the second-order restrictions are finite: the
first-order moment ranges over the full class $\cN$, and all
first-order restrictions are kept. This differs from constructions for parametric models \citep{BJN2026,CST2026}.

\begin{theorem}[Finite restrictions and infinite-dimensional cost]
\label{thm:limit}
For each fixed $n\ge2$, a normalized compatible pair exists if and
only if $\sup_d V_{2,n,d}^*<\infty$. In that case, as $d\to\infty$,
\[
 V_{2,n,d}^*\uparrow V_{2,n}^*,
 \qquad
 (\phi_{n,d}^*,k_{n,d}^*)\longrightarrow(\phi_n^*,k_n^*)
 \quad\text{in }\|\cdot\|_n.
\]
Let $\cD=\{\phi\in\cN:c_\phi<\infty\}$ be the linear space of
correctable first-order moments. Then, as $n\to\infty$,
\begin{equation}
 I_{2,n}\uparrow\|\Pi_{\overline{\cD}}\ell_{\eff}\|^2 .
 \label{eq:information-recovery}
\end{equation}
Hence the full first-order information $I_1=\|\ell_{\eff}\|^2$ is
recovered if and only if $\ell_{\eff}\in\overline{\cD}$; this need not
imply $\ell_{\eff}\in\cD$.
\end{theorem}

The first part shows that the full problem is the limit of finite
ones, and that finite minima growing without bound signal that no
exact correction exists. The second part concerns large samples. The
weight $2/(n-1)$ on the correction cost vanishes, so the
recoverable information is that of the projection of the efficient
score on $\overline{\cD}$, the first-order moments that are
mean-square limits of correctable ones. The efficient moment may have
infinite cost and still be such a limit, as in binary choice with a
covariate taking infinitely many values (Section~\ref{sec:binary}).
The next subsection solves the finite problems explicitly and allows
some curvature to remain.

\subsection{Finite optimization and the variance--bias frontier}

Choose nuisance directions $h_1,\ldots,h_d$ with linearly
independent scores $s=(s_{h_1},\ldots,s_{h_d})'$, and let
$\Gamma_d=\E[ss']$ be a positive definite score covariance matrix.
All first-order restrictions remain in force.
Write $B_\phi$ also for the matrix with entries
$B_\phi(h_j,h_l)$, and define
\[
 C_\phi=\Gamma_d^{-1/2}B_\phi\Gamma_d^{-1/2}.
\]
The matrix $C_\phi$ collects the curvature of $\phi$ relative to the information in the selected nuisance directions. In fact, if the scores are orthogonal, elements of $C_\phi$ are analogous to the curvature regression coefficients in \eqref{eq:spectral-regression-coefficients}. Denote the normalized scores by $z=\Gamma_d^{-1/2}s$, so $\E[zz']=I_d$. 

\begin{proposition}[Full-class finite representer]
\label{prop:finite}
For fixed $\phi$, the minimum kernel satisfying these
second-order restrictions is
\[
 k_{\phi,d}(w_i,w_j)
 =z(w_i)'C_\phi z(w_j),
 \qquad
 \|k_{\phi,d}\|^2=\|C_\phi\|_F^2.
\]
Let $C_{\rm eff}=C_{\phi_{\eff}}$ and define
$L:\cA\to\mathbb S^d$ by $La=C_a$, where $\mathbb S^d$
is the space of symmetric matrices with Frobenius inner
product. The minimizer and value at the minimum of
\begin{align*}
nV_{2,n,d}^* &= I_1^{-1} + \frac{2}{n-1} \min_{a \in \cA} \{\|-C_{\rm eff} - La\|_F^2 +  \frac{n-1}{2} \|a\|^2 \},
\end{align*}
are
\begin{align}
 a_{n,d}^*
 &=-\frac{2}{n-1}(I+\frac{2}{n-1}L^*L)^{-1}L^*C_{\rm eff},
 \label{eq:finite-a}\\
 nV_{2,n,d}^*
 &=I_1^{-1}
 +\frac{2}{n-1}
 \langle C_{\rm eff},(I+\frac{2}{n-1}LL^*)^{-1}C_{\rm eff}\rangle_F,
 \label{eq:finite-value}
\end{align}
where $L^*$ is the adjoint of $L$ for the $L_2$ and Frobenius inner
products. 
\end{proposition}
The adjustment $a_{n,d}^*$ has the form of a ridge regression solution: we approximate $-C_{\rm eff}$ through $La$ penalizing departures from the first-order efficient moment. The ridge penalty, $(n-1)/2$, increases with sample size. Consequently, for a fixed finite collection of nuisance directions, the optimal moment approaches the first-order efficient moment. Optimization over the full first-order class therefore
reduces to a finite-dimensional quadratic problem.
This criterion adds full first-order orthogonality with nonparametric nuisances and joint variance optimization to
finite-dimensional orthogonalization constructions such
as \citet{BJN2026}.

The case $L^*C_{\rm eff}=0$ has a simple interpretation. Since
$\langle La,C_{\rm eff}\rangle_F=\E[a\,L^*C_{\rm eff}]$, it holds exactly when every
change of the first-order moment alters its information-normalized
curvature only in directions orthogonal to $C_{\rm eff}$: the change can add
curvature but never offset it. Then $a_{n,d}^*=0$, the efficient first-order moment is
optimal, and $nV_{2,n,d}^*=I_1^{-1}+\frac{2}{n-1}\|C_{\rm eff}\|_F^2$ is the cost of
correcting it. This is the case of shared sensitivity discussed after
Corollary~\ref{cor:plm-impossible}. It holds in particular when every
second-order score $u_{h_j,h_l}$ lies in the first-order tangent space
$\cT$, so that $L=0$ and all moments have the same curvature. When
$L^*C_{\rm eff}\ne0$, some moments are less curved where correction is costly,
and the optimum departs from the efficient moment, as in binary choice.

We next allow some curvature to remain, reducing correction variance at the cost of allowing for some quadratic bias.
A symmetric matrix $M$ defines a kernel
$k_M(w_i,w_j)=z(w_i)'Mz(w_j)$ with squared norm
$\|M\|_F^2$. Proposition \ref{prop:finite} shows that exact correction is achieved by $M=C_\phi$. More generally, we can use any symmetric matrix $M$ instead of $C_\phi$ to define a kernel. Then, $M \neq C_\phi$ leaves some curvature uncorrected in exchange for potentially lower correction variance. 

Consider a nuisance displacement in the span of $h_1,...,h_d$, $h = \sum_j t_j h_j$, with score $s_h = \sum_j t_j s_{h_j} \equiv \mathbf t' s$ and information $\mathbf t'\Gamma_d \mathbf t$. Define its information coordinates $v=\Gamma_d^{1/2}\mathbf t$, so that $s_h = v'z$ and $\E[s_h^2] = \|v\|_2^2$. By these definitions, bilinearity of $u_{h,g}$, $\E[zz'] = I_d$ and the definition of $C_\phi$
\[
B_\phi(h,h) = \mathbf t'B_\phi \mathbf t, \quad (\mathcal{R}k_M)(h,h) = v'Mv, \quad B_\phi(h,h) - (\mathcal{R}k_M)(h,h) = v'(C_\phi-M)v. 
\]
We care about the remaining curvature, $v'(C_\phi-M)v$, relative to information. Since $C_\phi-M$ is symmetric, the maximum relative curvature is
\begin{equation}
\sup_{v \neq 0} \frac{|v'(C_\phi-M)v|}{\|v\|_2^2}
 =\|C_\phi-M\|_{\op}.
 \label{eq:quadraticbudget}
\end{equation}
For a curvature allowance $\delta\ge0$, consider
\begin{equation}
 \min_{a\in\cA,\ M\in\mathbb S^d}
 \{I_1^{-1}+\|a\|^2+\frac{2}{n-1}\|M\|_F^2\}
 \quad\text{subject to}\quad
 \|C_{\rm eff}+La-M\|_{\op}\le\delta.
 \label{eq:approxchoice}
\end{equation}
The objective is $n$ times the oracle variance of the pair $(\phi,k_M)$, while the constraint bounds quadratic bias relative to Fisher information. Now we show that the solution is unique and can be computed by soft-thresholding the eigenvalues of the information normalized curvature matrix at budget $\delta$.

\begin{proposition}[The finite variance--bias frontier]
\label{prop:frontier}
Problem \eqref{eq:approxchoice} has a unique minimizer
$(a_{n,d,\delta}^*,\allowbreak M_{n,d,\delta}^*)$, and
$a_{n,d,\delta}^*\in\operatorname{range}(L^*)$. If
$C_{\rm eff}+La_{n,d,\delta}^*=U\operatorname{diag}(\nu_j)U'$, then
\begin{equation}
 M_{n,d,\delta}^*=U\operatorname{diag}
 \{\operatorname{sign}(\nu_j)
   (|\nu_j|-\delta)_+\}U'.
 \label{eq:soft}
\end{equation}
The minimum value is continuous and nonincreasing in $\delta$.
At $\delta=0$, $a_{n,d,0}^*=a_{n,d}^*$ and the minimum value is
$nV_{2,n,d}^*$; once $\delta\ge\|C_{\rm eff}\|_{\op}$, it is $I_1^{-1}$.
\end{proposition}
Here $(x)_+=\max\{x,0\}$ and $\operatorname{sign}(0)=0$.
A curvature eigenvalue with magnitude at most $\delta$
requires no correction; larger eigenvalues are corrected
only enough to leave magnitude $\delta$.
The frontier is exact for the specified restrictions and
characterizes the minimum oracle variance for each bound $\delta$ on information-normalized curvature. Quadratic bias outside the nuisance span and higher-order Taylor terms require separate bounds. For the PLM, in the next section, we express the remaining
bias as an expected product involving the regression
error left uncorrected. That section constructs
feasible kernels and controls the additional errors
introduced by estimation.
%%%%%%%% SECTION 4 %%%%%%%%%%%
\section{Automatic Approximate Correction and Inference}
\label{sec:approx}

Section~\ref{sec:theory} characterized when second-order
corrections exist and what they cost. We now construct
feasible corrections when the nuisance parameters are
nonparametric regression functions, or more generally, orthogonal projections. We approximate the
nuisance functions in finite-dimensional spaces, while
their true values may lie outside these spaces. The
approximation space and the amount of regularization
determine which components of the second-order bias are
removed and at what variance cost. Motivated by the PLM example, here we take the first-order
moment as given and select its approximate correction.
The applications in Section~\ref{sec:applications}
examine the additional role of first-order moment choice. The partially linear model is the
running example; Procedure~\ref{proc:plm} summarizes the resulting
estimator and confidence interval, and Section~\ref{sec:numerical}
reports simulations for it. Appendix~\ref{app:auto} contains the
proofs and follows the order of this section;
Appendix~\ref{app:plm-sim} gives the simulation
specifications.

\subsection{First-order orthogonal moments and their remaining bias}
\label{sec:auto-bias}

Suppose $\theta_0$ is identified as the unique solution of $\E[m(W,\gamma_0,\theta)]=0$ in $\theta\in\Theta$, where $\gamma_0$ is the
orthogonal projection of a square-integrable variable $Z$
onto a closed linear space $\Gamma\subseteq L_2(P_{0,X})$:
\begin{equation}
\E[\{Z-\gamma_0(X)\}a(X)]=0,
\qquad a\in\Gamma.
\label{eq:auto-projection}
\end{equation}
Conditional expectations correspond to $\Gamma=L_2(P_{0,X})$.
We treat one projection for simplicity; the construction
extends to finitely many.

Let $F(\gamma)=\E[m(W,\gamma,\theta_0)]$ be twice Fr\'echet
differentiable at $\gamma_0$ with respect to the norm of $L_2(P_{0,X})$:
there are a bounded linear functional $F'(\gamma_0)$ and a bounded
symmetric bilinear form $F''(\gamma_0)$ on $\Gamma$ such that
\[
 F(\gamma_0+h)=F(\gamma_0)+F'(\gamma_0)[h]
 +\tfrac12F''(\gamma_0)[h,h]+o(\|h\|^2)
 \qquad(\|h\|\to0).
\]
By the Riesz representation theorem,
$F'(\gamma_0)[h]=\E[\alpha_0(X)h(X)]$ for a unique $\alpha_0\in\Gamma$. The first-order
orthogonal moment is \citep{LR2022}
\begin{equation}
\phi(W,\gamma,\alpha,\theta)
=m(W,\gamma,\theta)+\alpha(X)\{Z-\gamma(X)\}.
\label{eq:auto-first-moment}
\end{equation}
The term involving $\alpha$ removes the linear effect of
estimating $\gamma$, and \eqref{eq:auto-projection} gives
orthogonality with respect to $\alpha$. Write
$\phi_0=\phi(W,\gamma_0,\alpha_0,\theta_0)$ and
$J_0=-\left.\partial_\theta\E[\phi(W,\gamma_0,\alpha_0,\theta)]\right|_{\theta=\theta_0}$.

To see what remains, let $h_\gamma=\gamma-\gamma_0$,
$h_\alpha=\alpha-\alpha_0$, and $h=(h_\gamma,h_\alpha)$.
The projection equation and a second-order expansion give
\begin{equation}
\begin{split}
\E[\phi(W,\gamma,\alpha,\theta_0)]
&=F(\gamma)-F(\gamma_0)-\E[\alpha(X)h_\gamma(X)]\\
&=\underbrace{\tfrac12F''(\gamma_0)[h_\gamma,h_\gamma]
 -\E[h_\alpha(X)h_\gamma(X)]}_{B_{\phi_0}(h,h)}
 +o(\|h_\gamma\|^2).
\end{split}
\label{eq:auto-curvature}
\end{equation}
There is no second-order component in $h_\alpha$ because $\alpha$ enters linearly in the first-order moment. The remaining bias combines the curvature of the original
moment with the product of the errors in $\gamma$ and in
its Riesz representer $\alpha$. The second-order
correction must account for both.

For pointwise moments
$m(W,\gamma,\theta)=\bar m(W,\gamma(X),\theta)$, define
\[
 \dot m_0(W)
 =\left.\partial_z\bar m(W,z,\theta_0)\right|_{z=\gamma_0(X)},
 \qquad
 \ddot m_0(W)
 =\left.\partial_z^2\bar m(W,z,\theta_0)\right|_{z=\gamma_0(X)}.
\]
Write $V_Z=Z-\gamma_0(X)$ and
$a_0(X)=\E[\ddot m_0\mid X]$. When
$\Gamma=L_2(P_{0,X})$, the representer is
$\alpha_0(X)=\E[\dot m_0\mid X]$.

\begin{proposition}[Pointwise moments]\label{prop:pointwise-cost}
Suppose $\Gamma=L_2(P_{0,X})$, $P_{0,X}$ is nonatomic,
and the law is locally unrestricted,
$\mathcal T=L_2^0(P_0)$. Under the maintained admissibility
and pathwise regularity conditions, assume $J_0\ne0$,
$0<\E[\phi_0^2]<\infty$, and that $a_0$ and
$\E[V_Z^2+(\dot m_0-\alpha_0)^2\mid X]$ are bounded,
with $\E[V_Z^2\mid X]$ bounded away from zero.
For nuisance directions induced by target-preserving
law paths,
\[
 c_{\phi_0}<\infty
 \quad\Longleftrightarrow\quad
 \dot m_0-\alpha_0(X)=-\tfrac12a_0(X)V_Z
 \quad\text{a.s.},
\]
in which case $c_{\phi_0}=0$.
\end{proposition}

\begin{corollary}[Nonlinear regression functionals]
\label{cor:nonlinear-regression-cost}
Under the conditions of Proposition~\ref{prop:pointwise-cost},
let
\[
 m(W,\gamma,\theta)=f(\gamma(X))-\theta.
\]
Then $c_{\phi_0}<\infty$ if and only if
$f''(\gamma_0(X))=0$ almost surely, and the finite cost
is zero. 
\end{corollary}
\paragraph{Example: PLM, cont.}
In the PLM \eqref{eq:plm-model}, take $Z=D$, so that
$\gamma_0(X)=\E[D\mid X]$, and
$m(W,\gamma,\theta)=(Y-\theta D)\{D-\gamma(X)\}$. Then
$F'(\gamma_0)[h]=-\E[(Y-\theta_0D)h(X)]$, so
$\alpha_0(X)=-\E[Y-\theta_0D\mid X]=-c_0(X)$ and
\[
\phi(W,\gamma,\alpha,\theta)
=\{Y-\theta D+\alpha(X)\}\{D-\gamma(X)\}.
\]
Since $\alpha_0=\theta_0\gamma_0-\mu_0$ with $\mu_0(X)=\E[Y\mid X]$, we
estimate the regressions $\mu_0$ and $\gamma_0$, which do not depend on
$\theta$, and evaluate the representer as $\alpha(\cdot,\theta)=\theta\gamma-\mu$.
The moment then becomes
$\{Y-\mu(X)-\theta(D-\gamma(X))\}\{D-\gamma(X)\}$, the partialling-out
moment of \citet{Robinson1988}, and the estimating
equation is linear in $\theta$.
Since $F$ is affine, $F''=0$ and \eqref{eq:auto-curvature}
is exact:
\[
\E[\phi(W,\gamma,\alpha,\theta_0)]
=B_{\phi_0}(h,h)
=-\E[h_\alpha(X)h_\gamma(X)].
\]
The remaining bias is thus the inner product of the
errors in two nonparametric regressions. 

\subsection{Constructing the pairwise correction}
\label{sec:auto-kernel}

We construct a pairwise correction that matches the curvature in
\eqref{eq:auto-curvature} using the projection equations
that identify $\gamma_0$ and $\alpha_0$ in place of
likelihood scores. Choose linearly independent functions
$p=(p_1,\ldots,p_d)'$ in $\Gamma$, with Gram matrix
$G_0=\E[pp']$. For $a\in\Gamma$, let
$D_\gamma m(W,\gamma,\theta)[a]$ be the derivative of
$m(W,\gamma+ta,\theta)$ at $t=0$, and let $D_\gamma m[p]$
be the vector with components $D_\gamma m[p_j]$. For targets near $\theta_0$, write $\alpha_0(\cdot,\theta)$ for the
Riesz representer of $h\mapsto\E[D_\gamma m(W,\gamma_0,\theta)[h]]$,
with $\alpha_0(\cdot)=\alpha_0(\cdot,\theta_0)$.
For $\eta=(\gamma,\alpha)$, define
\[
u(W,\gamma)=p(X)\{Z-\gamma(X)\},\qquad
v(W,\gamma,\alpha,\theta)=D_\gamma m(W,\gamma,\theta)[p]-p(X)\alpha(X),
\]
and the $d\times d$ matrix $H_0$ with entries
$(H_0)_{jl}=F''(\gamma_0)[p_j,p_l]$. Both vectors are residuals from the equations that identify $\gamma_0$
and $\alpha_0$, weighted by $p$. They have mean zero at the truth, and
their means away from it reveal the nuisance errors along $p$; in the
PLM, they are the treatment and outcome residuals multiplied by $p(X)$.
For a symmetric matrix $M$, the pairwise kernel is
\begin{equation}
k(W_i,W_j,\eta,\theta;M)
=-\tfrac12\{u_i'Mv_j+v_i'Mu_j+u_i'MH_0Mu_j\},
\label{eq:auto-general-kernel}
\end{equation}
with $u_i=u(W_i,\gamma)$ and $v_i=v(W_i,\gamma,\alpha,\theta)$, and
$k_0(M)=k(\cdot,\cdot,\eta_0,\theta_0;M)$. It is
symmetric and degenerate at the truth. The first two
terms correct the interaction between the errors in
$\gamma$ and $\alpha$; the third corrects the curvature
of the original moment. There is no quadratic term in $v$ because $\alpha$ enters linearly.

The matrix $M$ determines how much bias is removed. Let
$T_Ma=p'M\E[pa]$ and $\widetilde h_{\gamma,M}=(I-T_M)h_\gamma$. With $M=G_0^{-1}$, $T_M$ is the population least-squares
projection onto $\operatorname{span}(p)$, and only the
regression error outside this space contributes to the
remaining bias. Regularizing $M$ can trade additional
bias for a lower variance cost.

\begin{proposition}[Remaining curvature]
\label{prop:auto-curvature}
Under the differentiability and admissibility conditions
of Section~\ref{sec:theory},
\begin{equation}
(B_{\phi_0}-\mathcal Rk_0(M))(h,h)
=\tfrac12F''(\gamma_0)[\widetilde h_{\gamma,M},\widetilde h_{\gamma,M}]
-\langle h_\alpha,\widetilde h_{\gamma,M}\rangle.
\label{eq:auto-remaining}
\end{equation}
In particular, $M=G_0^{-1}$ removes the quadratic bias
for every direction with
$h_\gamma\in\operatorname{span}(p)$ (i.e., $\widetilde h_{\gamma,M} \equiv 0$).
\end{proposition}

\paragraph{Example: PLM, cont.}
Here $u(W,\gamma)=p(X)\{D-\gamma(X)\}$,
$v(W,\gamma,\alpha,\theta)=-p(X)\{Y-\theta D+\alpha(X)\}$, and $H_0=0$, so
\eqref{eq:auto-general-kernel} is,  with $K_M(x,x')=p(x)'Mp(x')$, given by:
\[
k(W_i,W_j,\eta,\theta;M)
=\tfrac12K_M(X_i,X_j)\{e_i\widetilde D_j+e_j\widetilde D_i\},
\]
where $e_i=Y_i-\theta D_i+\alpha(X_i)$ and
$\widetilde D_i=D_i-\gamma(X_i)$. The remaining bias is
exactly
\begin{equation}
(B_{\phi_0}-\mathcal Rk_0(M))(h,h)
=-\E[h_\alpha(X)(I-T_M)h_\gamma(X)].
\label{eq:plm-remaining-bias}
\end{equation}
With $M=G_0^{-1}$, the correction removes the product of
the least-squares projections of the two errors onto
$\operatorname{span}(p)$, and the remaining bias is the
product of their approximation errors.

\subsection{Estimating the projection correction and other nuisances by cross-fitting}
\label{sec:auto-estimate}

Estimating $G_0^{-1}$ is difficult when the regressors are
numerous or highly correlated. We use two regularized
estimators: ridge and regressor selection. To estimate this and other nuisances we use cross-fitting, which avoids own-observation bias and permits nuisance estimation over non-Donsker classes. 

\paragraph{Cross-fitting.}
Following \citet{ET2026}, the
fits used for a pair $(i,j)$ are trained without either observation;
all pairs in the same block share these fits. For $I\subseteq\{1,\ldots,n\}$, let $I^c$ denote its complement.
Write $\mathcal P_b$ for pair block $b$ and $I_b$ for all observation
indices appearing in that block; its training sample is $I_b^c$.
The following constructions partition all distinct ordered pairs,
including both orientations, into a fixed number $B$ of blocks.

\begin{figure}[tbp]
\centering
\includegraphics[width=\textwidth]{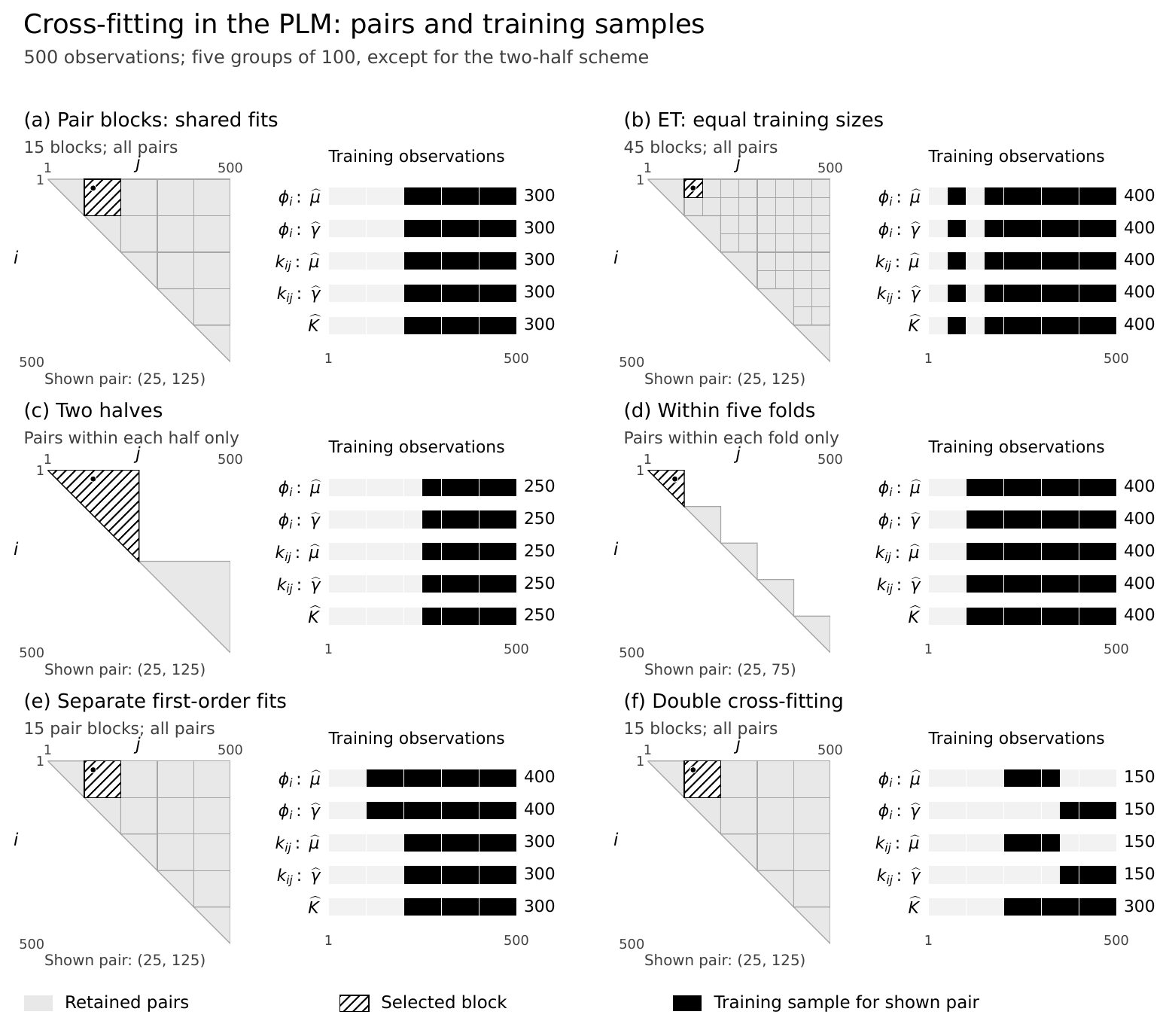}
\caption{Cross-fitting schemes for the PLM, with $n=500$.
Each triangular plot displays retained pairs; hatching identifies
one evaluation block. The black bars show the observations used
to train its outcome regression $\mu$, treatment regression
$\gamma$, and correction kernel $K$. Rows labeled $\phi_i$ refer
to the first-order moment at $i$; rows labeled $k_{ij}$ refer to
the pairwise correction. Numbers at the right are training sizes.
Panels (a), (b), (e), and (f) retain all pairs; (c) and (d) retain
only pairs within halves or folds. Panel (b) equalizes training
sizes at 400 across all blocks. In (e), the first-order fit at $i$
excludes its own fold, while the correction excludes both
participating folds. In (f), the two regressions use disjoint
halves of the available training sample; the kernel uses that
full sample. Pairs are displayed once, after arranging observations
by group; the estimator includes both orientations.}
\label{fig:cross-fitting-schemes}
\end{figure}

\emph{Group construction.}
Independently of the data, partition the observations into a fixed
number $L\ge3$ of approximately equal groups $\mathcal I_1,\ldots,\mathcal I_L$.
Form one block of distinct pairs within each group and one block
between each two groups. There are $B=L(L+1)/2$ blocks (see Figure~\ref{fig:cross-fitting-schemes}).
A within-group block has $I_b=\mathcal I_g$ and trains on $\mathcal I_g^c$ (triangles);
a between-group block has $I_b=\mathcal I_g\cup \mathcal I_h$ and trains on
$(\mathcal I_g\cup \mathcal I_h)^c$ (squares). The respective training sizes are approximately
$n(1-1/L)$ and $n(1-2/L)$.

\emph{Equal training sizes.}
The refinement of \citet{ET2026} divides each between-group block
into four. Split $\mathcal I_g$ and $\mathcal I_h$ into halves
$\mathcal I_g^{(1)},\mathcal I_g^{(2)}$ and $\mathcal I_h^{(1)},\mathcal I_h^{(2)}$, and form a block
for each combination of halves. Such a block has
$I_b=\mathcal I_g^{(a)}\cup \mathcal I_h^{(c)}$, with $a,c\in\{1,2\}$.
Leave the within-group blocks unchanged. This gives
$B=L+4\binom{L}{2}=L(2L-1)$ blocks, each involving approximately
$n/L$ observations and leaving approximately $n(1-1/L)$ for
training. These sizes are exact when $n$ is divisible by $2L$.
The numbers of pairs per block need not be equal.

Panels (a) and (b) of Figure~\ref{fig:cross-fitting-schemes}
illustrate these constructions.
The simulations use the group construction with $L=5$ and $B=15$.

\paragraph{Example: PLM, cont.}
For the highlighted block in panel (a),
$I_b=\mathcal I_1\cup \mathcal I_2$. Estimate $\mu_0(x)=\E[Y\mid X=x]$,
$\gamma_0(x)=\E[D\mid X=x]$, and the correction kernel
$\widehat K_b$ using only the 300 observations in $I_b^c$.
For $i\in I_b$, define
\[
\widehat V_{i,b}=D_i-\widehat\gamma_b(X_i),
\qquad
\widehat e_{i,b}(\theta)
=Y_i-\widehat\mu_b(X_i)-\theta\widehat V_{i,b}.
\]
Every pair $(i,j)$ with $i\in \mathcal I_1$ and $j\in \mathcal I_2$ uses these
same fits in its first-order moment and correction:
\begin{align*}
\widehat\phi_{i,b}(\theta)
&=\widehat e_{i,b}(\theta)\widehat V_{i,b},\\
\widehat k_b(W_i,W_j;\theta)
&=\tfrac12\widehat K_b(X_i,X_j)
\{\widehat e_{i,b}(\theta)\widehat V_{j,b}
+\widehat e_{j,b}(\theta)\widehat V_{i,b}\}.
\end{align*}
One set of fits serves the $100\times100$ unordered pairs in this
rectangle, each included in both orientations. The correction targets
the quadratic bias of the fitted moment it accompanies.
In panel (b), $I_b=\mathcal I_1^{(1)}\cup \mathcal I_2^{(1)}$;
the fits use 400 observations, including the unused halves of
$\mathcal I_1$ and $\mathcal I_2$, and serve $50\times50$ unordered pairs.

Panels (c) and (d) retain only pairs within two halves or five folds.
Panel (e) uses ordinary five-fold fits for the first-order moment
and pair-block fits for the correction, which then targets the bias
of a different fit. Panel (f) trains $\widehat\mu_b$ and
$\widehat\gamma_b$ on disjoint halves of $I_b^c$; both moment terms
share these fits, and the kernel uses all of $I_b^c$.

For the two all-pairs constructions, define
$\mathcal H_b=\mathcal H_n\vee\sigma(W_i:i\in I_b^c)$, where
$\mathcal H_n$, independent of the main sample, collects the
partition, algorithmic randomization, and any auxiliary data
used in the construction. Auxiliary data are not required. The block's first-step estimators and correction matrices are $\mathcal H_b$-measurable,
and its main-sample Gram matrix is
\[
\widehat G_b=\frac{1}{|I_b^c|}
\sum_{i\in I_b^c}p(X_i)p(X_i)'.
\]
If an independent auxiliary sample is available, it may
instead be used to learn the correction matrices.

\paragraph{Ridge.}
Let $\widehat U_{+,b}$ contain the eigenvectors of
$\widehat G_b$ with eigenvalues above a numerical
threshold, and let $\widehat\Lambda_{+,b}$ be the diagonal
matrix of these eigenvalues. For $\zeta\ge0$,
\begin{equation}
\widehat M_b(\zeta)
=\widehat U_{+,b}(\widehat\Lambda_{+,b}+\zeta I)^{-1}
\widehat U_{+,b}',
\qquad
\widehat K_b(x,x';\zeta)=p(x)'\widehat M_b(\zeta)p(x').
\label{eq:auto-ridge}
\end{equation}
This is the regularized inverse Gram matrix of ridge
regression. Ridge thus shrinks the correction
in directions with small eigenvalues, trading some
quadratic bias for lower variance.

\paragraph{Selected projection.}
Alternatively, select a subset $S_b$ of the regressors
using $I_b^c$, with selected vector $p_{S_b}$ and
sample Gram matrix $\widehat G_{S_b,b}$, and use the
least-squares projection kernel
\begin{equation}
\widehat K_{S,b}(x,x')
=p_{S_b}(x)'\widehat G_{S_b,b}^{-1}p_{S_b}(x').
\label{eq:auto-selected-kernel}
\end{equation}
Selection corrects fully along the retained directions
and leaves the others uncorrected, whereas ridge shrinks
all directions continuously. In the PLM, candidate
subsets come from the least-angle regression (LAR) paths
for $Y$ and $D$ on $I_b^c$.

\paragraph{Families and candidates.}
Varying the tuning parameter of each regularization gives a family of
candidate corrections: the ridge family varies $\zeta$, and the
selected-projection family varies the number of selected regressors.
Each candidate is one kernel, and ADA (Section~\ref{sec:auto-selection})
selects one candidate from the two families. Let $\tau$ index candidates
uniquely across families and $d_\tau$ denote their size parameter
(nominal for ridge), keeping $d$ for the full dictionary dimension.
In the PLM simulations of Section~\ref{sec:numerical}, for example,
$p$ contains scaled cosines of $x$; the ridge candidates use
$\zeta_\tau=(1+d_\tau)^{-2}$, which corrects roughly the first
$d_\tau$ cosines, for $d_\tau$ between 1 and 128, and the
selected-projection candidates project onto the first $d_\tau$ cosines
in the LAR order, for $d_\tau$ between 2 and 96. Write each candidate as
$\widehat K_{b,\tau}(x,x')=p(x)'\widehat M_{b,\tau}p(x')$;
for selected projection, write $S_{b,\tau}$ for its selected set;
$\widehat M_{b,\tau}$ contains $\widehat G_{S_{b,\tau},b}^{-1}$ in the selected coordinates and
zeros elsewhere. At the population level, both kernels
define positive contractions
(Lemma~\ref{lem:auto-operators}); the inference theory
below accounts separately for the error from estimating
them.

\subsection{Cross-fitted estimator}
\label{sec:auto-estimator}

Let $b(i,j)$ denote the block containing $(i,j)$, and write
\[
 \widehat\phi_{i,b}(\theta)=\widehat\phi_b(W_i,\theta),\qquad
 \widehat k_{ij,\tau}(\theta)=\widehat k_{b(i,j),\tau}(W_i,W_j,\theta),
 \qquad \widehat K_{ij,\tau}=\widehat K_{b(i,j),\tau}(X_i,X_j).
\]
Here $\widehat\phi_b$ and $\widehat k_{b,\tau}$ are computed with
block $b$'s first-step estimators. The block index on
$\widehat\phi_{i,b}$ is retained because its fit can change with
observation $i$'s partner. As in the all-pairs cross-fitting of \citet{ET2026}, each
distinct pair is used once, with nuisance estimates trained without
either of its observations; both orientations are included to
simplify notation. The estimator $\widehat\theta_\tau$ solves
\begin{equation}
\widehat{\mathcal M}_\tau(\theta)
=\frac1{n(n-1)}\sum_{b=1}^B\sum_{(i,j)\in\mathcal P_b}
\bigl\{\widehat\phi_{i,b}(\theta)
-\widehat k_{ij,\tau}(\theta)\bigr\}=0.
\label{eq:auto-equation}
\end{equation}
The first-order term and the correction of each pair share the same
fit, so the correction targets the bias of the fitted moment that it
accompanies. Let $\widehat J_\tau=-\partial_\theta\widehat{\mathcal
M}_\tau(\widehat\theta_\tau)$ be the estimated sensitivity.

\paragraph{Example: PLM, cont.}
Let $\widehat\mu_b$ and $\widehat\gamma_b$ estimate $\E[Y\mid X]$ and
$\E[D\mid X]$ on $I_b^c$, and implement the representer as
$\widehat\alpha_b(\cdot,\theta)=\theta\widehat\gamma_b-\widehat\mu_b$. For
$i\in I_b$, put
\[
\widehat R_{i,b}=Y_i-\widehat\mu_b(X_i),
\qquad
\widehat V_{i,b}=D_i-\widehat\gamma_b(X_i),
\qquad
\widehat e_{i,b}(\theta)=\widehat R_{i,b}-\theta\widehat V_{i,b}.
\]
The estimated correction is
\begin{equation}
\widehat k_{ij,\tau}(\theta)
=\tfrac12\widehat K_{ij,\tau}
\{\widehat e_{i,b}(\theta)\widehat V_{j,b}
+\widehat e_{j,b}(\theta)\widehat V_{i,b}\},
\label{eq:plm-estimated-kernel}
\end{equation}
so ridge and selected projection differ only in $\widehat K_{b,\tau}$.
The estimating equation is linear in $\theta$, with solution
$\widehat\theta_\tau=\widehat A_\tau/\widehat J_\tau$, where
\begin{equation}
\begin{aligned}
\widehat A_\tau&=\frac1{n(n-1)}\sum_b\sum_{(i,j)\in\mathcal P_b}
\widehat R_{i,b}\bigl\{\widehat V_{i,b}
-\widehat K_{ij,\tau}\widehat V_{j,b}\bigr\},\\
\widehat J_\tau&=\frac1{n(n-1)}\sum_b\sum_{(i,j)\in\mathcal P_b}
\widehat V_{i,b}\bigl\{\widehat V_{i,b}
-\widehat K_{ij,\tau}\widehat V_{j,b}\bigr\}.
\end{aligned}
\label{eq:auto-plm-equation}
\end{equation}
This is a residual-on-residual regression coefficient in which the
treatment residual of each observation is adjusted by a
kernel-weighted average of the treatment residuals of the others.

\paragraph{Variance estimation.}
For observation $i$
and candidate $\tau$, define
\begin{align}
\widehat\psi_{i,\tau}&=\bar\phi_{i,\tau}
-\frac2{n-1}\sum_{j\ne i}
\widehat k_{ij,\tau}(\widehat\theta_\tau),
\qquad
\bar\phi_{i,\tau}=\frac1{n-1}\sum_{j\ne i}
\widehat\phi_{i,b(i,j)}(\widehat\theta_\tau),
\label{eq:auto-plm-row}\\
P_\tau&=\frac2{(n-1)^2}\sum_b\sum_{(i,j)\in\mathcal P_b}
\widehat k_{ij,\tau}(\widehat\theta_\tau)^2.
\label{eq:auto-plm-pair-variance}
\end{align}
The row $\widehat\psi_{i,\tau}$ is the estimated H\'ajek projection of observation
$i$: its first-order term and twice the average correction over its
partners. With $\bar\psi_\tau=n^{-1}\sum_i\widehat\psi_{i,\tau}$, the variance estimator
is
\begin{equation}
\widehat\sigma_\tau^{\star2}
=\frac{\max\bigl\{\sum_i(\widehat\psi_{i,\tau}-\bar\psi_\tau)^2-P_\tau,\
\sum_i\bar\phi_{i,\tau}^2\bigr\}}{n^2\widehat J_\tau^2}.
\label{eq:auto-plm-variance-star}
\end{equation}
It allows conditional heteroskedasticity. When the correction
contributes to first-order variance, each row also carries the noise
of the correction terms of its own pairs, so the row-square sum counts
the variance of the correction twice; subtracting $P_\tau$ counts it
once, and the lower bound keeps the estimator positive. Adding $P_\tau$
instead gives
\begin{equation}
\widehat\sigma_\tau^2
=\frac{\sum_i(\widehat\psi_{i,\tau}-\bar\psi_\tau)^2+P_\tau}{n^2\widehat J_\tau^2},
\label{eq:auto-plm-variance}
\end{equation}
which counts that variance three times and is conservative. Both are
consistent when the correction variance is negligible
(Section~\ref{sec:auto-inference}). We use $\widehat\sigma_\tau^\star$
for selection and for confidence intervals.\footnote{Estimating the
two oracle terms separately, by
$(\sum_i\bar\phi_{i,\tau}^2+P_\tau)/(n^2\widehat J_\tau^2)$, also gives a
positive and consistent estimator, including when the correction
contributes to first-order variance. The rows $\widehat\psi_{i,\tau}$ also account
for the part of the first-step estimation error in $\bar\phi_{i,\tau}$ that
the correction removes; the separate estimator can overstate the
variance when it omits this cancellation in finite samples.}

\subsection{Automatic selection of the correction}
\label{sec:auto-selection}

We select the correction that captures the most nuisance
estimation error within a variance budget. The criterion
follows from \eqref{eq:auto-remaining}. Consider a common
training sample and a population correction operator
$T_\tau$ with $0\le T_\tau\le I$, let $b_\tau$ be the remaining
quadratic bias \eqref{eq:auto-remaining} with $T_\tau$ in
place of $T_M$, and define the uncorrected errors
\[
d_{a,\tau}=\langle h_a,(I-T_\tau)h_a\rangle,
\qquad a\in\{\gamma,\alpha\}.
\]
For a projection, $d_{a,\tau}$ is the squared approximation
error outside the selected space; for ridge, it is the
regression error left uncorrected after shrinkage. By
Cauchy--Schwarz,
\begin{equation}
|b_\tau|\le\sqrt{d_{\gamma,\tau}d_{\alpha,\tau}}
+\tfrac12C_Fd_{\gamma,\tau},
\label{eq:auto-bias-control}
\end{equation}
where $C_F$ bounds the norm of $F''(\gamma_0)$; in the
PLM, $C_F=0$. Since
$d_{a,\tau}=\|h_a\|^2-\langle h_a,T_\tau h_a\rangle$ and
$\|h_a\|^2$ is common to all candidates, reducing the
uncorrected errors amounts to increasing the captured
errors $\langle h_a,T_\tau h_a\rangle$, which are estimable. 
For affine $F$, conditional on training,
$\E[u]=-\E[ph_\gamma]$ and $\E[v]=-\E[ph_\alpha]$, so
products of residual vectors from different observations
estimate the captured errors:
\begin{equation}
\widehat Q_{a,\tau}
=\frac1{n(n-1)}\sum_b\sum_{(i,j)\in\mathcal P_b}
\widehat z_{a,i,b}'\widehat M_{b,\tau}\widehat z_{a,j,b},
\label{eq:auto-captured}
\end{equation}
where $\widehat z_{\gamma,i,b}=\widehat u_{i,b}=p(X_i)\{Z_i-\widehat\gamma_b(X_i)\}$, $\widehat z_{\alpha,i,b}=\widehat v_{i,b}(\widetilde\theta)$, $\widehat v_{i,b}(\theta)=D_\gamma m(W_i,\widehat\gamma_b,\theta)[p]
-p(X_i)\widehat\alpha_b(X_i,\theta)$ and $\widetilde\theta$ is a preliminary estimate without correction,
such as the first-order DML estimate. Appendix~\ref{app:auto-select} gives
the Hessian adjustment for nonlinear $F$. In the PLM,
$\widehat z_{\gamma,i,b}=p(X_i)\widehat V_{i,b}$ and
$\widehat z_{\alpha,i,b}=-p(X_i)\widehat e_{i,b}(\widetilde\theta)$: weighted products of regression residuals from different
observations estimate how much of each nuisance error a
candidate captures. Only these statistics use $\widetilde\theta$; the estimating equation
\eqref{eq:auto-equation} uses $\widehat\alpha_b(\cdot,\theta)$.

\paragraph{Selection rule.}
Order the candidates of each family from less to more
correction: decreasing $\zeta_\tau$ for ridge, larger $S_{b,\tau}$
for selected projection. First (rule C), retain in each
family $f$ the largest index satisfying
\begin{equation}
\widehat\sigma_\tau^{\star2}\le c\,\widehat\sigma_{f,\mathrm{base}}^{\star2},\qquad c\ge1,
\label{eq:auto-budget}
\end{equation}
where $\widehat\sigma_{f,\mathrm{base}}^{\star2}$ is the variance estimate of the
first member of family $f$. Second, let
$\mathcal J_{\mathrm{sel}}$ contain the two retained
candidates, one for ridge and one for selected projection, put
$q_a=\max_{\tau\in\mathcal J_{\mathrm{sel}}}\widehat Q_{a,\tau}$,
and select
\begin{equation}
\widehat \tau\in\arg\max_{\tau\in\mathcal J_{\mathrm{sel}}}
\sum_{a\in\{\gamma,\alpha\}:q_a>0}
\frac{\widehat Q_{a,\tau}}{q_a},
\label{eq:auto-selection}
\end{equation}
with ties resolved by a fixed or independently randomized priority ordering. We call this
procedure ADA. The normalization weights the two
regression errors equally, whatever their scale. ADA
thus chooses, within the variance budget, the correction
that captures the most normalized regression error, and
thereby controls the bound \eqref{eq:auto-bias-control}.
Proposition~\ref{prop:auto-selection} in
Appendix~\ref{app:auto-select} shows that the selected
correction nearly minimizes a weighted version of this
bound, up to the estimation error in $\widehat Q_{a,\tau}$,
and allows different training samples across pair
blocks.

\begin{procedure}
\small
\textit{Input:} data $(Y_i,D_i,X_i)_{i=1}^n$; learners for $\E[Y\mid X]$ and
$\E[D\mid X]$; a dictionary $p(X)$; grids of ridge penalties $\zeta$ and of subset
sizes; a budget $c\ge1$.
\begin{enumerate}[leftmargin=*,label=\arabic*.,itemsep=2pt,topsep=3pt]
\item \emph{Blocks.} Split the observations at random into $L$ groups. Form one block
with the pairs within each group and one with the pairs between each two groups.
Block $b$ trains on the observations outside it.
\item \emph{First steps.} In each block $b$, fit $\widehat\mu_b$ and
$\widehat\gamma_b$ on the training sample, and compute $\widehat G_b$, the ridge
matrices \eqref{eq:auto-ridge}, and the projections on the regressors selected by the
LAR paths of $Y$ and $D$ \eqref{eq:auto-selected-kernel}. For each observation $i$ in
the block, compute $\widehat R_{i,b}=Y_i-\widehat\mu_b(X_i)$ and
$\widehat V_{i,b}=D_i-\widehat\gamma_b(X_i)$.
\item \emph{Candidate estimates.} For each candidate $\tau$, compute
$\widehat\theta_\tau=\widehat A_\tau/\widehat J_\tau$ from \eqref{eq:auto-plm-equation}
and its standard error $\widehat\sigma^\star_\tau$ from
\eqref{eq:auto-plm-variance-star}. Compute also the uncorrected estimate
$\widetilde\theta$, which sets $\widehat K_{ij,\tau}=0$ in
\eqref{eq:auto-plm-equation}.
\item \emph{Budget.} In each family, retain the candidate with the most correction
among those with $\widehat\sigma_\tau^{\star2}\le
c\,\widehat\sigma_{f,\mathrm{base}}^{\star2}$ \eqref{eq:auto-budget}.
\item \emph{Selection.} For the retained candidates, compute
$\widehat Q_{\gamma,\tau}$ and $\widehat Q_{\alpha,\tau}$ from \eqref{eq:auto-captured},
with $\widehat z_{\gamma,i,b}=p(X_i)\widehat V_{i,b}$ and
$\widehat z_{\alpha,i,b}=-p(X_i)\widehat e_{i,b}(\widetilde\theta)$, and choose
$\widehat\tau$ by \eqref{eq:auto-selection}.
\item \emph{Report} $\widehat\theta_{\widehat\tau}$ and the 95\% confidence interval
$\widehat\theta_{\widehat\tau}\pm1.96\,\widehat\sigma^\star_{\widehat\tau}$.
\end{enumerate}
\caption{ADA for the PLM: estimate and confidence interval}
\label{proc:plm}
\end{procedure}

\subsection{Inference after automatic selection}
\label{sec:auto-inference}

We consider two regimes, according to whether the correction
contributes to first-order variance. Throughout this subsection,
the number of pair blocks is fixed, and the total number of
candidate corrections is bounded by a constant independent of $n$.

\paragraph{Negligible correction variance.}
When the correction variance is $o(n^{-1})$, we
give conditions in the style of \citet{LR2022}, mean-square
consistency of the first steps and a condition on the remaining bias,
for moments that are affine in the regression nuisance. This is the
class of \citet{CNS2022}, which contains the PLM as a special case. Let $V_Z=Z-\gamma_0(X)$ and $\xi_d=\sup_x\|p(x)\|_2$, and
write the first-step estimation errors in block $b$ as
$h_{\gamma,b}=\widehat\gamma_b-\gamma_0$,
$h_{\alpha,b}=\widehat\alpha_b(\cdot,\theta_0)-\alpha_0$, and
$h_{\dot\alpha,b}=\partial_\theta(\widehat\alpha_b(\cdot,\theta)-\alpha_0(\cdot,\theta))$.

\begin{assumption}[Moments affine in the nuisance]
\label{ass:auto-model}
The observations are iid, $\gamma_0(X)=\E[Z\mid X]$, and:
(i) $m(W,\gamma,\theta)$ is affine in $\gamma$ and in $\theta$, so that
$D_\gamma m(W,\gamma_0,\theta)[a]$ does not depend on $\gamma$, and $\alpha_0(\cdot,\theta)$
represents $a\mapsto\E\{D_\gamma m(W,\gamma_0,\theta)[a]\}$, with
$\alpha_0=\alpha_0(\cdot,\theta_0)$;
(ii) for $\theta$ near $\theta_0$,
$\E[\{D_\gamma m(W,\gamma_0,\theta)[a]\}^2]\le C\|a\|^2$, $\alpha_0(\cdot,\theta)$ and
$\partial_\theta\alpha_0(\cdot,\theta)$ are bounded, and $\E[V_Z^2\mid X]\le C$;
(iii) $J_0\ne0$, $\Omega_1=\E[\phi_0^2]\in(0,\infty)$, and
$\E[\{\partial_\theta m(W,\gamma_0,\theta)\}^2]<\infty$.
\end{assumption}

\begin{assumption}[First-step estimators and candidates]
\label{ass:auto-first-step}
(i) For every block, $\widehat\gamma_b$ and
$\widehat\alpha_b(\cdot,\theta)$ are $\mathcal H_b$-measurable,
$\widehat\alpha_b(\cdot,\theta)$ is affine in $\theta$, and
$\|h_{a,b}\|_\infty=O_p(1)$ and $\|h_{a,b}\|=o_p(1)$ for
$a\in\{\gamma,\alpha,\dot\alpha\}$;
(ii) the eigenvalues of $G_0$ are bounded away from zero and
infinity, and $\xi_d^2\log(2d)/n\to0$;
(iii) the total number of candidates is bounded by a constant
independent of $n$, and each candidate is ridge or selected
projection, with its penalty, retained eigenspace, or selected
set measurable with respect to $\mathcal H_b$.
\end{assumption}
Condition (ii) of Assumption~\ref{ass:auto-model} is the mean-square
continuity of \citet{CNS2022}. For candidate $\tau$ in block $b$, the
comparison operator $T_{b,\tau}$ replaces $\widehat G_b$ by $G_0$ in
$\widehat M_{b,\tau}$, holding the retained eigenspace or selected set
fixed; it is a positive contraction
(Lemma~\ref{lem:auto-operators}). With
$\omega_b=|\mathcal P_b|/\{n(n-1)\}$, let
$\mathcal Q_{a,\tau}=\sum_b\omega_b\langle h_{a,b},T_{b,\tau}h_{a,b}\rangle$
and $d_{a,\tau}=\sum_b\omega_b\langle h_{a,b},(I-T_{b,\tau})h_{a,b}\rangle$
be the captured and uncaptured errors, and let
$r_n=\max_b\max_{a\in\{\gamma,\alpha,\dot\alpha\}}\|h_{a,b}\|$.

\begin{theorem}[Inference with negligible correction variance]
\label{thm:auto-inference}
Suppose Assumptions~\ref{ass:auto-model} and~\ref{ass:auto-first-step} hold.
\begin{enumerate}[label=(\alph*),leftmargin=*]
\item If candidate $\tau$ satisfies
\begin{align*}
&\text{(R1)}\quad
\sqrt n\,\max_b\|(I-T_{b,\tau})^{1/2}h_{\alpha,b}\|\,
\|(I-T_{b,\tau})^{1/2}h_{\gamma,b}\|\to_p0,\\
&\text{(R2)}\quad
\xi_d\sqrt{\log(2d)}\,\max_b\|h_{\alpha,b}\|\,\|h_{\gamma,b}\|\to_p0,
\end{align*}
then
\[
\sqrt n(\widehat\theta_\tau-\theta_0)
=\frac1{J_0\sqrt n}\sum_{i=1}^n\phi_0(W_i)+o_p(1),
\qquad
n\widehat\sigma_\tau^{\star2}\to_p\frac{\Omega_1}{J_0^2},
\qquad
n\widehat\sigma_\tau^{2}\to_p\frac{\Omega_1}{J_0^2}.
\]
\item Let $c>1$ and the comparison operators be nested along each
family. Suppose that (i)
$r_n/\min_{b,\,a\in\{\gamma,\alpha\}}\|h_{a,b}\|=O_p(1)$; (ii) for some
candidate $\tau^\ast$ and constant $\underline c>0$,
$\mathcal Q_{a,\tau^\ast}\ge\underline c\sum_b\omega_b\|h_{a,b}\|^2$,
$a\in\{\gamma,\alpha\}$, with probability approaching one; and (iii)
$\sqrt n(d_{\gamma,\tau^\ast}+d_{\alpha,\tau^\ast})\to_p0$,
$\xi_d\sqrt{\log(2d)}\,r_n^2\to_p0$, $\sqrt n\,r_n^4\to_p0$, and the
preliminary estimate in \eqref{eq:auto-captured} satisfies
$\widetilde\theta-\theta_0=O_p(n^{-1/2}+r_n^2)$. Then the conclusions
of (a) hold for $\widehat\theta_{\widehat \tau}$,
$\widehat\sigma_{\widehat \tau}^{\star}$, and $\widehat\sigma_{\widehat \tau}$.
\end{enumerate}
\end{theorem}

Part (a) has the structure of the results of \citet{LR2022}. Condition
(R1) replaces the product-rate condition of DML,
$\sqrt n\,\|h_\alpha\|\,\|h_\gamma\|\to0$, by the same condition on the
error components that the correction leaves out. Condition (R2)
controls the estimation of the Gram matrix and is not needed when $G_0$
is known. The dictionary condition also makes the correction variance
of order $d/n^2=o(n^{-1})$. In part (b), ADA needs a candidate that
leaves both errors small, measured by their sum rather than by the
geometric mean in (R1); the estimate without correction satisfies the
rate required of $\widetilde\theta$
(Appendix~\ref{app:auto-negligible}), and rule C eventually retains
the last member of each family. Affinity in $\theta$ simplifies the
linearization; for moments that are smooth in $\theta$, the same
arguments apply with a consistent root and local uniform convergence
of the derivative. For nonlinear $F$, the bias contains additional
Taylor and Hessian-estimation terms
(Appendix~\ref{app:auto-curvature}), which require cubic-rate
conditions.

\begin{corollary}[The PLM]
\label{cor:plm}
In the PLM \eqref{eq:plm-model}, suppose $\gamma_0$ and $c_0$
are bounded, $\E[V^2]>0$, $\E[\varepsilon^2V^2]>0$,
$\E[\varepsilon^2\mid D,X]$ and $\E[V^4\mid X]$ are bounded,
and $\widehat\alpha_b(\cdot,\theta)
=\theta\widehat\gamma_b-\widehat\mu_b$, where
$\widehat\mu_b$ and $\widehat\gamma_b$ are
$\mathcal H_b$-measurable, bounded, and mean-square consistent.
\begin{enumerate}[label=(\alph*),leftmargin=*]
\item Assumption~\ref{ass:auto-model} and
Assumption~\ref{ass:auto-first-step}(i) hold, with $Z=D$,
$D_\gamma m(W,\gamma_0,\theta)[a]=-(Y-\theta D)a(X)$,
$\alpha_0(\cdot,\theta)=\theta\gamma_0-\mu_0$, $\phi_0=\varepsilon V$,
$J_0=\E[V^2]$, and $h_{\dot\alpha,b}=h_{\gamma,b}$. Hence
Theorem~\ref{thm:auto-inference} applies with
$\Omega_1=\E[\varepsilon^2V^2]$.
\item Suppose further that $X\in[0,1]^q$ has a density bounded
away from zero and infinity, and that $\gamma_0$ and $c_0$
belong to a H\"older ball of order $s$. Let $p_{\rm fs}$ be a
normalized tensor-product B-spline basis of order exceeding
$s$, on a quasi-uniform partition, with
$d_{\rm fs}\asymp(n/\log n)^{q/(2s+q)}$ terms. On each training
sample, estimate $\mu_0$ and $\gamma_0$ by least squares over
\[
 \left\{f\in\operatorname{span}(p_{\rm fs}):
             \|f\|_\infty\le C_{\rm fs}\right\},
 \qquad
 C_{\rm fs}>\max\{\|\mu_0\|_\infty,\|\gamma_0\|_\infty\},
\]
where $C_{\rm fs}$ is fixed. Suppose the correction projects onto
a B-spline space with $d\ge d_{\rm fs}$ terms that contains the
first-step estimation space. If
$s>(1+\sqrt{17})q/16\approx0.32\,q$, there is a sequence
$d=d_n$ for which the conditions of
Theorem~\ref{thm:auto-inference}(a) hold. For comparison,
the usual DML product-rate condition based on these first-step
rates requires $s>q/2$.
\end{enumerate}
\end{corollary}

With a known Gram matrix, condition (R2) is unnecessary.
The same nested-spline construction then yields
oracle-equivalent inference for $s>q/4$: choose
$d\asymp n^\kappa$ with
\[
 \max\left\{\frac{q}{4s},\frac{q}{2s+q}\right\}
 <\kappa<1.
\]
The stronger threshold
$s>(1+\sqrt{17})q/16$ is sufficient for the estimated-Gram
construction analyzed here. We do not establish that
this threshold is sharp or that the gap from $q/4$ is
an unavoidable cost of estimating the covariate
distribution.

\paragraph{Nonnegligible correction variance.}
\label{sec:nonnegligible}
With many correction directions, the correction can contribute
to first-order variance. We study the PLM conditionally on the
covariates and external training information, as in
many-regressor asymptotics \citep{CJN2018}. Put
$\mathcal G_n=\mathcal H_n\vee\sigma(X_1,\ldots,X_n)$.
For candidate $\tau$, let $\mathbb B_\tau$ be symmetric with zero
diagonal and entries
$\widehat K_{ij,\tau}/(n-1)$ for $i\ne j$.
At the true nuisance functions, the estimating equation
multiplied by $n$ is
$\sum_i\varepsilon_iV_i-\varepsilon'\mathbb B_\tau V$.
Define
\[
\Omega_{1,n}=\frac1n\sum_i\E[\varepsilon_i^2V_i^2\mid X_i],
\qquad
\Omega_{2,\tau,n}=\frac1n\sum_{i\ne j}(\mathbb B_\tau)_{ij}^2
\E[\varepsilon_i^2\mid X_i]\E[V_j^2\mid X_j].
\]
For $\mathcal G_n$-measurable matrices, the two terms are
conditionally uncorrelated, with variances $n\Omega_{1,n}$
and $n\Omega_{2,\tau,n}$. More generally, require a symmetric,
zero-diagonal, $\mathcal G_n$-measurable reference
$\mathbb B_\tau^\circ$ satisfying
\begin{equation}
\|\mathbb B_\tau\|_{\op}+\|\mathbb B_\tau^\circ\|_{\op}=O_p(1),
\qquad
n^{-1}\|\mathbb B_\tau-\mathbb B_\tau^\circ\|_F^2=o_p(1).
\label{eq:nonneg-stability}
\end{equation}
The training Gram matrices need not approximate $G_0$, so we
measure the remaining bias in the sample:
\[
\bar b_{\tau,n}=-\frac1{n(n-1)}\sum_{i\ne j}h_{\alpha,b}(X_i)
\{h_{\gamma,b}(X_i)
-\widehat K_{ij,\tau}h_{\gamma,b}(X_j)\},
\qquad b=b(i,j),
\]
the sample analog of \eqref{eq:plm-remaining-bias}.

\begin{theorem}[Inference with nonnegligible correction variance]
\label{thm:nonneg-inference}
In the PLM, suppose the conditions in the opening paragraph
of Corollary~\ref{cor:plm} hold, with
$\E[\varepsilon^4\mid D,X]$ bounded. The first-step estimators are
$\mathcal H_b$-measurable, and the correction matrices of
block $b$ are measurable with respect to
$\mathcal F_b=\mathcal G_n\vee\mathcal H_b$.
\begin{enumerate}[label=(\alph*),leftmargin=*]
\item If candidate $\tau$ satisfies \eqref{eq:nonneg-stability}
and $\sqrt n\,\bar b_{\tau,n}\to_p0$, then
\[
\frac{\widehat\theta_\tau-\theta_0}{\widehat\sigma_\tau^\star}
\dto N(0,1),\qquad
\frac{nJ_0^2\widehat\sigma_\tau^{\star2}}
{\Omega_{1,n}+\Omega_{2,\tau,n}}\to_p1,\qquad
\frac{nJ_0^2\widehat\sigma_\tau^2}
{\Omega_{1,n}+3\Omega_{2,\tau,n}}\to_p1.
\]
\item Suppose \eqref{eq:nonneg-stability} holds for every
candidate, $\sqrt n\,\bar b_{\widehat \tau,n}\to_p0$,
$\sqrt n\,r_n^4\to_p0$, and
$\widetilde\theta-\theta_0=O_p(n^{-1/2}+r_n^2)$.
If Assumption~\ref{ass:nonneg-ties} holds with the block-specific
signals and reference noises in Appendix~\ref{app:nonneg},
the conclusions of (a) hold for $\widehat \tau$.
The preliminary estimator without correction has the stated rate.
\end{enumerate}
\end{theorem}

The standard errors $\widehat\sigma_\tau^\star$ and
$\widehat\sigma_\tau$ count the correction variance once and
three times, respectively. The computable norm
$\|\mathbb B_\tau\|_{\op}$ bounds the weight placed on any
sample direction while allowing $\Omega_{2,\tau,n}>0$.
Subject to the operator-norm bound,
\eqref{eq:nonneg-stability} holds with
$\mathbb B_\tau^\circ=\mathbb B_\tau$ for covariate-only matrices
and matrices learned on an independent auxiliary sample,
and with $\mathbb B_\tau^\circ=0$ when
$\|\mathbb B_\tau\|_F^2=o_p(n)$.
For selected projection it also holds if, with probability
approaching one, each selected set $S_{b,\tau}$ equals a
$\mathcal G_n$-measurable set $S_{b,\tau}^\circ$:
construct $\mathbb B_\tau^\circ$ from those sets and the same
training Gram matrices and inversion rule. Then
$\mathbb B_\tau-\mathbb B_\tau^\circ=0$ with probability
approaching one, allowing stable response-selected projections
to contribute to first-order variance.
Since $\E[\varepsilon V\mid X]=0$, the reference estimation
noise is uncorrelated with the reference selection noises.
Their joint limit is Gaussian, so they are asymptotically
independent. Under the selection-limit conditions, the selected
studentized estimator is asymptotically normal even when
the selected candidate remains random in the limit.

\subsection{Numerical evaluation}
\label{sec:numerical}

We evaluate the estimator in \eqref{eq:auto-plm-equation} with ADA in the PLM. The
experiments measure the bias that first-order DML leaves when the first-step estimation errors are
correlated, whether ADA selects the appropriate correction, the variance cost of the
correction, and the effect of the cross-fitting scheme. Competing methods are
implemented as they are used in practice.

\subsubsection{Designs}

Data follow \eqref{eq:plm-model} with $\theta_0=1$, so that $Y=D+c_0(X)+\varepsilon$
and $D=\gamma_0(X)+V$, and $n\in\{500,1000\}$. In the six low-dimensional designs,
the coordinates of $X$ are independent with density $1.9-1.8x$ on $[0,1]$. The errors are
\begin{equation}
 V=\sigma_{V,0}(X)Z_1,\qquad
 \varepsilon=\sigma_{\varepsilon,0}(X)\sqrt{(1+Z_1^2)/2}\,Z_2,
 \label{eq:plm-sim-errors}
\end{equation}
where $Z_1,Z_2$ are independent standard normal variables, independent of $X$, and
$\sigma_{V,0}(x)=(1+x_1)/2$. Thus $\E[\varepsilon\mid D,X]=0$, while the conditional variance
of $\varepsilon$ depends on $D$ and $X$. The two regressions share an oscillating
component $h$, which the first-step estimators estimate poorly:
\[
 c_0(x)=A_c h(x)+\sin(2x_1),\qquad \gamma_0(x)=A_\gamma h(x)+x_1/2,
 \qquad
 h_s(t)=\sum_{j=1}^{11}2^{-j s}\cos(2^j\pi t+j).
\]
Designs A and B use $h(x)=h_{0.6}(x_1)$. A has $\sigma_{\varepsilon,0}(x)=0.8$; the other
low-dimensional designs have $\sigma_{\varepsilon,0}(x)=1.2-0.8x_1$. Local uses
$h(x)=h_{0.5}(x_1)/[1+\exp\{-(x_1-0.6)/0.04\}]$, which oscillates only where the
density is low. Sparse uses four frequencies,
$h(x)=\sqrt2\sum_{j\in\{4,13,37,97\}}j^{-0.3}\cos(j\pi x_1)$. Two has two covariates,
$h(x)=h_{0.6}(x_1)+h_{0.6}(x_2)+h_{0.6}(x_1)h_{0.6}(x_2)$ and
$\sigma_{\varepsilon,0}(x)=1.2-0.4(x_1+x_2)$. Rough uses $h_{0.35}$, whose high-frequency terms lie
beyond the dictionaries at these sample sizes; it serves as a stress test. The
scales $(A_c,A_\gamma)$ are $(2.4,1.6)$ in A and B, $(3,2)$ in Local, $(0.9,0.6)$ in
Sparse, $(1,0.7)$ in Two, and $(1.6,1.1)$ in Rough.

Designs C1 and C2 follow \citet{CST2026}: $X\sim N(0,I_p)$ with $p=n/2$ and
$c_0(X)=\gamma_0(X)=X'\beta$, $\beta_j=2^{1-j}$. C1 has $V=Z_1$ and
$\varepsilon=Z_2$. C2 uses \eqref{eq:plm-sim-errors} with $\sigma_{V,0}^2(X)=(1+X_1^2)/2$ and
$\sigma_{\varepsilon,0}^2(X)=(1+X_2^2)/2$.

Each design and sample size has 500 replications, and all methods use the same
samples. At coverage .95 the Monte Carlo standard error is .010. The main text
reports A, Local, Sparse, Two, C1, and Rough. Appendix~\ref{app:plm-sim} reports B
and C2, which behave like A and C1, together with bias, standard deviation, and
interval length for every method.

\subsubsection{Methods}

\emph{First-step estimators.} In the low-dimensional designs a single random forest with 100 trees
and minimum leaf size five estimates $\mu_0$ and $\gamma_0$ jointly. In C1 and C2 the
first-step estimators are heteroskedastic plug-in Lasso fits. All methods use the same first-step
learners.

\emph{Cross-fitting.} We split the sample into five groups and form the 15 pair
blocks of \citet{ET2026}: the pairs within each group and the pairs between each two
groups. Block $b$ is trained on the groups that it does not contain. Each pair uses
the block fits $\widehat\mu_b$ and $\widehat\gamma_b$ both for its correction and for
its first-order terms (shared fits), as in \eqref{eq:auto-equation}.

\emph{Corrections.} The ridge dictionary consists of a constant and the scaled
cosines $\sqrt2\cos(j\pi t)/(1+j)$, $j<n/4$; Two uses their tensor products in
hyperbolic-cross order, and C1 and C2 use the covariates. The ridge penalties are
$\zeta_\tau=(1+d_\tau)^{-2}$ for $d_\tau\in\{1,2,3,4,6,8,11,16,23,32,45,64,90,128\}$, $d_\tau\le n/6$;
in C1 and C2 they are set to attain the same values of $\tr(\widehat G_b\widehat M_{b,\tau})$. Selected
projection orders $n/4$ unscaled cosines (the covariates in C1 and C2) by the
training LAR paths of $Y$ and $D$ and projects onto the first
$d_\tau\in\{2,4,6,8,12,16,24,32,48,64,96\}$, $d_\tau\le n/8$, of them. Dictionary sizes and
grids thus grow with $n$. Rule C \eqref{eq:auto-budget} uses $c=2$, and ADA
\eqref{eq:auto-selection} chooses between the two retained candidates. Standard
errors use \eqref{eq:auto-plm-variance-star}, in rule C and for the intervals, with
no adjustment for selection.

\emph{Methods.} DML is conventional five-fold DML \citep{DML2018}. We consider two implementations of the second-order projection correction of \citet{Robins2008,Robins2017} (HOIF) onto the first $d_{\mathrm H}$
dictionary functions: cosines, tensor cosines in Two, and covariates in C1 and C2. As
in the HOIF literature, $d_{\mathrm H}$ is fixed in advance; we take
$d_{\mathrm H}\approx\sqrt n$ ($d_{\mathrm H}=24$ and $32$). The Gram matrix is inverted by Cholesky factorization, and the standard error
is the first-order one, which the HOIF theory justifies when $d_{\mathrm H}=o(n)$. The first HOIF method, eHOIF,
computes the first-step estimators and the Gram matrix in one half of the sample, computes the
U-statistic in the other, and averages over the two assignments \citep{LMNR2026}.
The second, sHOIF uses five-fold cross-fitted residuals and estimates the Gram matrix from the
same sample as the U-statistic \citep{Liu2026}; this is also the default of the HOIF
software. In C1 and C2 the covariates are ordered by the true coefficients, which
favors HOIF. Table \ref{tab:plm-coverage} shows the results for sHOIF while the results of eHOIF are in Appendix \ref{app:plm-sim}. TL is the triple-debiased Lasso of \citet{CST2026} with its own five-fold
scheme and plug-in penalties, applied to the cosine dictionary in the
low-dimensional designs, its results are in Appendix \ref{app:plm-sim}; TL-CV cross-validates the two regression penalties.

\subsubsection{Bias reduction and selection}

Table~\ref{tab:plm-coverage} reports coverage and RMSE. Outside Rough, first-order DML
covers $\theta_0$ in 32--80\% of replications, and its bias is between 0.95 and 2.44
standard deviations. The two corrections succeed in different designs. Ridge covers
.45--.74 in C1 and C2, and selected projection covers .77--.88 in Local and Two. ADA
chooses selected projection in at least 99.4\% of the C1 and C2 replications and
ridge in at least 93\% of the A, B, Local, and Two replications. Outside Rough, its
coverage is .94--.97, its absolute bias is at most 0.56 standard deviations, and its
RMSE is 24--47\% below that of DML and within 3\% of the better of its two
candidates.

\begin{table}[htbp]
\centering
\small
\caption{Coverage of nominal 95\% intervals (RMSE$\times100$)}
\label{tab:plm-coverage}
\begin{tabular}{llcccccc}
\toprule
Design & $n$ & DML & sHOIF & TL-CV & Ridge & SP & ADA\\
\midrule
A      & 500  & .71 (12.2) & .80 (10.5) & .55 (250)  & .96 (7.9)  & .90 (9.6)  & .96 (8.0)\\
       & 1000 & .74 (7.9)  & .79 (7.4)  & .82 (7.3)  & .95 (5.7)  & .93 (6.1)  & .95 (5.7)\\
Local  & 500  & .57 (17.2) & .76 (13.2) & .51 (69.4) & .96 (9.4)  & .82 (14.0) & .96 (9.5)\\
       & 1000 & .65 (10.5) & .75 (9.0)  & .72 (34.0) & .97 (6.3)  & .88 (7.7)  & .97 (6.3)\\
Sparse & 500  & .72 (12.6) & .73 (12.3) & .88 (10.0) & .95 (8.9)  & .95 (9.3)  & .95 (9.2)\\
       & 1000 & .80 (8.2)  & .80 (8.2)  & .92 (6.2)  & .94 (6.2)  & .94 (6.3)  & .94 (6.2)\\
Two    & 500  & .50 (17.4) & .56 (16.0) & .61 (24.8) & .96 (10.0) & .77 (13.4) & .96 (10.1)\\
       & 1000 & .48 (12.7) & .52 (11.9) & .67 (10.5) & .94 (7.4)  & .80 (9.0)  & .94 (7.4)\\
C1     & 500  & .32 (11.1) & .94 (4.4)  & .92 (5.1)  & .45 (10.3) & .96 (5.8)  & .96 (5.9)\\
       & 1000 & .53 (6.5)  & .93 (3.2)  & .93 (3.5)  & .65 (5.9)  & .95 (3.7)  & .95 (3.8)\\
Rough  & 500  & .17 (24.1) & .25 (22.0) & .33 (563)  & .73 (14.9) & .65 (16.5) & .74 (14.7)\\
       & 1000 & .22 (16.2) & .27 (15.2) & .50 (13.4) & .83 (9.1)  & .74 (10.4) & .84 (9.0)\\
\bottomrule
\end{tabular}

\begin{minipage}{.97\textwidth}
\footnotesize
\medskip
SP: selected projection. Ridge and SP are the candidates retained by rule C; ADA
chooses between them. eHOIF and TL are discussed in the text; all methods are
reported in Appendix~\ref{app:plm-sim}.
\end{minipage}
\end{table}

The implemented eHOIF specification is unstable in the
low-dimensional designs: its RMSE exceeds 0.25 in every
such cell and reaches 82, whereas ADA's is at most 0.15.
Its empirical sampling variability substantially exceeds
the uncertainty reflected in its reported intervals;
see Tables~\ref{tab:plm-extra} and~\ref{tab:plm-bias}
in Appendix~\ref{app:plm-sim}. The Gram matrix of $d_{\mathrm H}\approx\sqrt n$ cosines, estimated from a
half sample with few observations near $x_1=1$, is poorly conditioned, and its
inverse is applied to the other half, where the kernel is not a projection. sHOIF is
stable but corrects little at $d_{\mathrm H}\approx\sqrt n$; its coverage in the low-dimensional
designs other than Rough is .52--.81. TL covers at most .04 in A, B, Local, and Two,
and TL-CV covers .50--.92. In C1 and C2, the designs for which TL is built and in
which HOIF uses the true ordering, sHOIF and TL-CV cover .90--.94 with RMSEs 7--25\%
below that of ADA. The large RMSEs of TL-CV come from a few replications in which the
triple-Lasso denominator of one of the five folds is close to zero; in
A at $n=500$ its median absolute error is .12. Rough remains a failure case for every method: ADA covers .74 and
.84, with bias of 1.39 and 1.05 standard deviations. Its regressions
contain oscillations of frequency up to $2^{11}\pi$ whose amplitudes
decay slowly, so the errors of the first-step forests are spread over
many directions, several of them beyond the dictionaries at these
sample sizes. Removing enough of this bias would require more
correction variance than the budget allows, which is the cost studied
in Section~\ref{sec:theory}; the remaining bias falls as $n$ grows and
the dictionaries expand.

\subsubsection{The cost of the correction}

\begin{figure}[htbp]
\centering
\includegraphics[width=\textwidth]{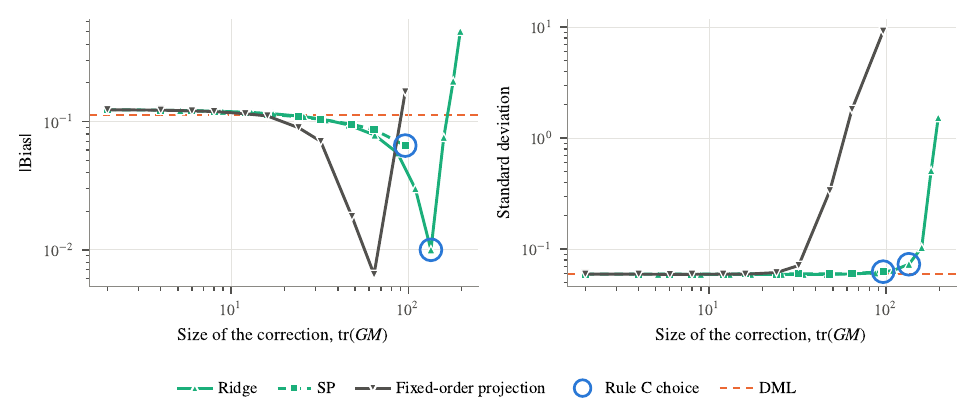}
\caption{Absolute bias (left) and standard deviation (right) along the tuning
paths in design Two, $n=1000$, against the size of the correction $\tr(G_0M)$. Circles
mark the member that rule C selects most often. Dashed lines: DML. Logarithmic
scales. Each curve ends at the largest candidate considered
within that family.}
\label{fig:plm-cost}
\end{figure}

Figure~\ref{fig:plm-cost} traces bias and standard deviation along the tuning paths
in Two at $n=1000$. As the ridge correction grows to $\tr(G_0M)\approx134$, the bias
falls from .122 to $-.010$ while the standard deviation rises from .060 to .073; the
correction then accounts for 21\% of the estimated variance. One grid step further,
the standard deviation is .103, and it reaches 1.53 at the end of the grid. The
fixed-order projection reaches a standard deviation of 0.34 with 48 retained
terms and 9.2 with 96. Rule C selects the ridge members with $\tr(G_0M)\approx134$ or $158$ in 84\% of
the replications and one of the last two members in 4\%. Across designs, the
correction accounts for 9--34\% of ADA's estimated variance, and ADA's intervals are
15--47\% longer than those of DML. The excess length falls from $n=500$ to $n=1000$
in every design except Two, where it remains 28\%.

\subsubsection{Cross-fitting}

Table~\ref{tab:plm-crossfit} compares six cross-fitting schemes. Pair blocks with
shared fits give the smallest RMSE in most cells. Their equal-size version, which
trains every block on about $4n/5$ observations, raises training size for
between-group blocks from $3n/5$ and lowers ADA's MSE
by a median of 3--6\% over the designs other than Rough, at three times the number of
first-step fits. The other schemes raise the median MSE by 9--48\%. Two halves and
within-fold pairs use fewer pairs, so their corrections are noisier; separate
first-order fits correct the bias of a different fit; and double cross-fitting halves
each training sample, which weakens the Lasso first-step estimators in C1 (coverage .54 at $n=500$).
In Rough, separate first-order fits, which use $4n/5$ observations, perform best.

\begin{table}[H]
\centering
\small
\caption{ADA under alternative cross-fitting schemes: coverage (RMSE$\times100$)}
\label{tab:plm-crossfit}
\begin{tabular}{llcccccc}
\toprule
& & \multicolumn{3}{c}{Pair blocks} & & & \\
\cmidrule(lr){3-5}
Design & $n$ & Shared fits & Equal size & Separate fits & Two halves & Within folds & Double\\
\midrule
A      & 500  & .96 (8.0)  & .96 (8.0)  & .97 (8.3)  & .94 (9.1)  & .92 (9.8)  & .94 (9.4)\\
       & 1000 & .95 (5.7)  & .96 (5.6)  & .95 (6.0)  & .94 (6.3)  & .92 (6.8)  & .95 (6.5)\\
Local  & 500  & .96 (9.5)  & .98 (9.0)  & .98 (9.5)  & .93 (11.6) & .91 (12.1) & .94 (11.3)\\
       & 1000 & .97 (6.3)  & .96 (6.2)  & .96 (6.7)  & .94 (7.4)  & .91 (7.7)  & .97 (7.0)\\
Sparse & 500  & .95 (9.2)  & .95 (8.9)  & .93 (10.0) & .95 (9.5)  & .93 (9.8)  & .91 (10.4)\\
       & 1000 & .94 (6.2)  & .94 (6.1)  & .94 (6.6)  & .94 (7.1)  & .94 (7.0)  & .93 (7.1)\\
Two    & 500  & .96 (10.1) & .95 (10.0) & .96 (10.2) & .92 (11.5) & .86 (13.1) & .93 (11.5)\\
       & 1000 & .94 (7.4)  & .93 (7.5)  & .94 (7.6)  & .94 (7.6)  & .90 (8.5)  & .94 (7.3)\\
C1     & 500  & .96 (5.9)  & .96 (5.4)  & .82 (9.1)  & .97 (5.7)  & .94 (6.0)  & .54 (14.2)\\
       & 1000 & .95 (3.8)  & .94 (3.7)  & .87 (4.8)  & .93 (4.5)  & .93 (4.3)  & .93 (4.7)\\
Rough  & 500  & .74 (14.7) & .74 (14.6) & .85 (12.5) & .65 (17.4) & .64 (17.8) & .59 (18.7)\\
       & 1000 & .84 (9.0)  & .80 (9.4)  & .92 (7.3)  & .75 (11.3) & .73 (11.7) & .74 (11.6)\\
\bottomrule
\end{tabular}

\begin{minipage}{.97\textwidth}
\footnotesize
\medskip
Shared fits: the scheme of Table~\ref{tab:plm-coverage}. Equal size: as shared fits,
but each between-group block is split into the four blocks between the halves of its
two groups, so that all 45 blocks are trained on about $4n/5$ observations
\citep{ET2026}. Separate fits: as shared fits, but the first-order terms use five-fold
fits. Two halves: first-step estimators and matrices from the other half, pairs within each half.
Within folds: five folds, first-step estimators and matrices from the complement, pairs within each
fold. Double: as shared fits, with $\widehat\mu_b$ and $\widehat\gamma_b$ fitted on
disjoint halves of $I_b^c$. Rule C and ADA are applied within each scheme, and
the standard error replaces $n-1$ by the number of partners of each observation.
\end{minipage}
\end{table}
%%%%%%%%%%%%%%% SECTION 5 %%%%%%%

\section{Further Applications}\label{sec:applications}

The binary-choice model makes joint moment choice explicit.
The DiD model shows how exclusion restrictions affect exact
correctability, and when correction cost persists after
optimization over all first-order moments.

\subsection{Joint moment choice in a binary-choice model}
\label{sec:binary}

Let $W=(Y_1,Y_2,X)$, where the two binary outcomes are
conditionally independent given $X$ and
\begin{equation}
 \PP(Y_1=1\mid X)=\Lambda(\gamma_0(X)),\qquad
 \PP(Y_2=1\mid X)=\Lambda(\gamma_0(X)+\theta_0),\qquad
 \Lambda(t)=\frac{e^t}{1+e^t}.
 \label{eq:binary-early}
\end{equation}
The nuisance function $\gamma_0$ determines baseline choice
probabilities; $\theta_0$ is a common change in log odds.
Both $\gamma_0$ and $P_{0,X}$ are unknown and locally unrestricted.
Assume the two probabilities are bounded away from zero
and one. Each sampling unit contains both outcomes; the
pairwise correction uses two independent sampling units.
For example, two independently sampled consumers may face
the same product characteristics and prices differing by
a fixed amount. A common logit price coefficient then gives
the shift $\theta_0$.

Conditioning on $Y_1+Y_2=1$ removes $\gamma_0$, giving the moment
\[
 C(W,\theta)
 =\ind\{Y_1+Y_2=1\}\{Y_2-\Lambda(\theta)\}.
\]
This is the conditional-likelihood construction of
\citet{Rasch1960} and \citet{Andersen1970}, introduced in
econometrics by \citet{Chamberlain1980}. Its expectation at $\theta_0$
is zero for every baseline $\gamma_0$. It is therefore orthogonal
of every order with respect to that nuisance, although
it need not use all first-order information.

At the true law, for $j=1,2$, define
\[
 \begin{aligned}
 p_{j,0}&=\Lambda(\gamma_0+(j-1)\theta_0),
 &\varepsilon_j&=Y_j-p_{j,0},
 &v_{j,0}&=p_{j,0}(1-p_{j,0}),\\
 R_B&=\varepsilon_1+\varepsilon_2,
 &D_B&=\varepsilon_1\varepsilon_2,
 &V_0&=v_{1,0}+v_{2,0},\quad w_0=v_{1,0}v_{2,0}.
 \end{aligned}
\]
Put $C_0(W)=C(W,\theta_0)$,
$\kappa_0=2\Lambda(\theta_0)-1$,
$e_{B,0}=w_0/V_0$, and
\[
 J_0(X)=\E[C_0^2\mid X]
 =\PP(Y_1+Y_2=1\mid X)
   \Lambda(\theta_0)\{1-\Lambda(\theta_0)\}.
\]
The nuisance scores are $R_Bh(X)$ and centered functions
of $X$, while the target score is $\varepsilon_2$.
The first-order information is $I_1=\E[e_{B,0}]$.

\begin{proposition}[The full binary first-order class]
\label{prop:binaryclass}
Every normalized first-order value has the form
\begin{equation}
 \phi_0(W;a,b)=a(X)C_0+b(X)D_B,\qquad \E[aJ_0]=1,
 \label{eq:binaryclass}
\end{equation}
with square-integrable coefficients. Its variance and
nuisance curvature are
\begin{align}
 \E[\phi_0(W;a,b)^2]
 &=\E[a^2J_0+2ab\kappa_0 w_0+b^2w_0],
 \label{eq:binaryvar}\\
 B_{\phi_0}(h,\widetilde h)
 &=\E[bw_0\,h(X)\widetilde h(X)].
 \label{eq:binarycurv}
\end{align}
Curvatures involving a covariate-law direction are zero.
\end{proposition}

There are four conditional outcomes. Centering and
orthogonality to $R_B$ leave a two-dimensional space,
spanned by $C_0$ and $D_B$. The latter has zero target
sensitivity but can reduce variance through its covariance
with $C_0$. The same addition creates the curvature in
\eqref{eq:binarycurv}.

Let $\mathcal X_a$ be the atoms of $P_{0,X}$ and
$\pi_{x,0}=P_{0,X}\{x\}$, with $\pi_{x,0}=0$ elsewhere. Define
\begin{equation}
 \rho_{n,0}(x)=
 \frac{(n-1)\pi_{x,0}}{(n-1)\pi_{x,0}+2w_0(x)/V_0(x)^2},
 \qquad
 H_{n,0}=J_0-\kappa_0^2w_0\rho_{n,0},\qquad
 Z_{n,0}=\E[J_0^2/H_{n,0}].
 \label{eq:binaryweights}
\end{equation}

\begin{theorem}[Binary correctability and joint optimum]
\label{thm:binary}
A value in \eqref{eq:binaryclass} admits an exact $L_2$
correction if and only if $b=0$ on the nonatomic part and
$\sum_{x\in\mathcal X_a}b(x)^2e_{B,0}(x)^2<\infty$.
Its minimum kernel is
\begin{equation}
 k(W_i,W_j;b)=
 \sum_{x\in\mathcal X_a}
 \frac{\ind\{X_i=X_j=x\}}{\pi_{x,0}}
 \frac{b(x)w_0(x)}{V_0(x)^2}R_B(W_i)R_B(W_j),
 \label{eq:binarykernel}
\end{equation}
with squared norm $\sum_x b(x)^2e_{B,0}(x)^2$.
The jointly optimal normalized pair is
\begin{equation}
 a_n^*=\frac{J_0}{H_{n,0}Z_{n,0}},\qquad
 b_n^*=-\kappa_0\rho_{n,0} a_n^*,\qquad
 \phi_n^*=\phi_0(\cdot;a_n^*,b_n^*),\qquad
 k_n^*=k(\cdot,\cdot;b_n^*),
 \label{eq:binaryopt}
\end{equation}
with $I_{2,n}=Z_{n,0}$ and $V_{2,n}^*=1/(nZ_{n,0})$.
These values admit admissible local extensions. For a
fixed covariate law,
\begin{equation}
 Z_{n,0}\longrightarrow
 \E[e_{B,0}\ind\{X\in\mathcal X_a\}]
 +\E[J_0\ind\{X\notin\mathcal X_a\}].
 \label{eq:binarylimit}
\end{equation}
\end{theorem}

The fraction $\rho_{n,0}$ governs how much of $D_B$ is retained.
It approaches one in well-populated cells and equals zero
on the nonatomic part. Thus moment choice responds directly
to correction cost. For a purely atomic law, the optimum
approaches the full first-order information, even when the
fixed efficient moment has infinite correction cost.
On a nonatomic component, exact correction instead requires
using the conditional moment alone. If $\theta_0=0$,
then $\kappa_0=0$ and that moment is already efficient. Appendix~\ref{app:binary}
contains the proofs.

\subsection{Exact correction and exclusions in DiD}
\label{ssec:did}

Let $W=(D,Y,X)$, where $D$ indicates treatment, $Y$ is the change in
the outcome between two periods, and $X$ contains pretreatment
covariates. Under conditional parallel trends and no anticipation, the
average treatment effect on the treated is
$\theta_0=\E[Y-\E(Y\mid D=0,X)\mid D=1]$ \citep{Abadie2005,SZ2020}.
Applications often restrict which covariates drive each nuisance
function. Treatment may be assigned on a subset $U$ of the covariates,
as when eligibility or a randomization stratum is determined by
observed characteristics, while untreated outcome trends may depend
only on another subset $V$, such as region or industry. With $U$ and
$V$ specified functions of $X$, these exclusions read
\[
 \PP(D=1\mid X)=e_0(U),\qquad
 \E[Y\mid D=0,X]=\mu_0(V).
\]
Write $a_0(U)=e_0(U)/\{1-e_0(U)\}$ for the assignment odds and
$\sigma_0^2(X)=\Var(Y\mid D=0,X)$. Ordinary conditional DiD has
$U=V=X$. When $U$ and $V$ differ, the exclusions are overidentifying
restrictions that efficient estimators exploit, as the efficient DiD
estimators of \citet{CSX2025} exploit those generated by multiple
periods and comparison groups.

\begin{assumption}[DiD model]\label{ass:did}
The model consists of all laws of $W$, with $X$ valued in a standard
Borel space, whose propensity $e$, odds $a$, control mean $\mu$, and
control variance $\sigma^2$ satisfy the two exclusions and, for fixed constants
$M<\infty$, $0<a_-<a_+<\infty$, and $v_->0$, satisfy $|Y|\le M$,
$a_-\le a(U)\le a_+$, and $\sigma^2(X)\ge v_-$. At the true law, there is a constant $\delta>0$ such that,
almost surely,
\[
 a_-+\delta\le a_0(U)\le a_+-\delta,
 \qquad
 \sigma_0^2(X)\ge v_-+\delta.
\]
\end{assumption}

The model imposes no other restriction: the covariate law, the treated
outcome law, and the control outcome law beyond its mean are
unrestricted. Bounded likelihood-ratio perturbations that preserve the
exclusions therefore remain in the model, which delivers the
regularity conditions of Section~\ref{sec:theory}
(Appendix~\ref{app:didexact}).

Write $Q_0=\mathcal L(X\mid D=0)$, $c_0=\PP(D=0)$, and $\pi_0=1-c_0$. The usual
doubly robust moment is
\begin{equation}
 \psi(W,a,\mu,\theta)
 =D\{Y-\mu(V)-\theta\}
 -(1-D)a(U)\{Y-\mu(V)\}.
 \label{eq:did-early-moment}
\end{equation}
Its target sensitivity is $\pi_0$. For candidate functions $a',\mu'$,
its bias is exactly
\begin{equation}
 \E[\psi(W,a',\mu',\theta_0)]
 =c_0\E_{Q_0}[(a'-a_0)(\mu'-\mu_0)].
 \label{eq:did-early-bias}
\end{equation}
Define $\varepsilon=Y-\mu_0(V)$ and the normalized value
$\phi_{D,0}=\psi(W,a_0,\mu_0,\theta_0)/\pi_0$.

\begin{proposition}[All first-order DiD moments]
\label{prop:didclass}
Under Assumption~\ref{ass:did}, the full normalized first-order class
is
\begin{equation}
 \begin{split}
 \phi_0(W;\rho,\beta)
 &=\phi_{D,0}+\pi_0^{-1}
   \{(D-e_0(U))\rho(X)+(1-D)\varepsilon\beta(X)\},\\
 \E_{Q_0}[\rho\mid U]&=0,\qquad
 \E_{Q_0}[\beta\mid V]=0,
 \end{split}
 \label{eq:didclass}
\end{equation}
where $\rho,\beta\in L_2(Q_0)$. Every value has an admissible local
extension.
\end{proposition}

The assignment exclusion permits the first adjustment; the outcome
exclusion permits the second. Let $\phi_{E,0}$ be the minimum-variance
member, and write $V_{E,0}=\E[\phi_{E,0}^2]$ and $V_{D,0}=\E[\phi_{D,0}^2]$. Its
coefficients and the variance decomposition are given in
Appendix~\ref{app:didexact}. When $U=V=X$, both centering restrictions
force zero adjustments and $\phi_{D,0}$ is efficient, as in
\citet{SZ2020}.

Let $T:L_2(Q_{V,0})\to L_2(Q_{U,0})$ be the conditional-expectation operator
$Tf=\E_{Q_0}[f(V)\mid U]$. The next result characterizes exact correction
over the entire class in \eqref{eq:didclass}.

\begin{theorem}[Full exact DiD moment choice]
\label{thm:didexact}
Suppose Assumption~\ref{ass:did} holds and $Q_{U,0},Q_{V,0}$ are nonatomic.
Any normalized first-order value admitting an admissible exact $L_2$
correction must equal $\phi_{D,0}$. Such a correction exists if and only if
\begin{equation}
 Q_{UV,0}\ll Q_{U,0}\otimes Q_{V,0},\qquad
 R_0=\frac{dQ_{UV,0}}{d(Q_{U,0}\otimes Q_{V,0})}
   \in L_2(Q_{U,0}\otimes Q_{V,0}),
 \label{eq:didexist}
\end{equation}
equivalently, if $T$ is Hilbert--Schmidt. When this condition holds,
put $A_D=D-(1-D)a_0(U)$ and $B_D=(1-D)\varepsilon$. A correcting kernel
is
\begin{equation}
 k_0(W_i,W_j)=
 \frac{A_D(W_i)R_0(U_i,V_j)B_D(W_j)
       +A_D(W_j)R_0(U_j,V_i)B_D(W_i)}{2\pi_0 c_0}.
 \label{eq:didK0}
\end{equation}
With $k_{\min,0}=\Pi_{\mathcal Q}k_0$, where $\mathcal Q$ is the
score-product space in Section~\ref{sec:theory}, the full-class
optimum is
\begin{equation}
 V_{2,n}^*
 =\frac{V_{D,0}}{n}+\frac{2\|k_{\min,0}\|^2}{n(n-1)}.
 \label{eq:didopt}
\end{equation}
The minimum is attained by a locally valid, degenerate moment family.
\end{theorem}

The unknown covariate law creates additional restrictions. A nonzero
$\rho$ or $\beta$ introduces mixed curvature between a regression
perturbation and a perturbation of $Q_0$. On a nonatomic marginal this
curvature is a multiplication operator, which has no $L_2$ kernel.
Hence exact correction removes every efficiency-improving adjustment.

Condition \eqref{eq:didexist} can hold for dependent, continuously
distributed $U,V$: bounded joint densities and marginal densities
bounded away from zero on a compact product support give one
sufficient case. If $U$ and $V$ share a nonatomic coordinate, however,
their joint law is concentrated on equality of the two copies of that
coordinate, a set with zero product-marginal probability, and exact
correction is impossible throughout the class. Even when exact
correction exists, its cost relative to first-order efficiency is
\begin{equation}
 V_{2,n}^*-\frac{V_{E,0}}{n}
 =\frac{V_{D,0}-V_{E,0}}{n}
  +\frac{2\|k_{\min,0}\|^2}{n(n-1)}.
 \label{eq:didlosssplit}
\end{equation}
The first term is the information lost by restricting the moment; the
second is the variance of its correction.

\subsection{Discrete covariates and shared directions}
\label{ssec:did-discrete}

Covariates in DiD applications are often discrete: region, industry,
cohort, or demographic cells, frequently interacted. Discreteness
changes the conclusion of Theorem~\ref{thm:didexact}, as it does in
binary choice. The obstruction there is a multiplication operator on
$L_2(Q_0)$, which is Hilbert--Schmidt when $X$ takes finitely many
values.

\begin{corollary}[Finitely many covariate cells]
\label{cor:diddiscrete}
Suppose Assumption~\ref{ass:did} holds and $X$ takes finitely many
values. Then every first-order value in \eqref{eq:didclass} has finite
correction cost, and $nV_{2,n}^*\downarrow V_{E,0}$ as $n\to\infty$.
\end{corollary}

With few cells relative to the sample size, the efficient moment can
thus be corrected exactly at a negligible cost. The cost grows with
the number of cells, however, and with many cells, as in saturated
specifications with interactions, it becomes first order. Along cells
shared by the assignment and outcome covariates, no choice of moment
avoids it.

Write $X=(C,X_V,X_U)$, $U=(C,X_U)$, and $V=(C,X_V)$, where $C$
collects the shared covariates. Choose a bounded vector
$h(C)\in\R^d$ satisfying
\[
 \E_{Q_0}[hh']=I_d,\qquad
 \E_{Q_0}h=\E_{Q_0}[a_0h]=\E_{Q_0}[\Delta_0 h]=0,
 \qquad
 \Delta_0(X)=\E[Y\mid D=1,X]-\mu_0(V).
\]
Define $A_0=\E_{Q_0}a_0=\pi_0/c_0$ and
\[
 \begin{aligned}
 s_a&=D h(C)/a_0(U),
 &G_{a,0}&=c_0\E_{Q_0}[hh'/a_0],\\
 s_\mu&=(1-D)\varepsilon h(C)/\sigma_0^2(X),
 &G_{\mu,0}&=c_0\E_{Q_0}[hh'/\sigma_0^2].
 \end{aligned}
\]
These are target-preserving scores, with Gram matrices $G_{a,0},G_{\mu,0}$.
If $C$ takes $K$ values, $d$ can be as large as $K-3$.

\begin{theorem}[A sharp shared-direction cost]
\label{thm:shared}
Under Assumption~\ref{ass:did}, every normalized pair
$(\phi,k)\in\cN\times\cK$ correcting all second-order restrictions on
the span of these $2d$ scores satisfies
\begin{equation}
 \Var(\Pn\phi-\Un k)
 \ge\frac{V_{E,0}}{n}
 +\frac{\operatorname{tr}(G_{a,0}^{-1}G_{\mu,0}^{-1})}
       {A_0^2n(n-1)}
 \ge\frac{V_{E,0}}{n}
 +\frac{a_-v_-}{\pi_0^2}\frac d{n(n-1)}.
 \label{eq:sharedbound}
\end{equation}
The first bound is attained when these are the only second-order
restrictions, retaining all first-order restrictions. If $L_2(Q_{C,0})$ is
infinite dimensional, no relevant finite-variance pair corrects every
second nuisance derivative.
\end{theorem}

The curvature in these directions is identical for every normalized
first-order moment, so moment choice cannot reduce this part of the
cost, which grows linearly with the number of shared cells. The
attaining kernel is given in Appendix~\ref{app:shared}. 
\subsection{Approximation after choosing the DiD moment}
\label{sec:didapprox}

Approximate correction must also account for the nuisance functions
introduced by moment choice. Choose bounded $f(X),v(X)$ on a training
sample, and define
\[
 r_{f,0}(U)=\E_{Q_0}[f(X)\mid U],\qquad
 t_{v,0}(V)=\E_{Q_0}[v(X)\mid V].
\]
Let $\widehat a,\widehat\mu,\widehat r,\widehat t$ be fitted functions,
with errors $\delta_a=\widehat a-a_0$, $\delta_\mu=\widehat\mu-\mu_0$,
$\delta_r=\widehat r-r_{f,0}$, and $\delta_t=\widehat t-t_{v,0}$, and set
$\widehat e=\widehat a/(1+\widehat a)$. The unnormalized fitted moment
is
\begin{equation}
 \widehat\psi(W,\theta)=\psi(W,\widehat a,\widehat\mu,\theta)
 +(D-\widehat e)(f-\widehat r)
 +(1-D)(Y-\widehat\mu)(v-\widehat t).
 \label{eq:didfitmoment}
\end{equation}

\begin{proposition}[The exact three-product bias]
\label{prop:threebias}
Conditional on training,
\begin{equation}
 \E[\widehat\psi(W,\theta_0)]
 =c_0\left\{
 \E_{Q_0}[\delta_a\delta_\mu]
 +\E_{Q_0}\left[
   \frac{\delta_a}{1+\widehat a}\delta_r
 \right]
 +\E_{Q_0}[\delta_\mu\delta_t]
 \right\}.
 \label{eq:threebias}
\end{equation}
At the true nuisance values, $\widehat\psi(W,\theta_0)/\pi_0$ equals
\eqref{eq:didclass} with $\rho=f-r_{f,0}$ and $\beta=v-t_{v,0}$.
\end{proposition}

The original bias product is accompanied by two products from the
estimated centering regressions, and the correction must reproduce all
three. Appendix~\ref{app:didfeasible} constructs such a correction from
cell projections and an estimated covariate density, and gives rates
under which the estimator is asymptotically linear with influence
function $\phi_0(W;\rho,\beta)$. Approximate correction thus recovers
$V_{E,0}$ even when exact correction is impossible.
%%%%%%%%%%%%%%%%%%%%%%%%%%%
\section{Conclusion}
\label{sec:conclusion}

Second-order orthogonality has a variance cost that depends on the
moment and may be infinite. We characterize exact correctability and
the minimum oracle variance over the full first-order class,
connecting finite correction problems to their infinite-dimensional
limit. The first-order efficient moment need not remain optimal after
correction, and with continuous covariates exact correction can
prevent the use of information supplied by valid exclusions; with
discrete covariates it does not, but its cost grows with the number of
cells.

For projection nuisances, curvature matching yields feasible
approximate corrections. ADA compares the regression-error components
captured by ridge and selected-projection kernels within a variance
budget, and its oracle inequality controls a bound on the remaining
quadratic bias. When the correction variance is negligible, the
selected estimator is first-order equivalent to the oracle under a
bias condition on the error components that the correction leaves
out. When the correction contributes to first-order variance,
the standard error includes this contribution. In the PLM, validity after selection extends to overlapping
pair-block training under correction-matrix stability. This
condition covers covariate-only and auxiliary-sample matrices,
and selected projections with stable selection or asymptotically
negligible matrix size. The PLM
experiments illustrate the value of combining the two families and
the limits of a variance budget when substantial bias remains.

\clearpage
\appendix
\begin{center}
{\Large\bfseries Appendix}
\end{center}
\section{Proofs for Section~\ref{sec:theory}}
\label{app:pathwise}

\subsection{Notation and pathwise regularity}
\label{app:regularity_paths}

\paragraph{Derivatives.}
For a map $\eta\mapsto f(\cdot,\eta)$ into $L_2(P_0)$ or
$L_2(P_0^2)$ and directions $h,g\in\cH$, write
\[
 D_\eta f(\cdot,\eta)[h]=\frac{d}{dt}f(\cdot,\eta+th)\Big|_{t=0},
 \qquad
 D_\eta^2f(\cdot,\eta)[h,g]=\frac{\partial^2}{\partial t\,\partial v}
 f(\cdot,\eta+th+vg)\Big|_{t=v=0}.
\]
The notation $D_\eta\phi_0[h]$, $D_\eta^2\phi_0[h,g]$, and
$\partial_\theta\phi_0$ means that the family is differentiated first
and then evaluated at $\lambda_0$; the same convention applies to $k$.
With $r(\lambda)=dP_\lambda/dP_0$, the scores are
$s_h=D_\eta r(\lambda_0)[h]$,
$u_{h,g}=D_\eta^2r(\lambda_0)[h,g]$, and
$\ell=\partial_\theta r(\lambda_0)$, with the same evaluation convention.
At nearby laws, $s_h(\lambda)=D_\eta r(\lambda)[h]/r(\lambda)$
is the nuisance score of $P_\lambda$.

\paragraph{Pathwise regularity.} The conditions in Assumption~\ref{ass:regularity} apply along each
finite-dimensional parameter path
\[
 \lambda(\mathbf t)=
 \Bigl(\theta_0+t_\theta,\,\eta_0+\sum_{j=1}^d t_j h_j\Bigr),
 \qquad
 \mathbf t=(t_\theta,\ldots,t_d),\quad h_1,\ldots,h_d\in\cH.
\]
For every such collection of directions, there is an open neighborhood
of $\mathbf t=0$ on which the path lies in the parameter space and (i)
$P_{\lambda(\mathbf t)}$ is equivalent to $P_0$; (ii) the likelihood-ratio map
$\mathbf t\mapsto r(\lambda(\mathbf t))$ is twice continuously differentiable into
$L_2(P_0)$; and (iii) for each moment family considered, the maps
$\mathbf t\mapsto\phi(\cdot,\lambda(\mathbf t))$ and $\mathbf t\mapsto k(\cdot,\cdot,\lambda(\mathbf t))$ are twice
continuously differentiable into $L_2(P_0)$ and $L_2(P_0^2)$. Square
integrability under the corresponding laws, local validity,
degeneracy, and first-order orthogonality are required by
Definition~\ref{def:admissible}. The scores belong to $L_2^0(P_0)$
because the likelihood ratio integrates to one. Expectations under
$P_{\lambda(\mathbf t)}$ and its product law are continuous $L_2$ pairings with
$r(\lambda(\mathbf t))$ and $r(\lambda(\mathbf t))(W_i)r(\lambda(\mathbf t))(W_j)$, which
justifies differentiating the validity and degeneracy identities;
equivalence of the laws allows the conditional degeneracy identities
to be treated on the fixed reference space.

\subsection{Admissible extensions}
\label{app:extensions}

\begin{definition}[Admissible extension]
\label{def:extension}
An admissible extension of a pair
$(\phi,k)\in\cN\times\cK$ is a family
$(\phi(\cdot,\lambda),k(\cdot,\cdot,\lambda))$ defined on a neighborhood
of $\lambda_0$, satisfying
Definition~\ref{def:admissible} and the regularity
conditions in Appendix~\ref{app:regularity_paths},
with
\[
 \phi_0=\phi
 \quad P_0\text{-almost surely},
 \qquad
 k_0=k
 \quad P_0^2\text{-almost surely}.
\]
\end{definition}

For a compatible pair, Lemma~\ref{lem:pathwise}
implies that any admissible extension is
second-order orthogonal at $\lambda_0$.
Existence of an extension is required separately
when asserting attainment by a moment family.

\subsection{Proofs}
\label{app:general}

\begin{proof}[Proof of Lemma~\ref{lem:pathwise}]
We use the notation of Appendix~\ref{app:regularity_paths}; moment values and the displayed derivatives are evaluated at the truth. Along the two-parameter nuisance path, the likelihood ratio has first derivatives $s_h,s_g$ and mixed derivative $u_{h,g}$.

\emph{First-order value and sensitivity.} Differentiate validity, $\E_\lambda \phi(\cdot,\lambda)=0$, in direction $h$. The product rule gives
\[
 0=\E[D_\eta\phi_0[h]]+\E[\phi s_h].
\]
The first term is zero by local first-order orthogonality. Thus $\phi\in\cN$. A target derivative of the same validity identity gives $-\E[\partial_\theta\phi_0]=\E[\phi\ell]$. A first derivative of $\E_\lambda k(\cdot,\cdot,\lambda)=0$ gives
\[
 0=\E[D_\eta k_0[h]]+\E[k\{s_h(W_i)+s_h(W_j)\}].
\]
Each score term is zero by degeneracy. The argument with the target score also applies, so $\E[D_\eta k_0[h]]=\E[\partial_\theta k_0]=0$. This proves the relevance and first-order nuisance orthogonality of the correction.

\emph{Curvature of the first-order moment.} Differentiate the identity $\E_\lambda[D_\eta\phi(\cdot,\lambda)[h]]=0$ in direction $g$:
\[
 \E[D_\eta^2\phi_0[h,g]]+\E[D_\eta\phi_0[h]s_g]=0.
\]
Interchanging $h,g$ gives $\E[D_\eta^2\phi_0[h,g]]+\E[D_\eta\phi_0[g]s_h]=0$. Twice differentiating validity gives the four separate terms
\[
 0=\E[D_\eta^2\phi_0[h,g]]+\E[D_\eta\phi_0[h]s_g]+\E[D_\eta\phi_0[g]s_h]+\E[\phi u_{h,g}].
\]
Substitute the preceding two equalities. The right side becomes $-\E[D_\eta^2\phi_0[h,g]]+\E[\phi u_{h,g}]$. Hence $\E[D_\eta^2\phi_0[h,g]]=\E[\phi u_{h,g}]$.

\emph{Derivatives of degeneracy.} The identity in the second argument is
$\E_\lambda[k(w_i,W_j,\lambda)]=0$. Its derivative gives
\begin{equation}
 \E[D_\eta k_0[h](w_i,W_j)]=-\E[k(w_i,W_j)s_h(W_j)].\label{eq:degdiff1}
\end{equation}
The identity in the first argument gives, separately,
\begin{equation}
 \E[D_\eta k_0[h](W_i,w_j)]=-\E[k(W_i,w_j)s_h(W_i)].\label{eq:degdiff2}
\end{equation}
For example, multiply \eqref{eq:degdiff2} by $s_g(w_j)$ and integrate $w_j$. Fubini and Cauchy--Schwarz yield
\[
 \E[D_\eta k_0[h](W_i,W_j)s_g(W_j)]
 =-\E[k(W_i,W_j)s_h(W_i)s_g(W_j)]
 =-(\mathcal Rk)(h,g).
\]
Using \eqref{eq:degdiff1} instead gives the same value with $s_g(W_i)$. Symmetry of $k$ makes $\mathcal Rk$ symmetric in $h,g$. The two expressions with $D_\eta k_0[g]$ and $s_h$ therefore also equal $-\mathcal Rk(h,g)$.

\emph{Second derivative of the mean of the kernel.} The likelihood ratio for two independent observations is the product of their likelihood ratios. Its mixed derivative at zero is
\[
 u_{h,g}(W_i)+u_{h,g}(W_j)
 +s_h(W_i)s_g(W_j)+s_g(W_i)s_h(W_j).
\]
Consequently twice differentiating $\E_\lambda k(\cdot,\cdot,\lambda)=0$ gives
\begin{align*}
0={}&\E[D_\eta^2k_0[h,g]]\\
 &+\E[D_\eta k_0[h]\{s_g(W_i)+s_g(W_j)\}]\\
 &+\E[D_\eta k_0[g]\{s_h(W_i)+s_h(W_j)\}]\\
 &+\E[k\{u_{h,g}(W_i)+u_{h,g}(W_j)\}]\\
 &+\E[k\{s_h(W_i)s_g(W_j)+s_g(W_i)s_h(W_j)\}].
\end{align*}
The second and third lines each equal $-2\mathcal Rk(h,g)$. The fourth is zero by degeneracy. The last equals $2\mathcal Rk(h,g)$. Thus $\E[D_\eta^2k_0[h,g]]=2\mathcal Rk(h,g)$. Combining this with $\E[D_\eta^2\phi_0[h,g]] = \E[\phi u_{h,g}] = 2B_\phi(h,g)$ proves the Lemma by a second-order expansion of $\mathcal M_0(\theta_0,\eta_0+th)$.
\end{proof}

\begin{proposition}[A sufficient extension condition]\label{prop:extension}
Let $r(\lambda)=dP_\lambda/dP_0$ be bounded above and away from zero and twice continuously differentiable in $L_\infty$ along the paths. On $L_2(P_0)$, suppose the projections $\Pi_\lambda^\perp$ onto the orthogonal complement of
\[
 \cl{\spanof\{\sqrt{r(\lambda)},\sqrt{r(\lambda)}s_h(\lambda):h\in\cH\}}
\]
are twice differentiable in operator norm along those paths. Then every pair $(\phi,k) \in \cN \times \mathcal{K}$ admits an admissible extension in the sense of Definition \ref{def:extension}.
\end{proposition}
\begin{proof}
For $\phi\in\cN$, set $\phi(\cdot,\lambda)=r(\lambda)^{-1/2}\Pi_\lambda^\perp\phi$. At the truth it equals $\phi$. Orthogonality to $\sqrt r$ gives validity; orthogonality to $\sqrt r s_h$ gives $\E_\lambda[\phi(\cdot,\lambda) s_h(\lambda)]=0$. Differentiating validity then gives \eqref{eq:uniformfirst}. For $k\in\cK$, use double centering:
\[
 k(w_i,w_j,\lambda)=k(w_i,w_j)-\E_\lambda k(w_i,W)
                   -\E_\lambda k(W,w_j)+\E_\lambda k.
\]
This has value $k$ at the truth, is locally degenerate, and has the required differentiability by the bounded likelihood-ratio conditions. These are local oracle extensions; the argument asserts neither knowledge of $P_0$ nor a rate for estimating the projections.
\end{proof}

\begin{proof}[Proof of Theorem~\ref{thm:cost}]
Recall the closed span $\mathcal Q$ of $Q_{h,g}$ in $\cK$. A correcting kernel defines the functional $\sum a_j Q_{h_j,g_j}\mapsto\sum a_j B_\phi(h_j,g_j)$ by inner product with $k$, so Cauchy--Schwarz gives \eqref{eq:cost-criterion} with $C=\|k\|$. Conversely, finite $C$ makes this functional well-defined: any zero linear combination on the right must have zero value on the left. It is bounded on the algebraic span and extends continuously to $\mathcal Q$. Riesz representation gives a unique $k_\phi\in\mathcal Q$ with functional norm $c_\phi$. It is symmetric and degenerate because $\mathcal Q\subset\cK$ is closed. Any other correcting kernel differs by an element of $\mathcal Q^\perp$, proving minimality and \eqref{eq:mincost}. If $(\phi, k_\phi)$ admit an admissible extension, Lemma~\ref{lem:pathwise} establishes its second-order orthogonality. Relevance follows from the definition.
\end{proof}

\begin{proof}[Proof of Proposition~\ref{prop:spectral}]
For any correcting kernel, testing on $e_j,e_l$ gives
\[
 \ip{k}{f_j\otimes f_l}
 =\frac{b_{jl}}{\sqrt{I_{\eta,j}I_{\eta,l}}}.
\]
The tensors are orthonormal, so Bessel's inequality bounds $\|k\|^2$ below by the sum in \eqref{eq:spectral}. If the sum is finite, the series with these coefficients converges in $L_2(P_0^2)$ to a symmetric degenerate kernel attaining this bound. Its correction identity holds on finite linear combinations of the $e_j$ and extends to $(\ker S)^\perp$ by continuity of $S$ and $B_\phi$. The vanishing condition extends the identity to all of $\cH$. If that condition fails, $\mathcal Rk=B_\phi$ is impossible because $Sh=0$ implies $(\mathcal Rk)(h,g)=0$ for every $k$ and $g$. Theorem~\ref{thm:cost} gives \eqref{eq:spectral}.
\end{proof}

\begin{proof}[Proof of Proposition~\ref{prop:support}]
The map $f\mapsto R'\Sigma^{-1/2}f(X)$ is an isometry from $L_2(P_{0,X})^r$ into the nuisance score block, where $r$ is the length of $R$. In these coordinates the curvature operator is multiplication by $M(X)=\Sigma^{-1/2}C_\phi\Sigma^{-1/2}$. A correcting $L_2$ kernel exists exactly when this multiplication operator is Hilbert--Schmidt.

On a nonatomic set where an entry of $M$ is nonzero, localize further so this entry is bounded and bounded away from zero. The resulting multiplication map is bounded below on an infinite-dimensional $L_2$ subspace. The sum of its squared values on an orthonormal sequence is infinite. Thus $M$ must vanish on the nonatomic part. On the atoms, the normalized indicators $\ind\{X=x\}/\sqrt{\pi_x}$ form an orthonormal basis. Summing over atoms and vector coordinates gives exactly \eqref{eq:atomiccost}. Transforming back gives the displayed kernel. This also proves sufficiency and minimality.
\end{proof}

\begin{proof}[Proof of Corollary~\ref{cor:plm-impossible}]
Write $\sigma^2(D,X)=\E[\varepsilon^2\mid D,X]$.
By the conditional-moment characterization cited in the main
text,
\[
 \phi=\varepsilon r(D,X),\qquad
 \E[r\mid X]=0,\qquad \E[rV]\ne0.
\]
Since $\phi\in L_2$ and $\sigma^2$ is bounded away from zero,
$r\in L_2$. In particular, $r$ is not identically zero.

We first choose a bounded instrument for perturbing the law
of $D$ given $X$. For $M>0$, let
\[
 r_M=\max\{-M,\min\{r,M\}\},\qquad
 q=r_M-\E[r_M\mid X],\qquad
 b(X)=\E[rq\mid X].
\]
Then $\E[q\mid X]=0$ and
$b(X)=\E[rr_M\mid X]\ge0$, with $\E[b]>0$.
Moreover, $b>0$ implies $\E[q^2\mid X]>0$.
Hence there is a measurable set $A$ of positive probability
on which $b$ and $\E[q^2\mid X]$ are bounded above and away
from zero.

We next construct a bounded conditional outcome score.
Choose $L<\infty$ such that $C_4/L^2\le v_-/2$, and put
\[
 T_L=\max\{-L,\min\{\varepsilon,L\}\},\qquad
 d_L(D,X)=\E[\varepsilon T_L\mid D,X].
\]
Since
\[
 d_L(D,X)
 \ge \E[\varepsilon^2\ind\{|\varepsilon|\le L\}\mid D,X]
 \ge v_--C_4/L^2
 \ge v_-/2,
\]
the function
\[
 \zeta(W)=
 \frac{T_L-\E[T_L\mid D,X]}{d_L(D,X)}
\]
is bounded and satisfies
\[
 \E[\zeta\mid D,X]=0,\qquad
 \E[\varepsilon\zeta\mid D,X]=1.
\]

For bounded directions $h=(h_1,h_2)'$ supported on $A$,
define the likelihood ratio
\[
 \frac{dP_t}{dP_0}
 =\{1+t h_2(X)q(D,X)\}
  \{1+t h_1(X)\zeta(W)\}.
\]
Both factors are positive for sufficiently small $|t|$.
Successive conditioning shows that the ratio integrates
to one and preserves the marginal law of $X$. The
conditional law of $D$ given $X$ has likelihood ratio
$1+t h_2(X)q(D,X)$, and
\[
 \E_t[Y\mid D,X]=\theta_0D+c_0(X)+t h_1(X).
\]
Thus the path remains in the PLM with $\theta=\theta_0$.
Its likelihood ratio is polynomial with bounded
coefficients, giving twice continuous $L_2$
differentiability. The same construction applies jointly
to finite collections of directions. Its score is
\[
 s_h=R'h,\qquad
 R=\begin{pmatrix}\zeta(W)\\q(D,X)\end{pmatrix}.
\]

Keeping $\phi$ evaluated at the original law gives the
exact identity
\[
 \E_t[\phi]
 =t\,\E\!\left[h_1(X)\E_t[r\mid X]\right]
 =t^2\E[b(X)h_1(X)h_2(X)].
\]
Consequently, polarization yields
\[
 B_\phi(h,g)=\E[h'C_\phi g],\qquad
 C_\phi(X)=\frac{b(X)}2
 \begin{pmatrix}0&1\\1&0\end{pmatrix}.
\]
Also, $\E[R\mid X]=0$ and
\[
 \Sigma(X)=\operatorname{diag}\left\{
   \E[\zeta^2\mid X],\,\E[q^2\mid X]
 \right\}.
\]
Conditional Cauchy--Schwarz gives
$\E[\zeta^2\mid D,X]\ge1/\sigma^2(D,X)$.
Since $T_L^2\le\varepsilon T_L$,
\[
 \E[\zeta^2\mid D,X]
 =\frac{\Var(T_L\mid D,X)}{d_L(D,X)^2}
 \le\frac1{d_L(D,X)}
 \le\frac2{v_-}.
\]
Therefore
\[
 \frac1{\sqrt{C_4}}\le\E[\zeta^2\mid X]\le\frac2{v_-}.
\]
On $A$, the eigenvalues of $\Sigma$ are bounded above and
away from zero, while $C_\phi$ is nonzero.
The restriction of $P_{0,X}$ to $A$ is nonatomic.
Proposition~\ref{prop:support}, applied to directions
supported on $A$, rules out an $L_2$ correction for this
block. Any correction for the full model would also
correct this block, so $c_\phi^2=\infty$.
\end{proof}
\begin{proof}[Proof of Theorem~\ref{thm:joint}]
For each $h,g$, the functional $(\phi,k)\mapsto\ip{k}{Q_{h,g}}-\E[\phi u_{h,g}]/2$ is continuous. Its kernel is closed, and the intersection of these kernels with $\cN\times\cK$ is $\mathcal C$. On this Hilbert space the target functional is
\[
 J(\phi,k)=\E[\phi\ell]=\ip{(\ell_{\eff},0)}{(\phi,k)}_n=\ip{v_n}{(\phi,k)}_n.
\]
Cauchy--Schwarz gives $J^2\le\|v_n\|_n^2\|(\phi,k)\|_n^2$, with equality at a multiple of $v_n$. Hence $I_{2,n}=\|v_n\|_n^2$, and a value with $J=1$ exists exactly when it is positive. Its minimum norm is attained uniquely at $v_n/I_{2,n}$.

For completeness, the sample variance follows directly from degeneracy. Distinct pair terms have zero covariance even if they share one observation, and every pair term is uncorrelated with every single-observation term. Thus $\Var(\Un k)=\binom n2^{-1}\|k\|^2$, proving \eqref{eq:oracle_variance}. Every normalized $\phi$ has the form $\phi_{\eff}+a$, $a\in\cA$, and $\phi_{\eff}\perp\cA$. This proves \eqref{eq:joint-decomp}. Minimizing first over $k$ gives \eqref{eq:profile}. The minimizer is in the compatible space even if the range of $\mathcal R$ is not closed; no closed-range assumption has been used.
\end{proof}

\begin{proof}[Proof of Theorem~\ref{thm:limit}]
Fix $n$. The normalized feasible sets are nested closed affine subsets of the product Hilbert space. Their minimum squared norms increase. If they are bounded, choose their minimizers $x_d$. A weakly convergent subsequence exists. For every fixed $j$, the tail belongs to the $j$th closed convex feasible set, which is weakly closed. Its weak limit therefore satisfies every restriction and target normalization. Lower semicontinuity of the norm and the reverse inequality from feasibility of the infinite optimum imply that the limiting value is $nV_{2,n}^*$. Uniqueness identifies every weak subsequential limit with $x^*$. Also $\|x_d\|_n\to\|x^*\|_n$; weak convergence and convergence of norms imply strong convergence. If the infinite optimum exists, its norm bounds every finite optimum, proving both directions.

For the information limit, profile the kernel in \eqref{eq:information2}:
\[
 I_{2,n}=\sup_{\phi\in\cD\setminus\{0\}}
 \frac{\ip{\ell_{\eff}}{\phi}^2}{\|\phi\|^2+\frac{2}{n-1}c_\phi^2}.
\]
The quotient supremum is interpreted as zero if $\cD=\{0\}$.
It increases with $n$ and is bounded above by $\|\Pi_{\cl\cD}\ell_{\eff}\|^2$. For each fixed $\phi\in\cD$, its quotient converges to $\ip{\ell_{\eff}}{\phi}^2/\|\phi\|^2$. Taking the supremum after this lower bound gives the reverse limiting inequality.
\end{proof}

\begin{proof}[Proof of Proposition~\ref{prop:finite}]
Whiten the scores: $z=\Gamma_d^{-1/2}s$. The tensors $z_j\otimes z_l$ are orthonormal. The restrictions require their kernel coefficients to equal the entries of $C_\phi$. Their tensor expansion is the displayed $k_{\phi,d}$, whose squared norm is $\|C_\phi\|_F^2$; any orthogonal component only increases norm.

The remaining objective is $I_1^{-1}+\|a\|^2+\frac{2}{n-1}\|C_{\rm eff}+La\|_F^2$. Because $L$ is bounded and finite rank, its first-order equation is
$(I+\frac{2}{n-1}L^*L)a=-\frac{2}{n-1}L^*C_{\rm eff}$. The strictly positive operator on the left is invertible. Substitution, or completing the square, gives \eqref{eq:finite-value}. The range of $L^*$ is spanned by $\Pi_{\cA}u_{h_j,h_l}$, since the matrix entries are linear combinations of $\E[au_{h_j,h_l}]/2$.
\end{proof}

\begin{proof}[Proof of Proposition~\ref{prop:frontier}]
Decompose $a=a_R+a_{\rm null}$ with $a_R\in\operatorname{range}(L^*)$ and $a_{\rm null}\in\ker L$. Dropping $a_{\rm null}$ preserves the constraint and reduces the objective. The problem therefore has a finite-dimensional closed convex feasible set and a coercive strictly convex objective. It has a unique minimizer.

Fix $a$, let $C=C_{\rm eff}+La$, and change to its eigenbasis. Replacing $M$
by its diagonal part reduces its Frobenius norm and cannot increase
$\|C-M\|_{\op}$: diagonal pinching is a contraction for this norm.
Each diagonal entry then minimizes $m_j^2$ subject to
$|\nu_j-m_j|\le\delta$, so the minimizing $M$ soft-thresholds the
eigenvalues of $C$ at $\delta$. Taking $a=a_{n,d,\delta}^*$ gives
\eqref{eq:soft}. Continuity follows, for example, by substituting this continuous fixed-$a$ solution and using coercivity to confine minimizers on compact sets when $\delta$ varies on a compact interval. Monotonicity follows from nested constraints. At $\delta=0$ the constraint is exact; if $\delta\ge\|C_{\rm eff}\|_{\op}$, $a=M=0$ attains the unconstrained lower bound.
\end{proof}

\section{Proofs for Section~\ref{sec:approx}}
\label{app:auto}

This appendix follows the order of Section~\ref{sec:approx}; the
table lists where each result is proved.

\begin{center}
\small
\begin{tabular}{lll}
\toprule
Section & Result & Appendix\\
\midrule
\ref{sec:auto-bias}
 & Proposition~\ref{prop:pointwise-cost}; Corollary~\ref{cor:nonlinear-regression-cost}
 & \ref{app:auto-curvature}\\
\ref{sec:auto-kernel}
 & Proposition~\ref{prop:auto-curvature} (remaining curvature)
 & \ref{app:auto-curvature}\\
\ref{sec:auto-estimate}
 & Population ridge and projection operators
 & \ref{app:auto-operators}\\
\ref{sec:auto-estimator}
 & Pair-block cross-fitting; the PLM equation
 & \ref{app:auto-crossfit}\\
\ref{sec:auto-selection}
 & Bound \eqref{eq:auto-bias-control};
   Proposition~\ref{prop:auto-selection} (oracle bound)
 & \ref{app:auto-select}\\
\ref{sec:auto-inference}
 & Theorem~\ref{thm:auto-inference}; Corollary~\ref{cor:plm}
 & \ref{app:auto-negligible}\\
\ref{sec:auto-inference}
 & Theorem~\ref{thm:nonneg-inference}
 & \ref{app:nonneg}\\
\bottomrule
\end{tabular}
\end{center}

\subsection{Conventions}
\label{app:auto-conventions}

\paragraph{Expectations of fitted objects.}
An expectation of a fitted quantity holds its training fit
fixed and integrates over fresh observations. For instance,
$\E[\widehat k_{b,\tau}(W_i,W_j,\theta_0)]$ integrates
$(W_i,W_j)$ under $P_0^2$ with the block-$b$ fit fixed.
Different blocks have different, overlapping training
samples, so the arguments below are applied one block at a
time; independence across blocks is never used. Norms and
inner products are those of $L_2(P_0)$ or $L_2(P_0^2)$, as
in Section~\ref{sec:theory}; for functions of $X$,
$\langle f,g\rangle=\E[f(X)g(X)]$.

\paragraph{Nuisance errors.}
For the fit trained on $I_b^c$, write
$h_{\gamma,b}=\widehat\gamma_b-\gamma_0$ and
$h_{\alpha,b}=\widehat\alpha_b(\cdot,\theta_0)-\alpha_0$. In the PLM at $\theta_0$,
$h_\alpha=-h_e$ in the notation of Section~\ref{sec:motivation},
and the block residuals satisfy
\begin{equation}
\widehat V_{i,b}=V_i-h_{\gamma,b}(X_i),
\qquad
\widehat e_{i,b}(\theta_0)=\varepsilon_i+h_{\alpha,b}(X_i).
\label{eq:app-auto-residuals}
\end{equation}
Hence, for $i\in I_b$,
$\E[\widehat V_{i,b}\mid X_i]=-h_{\gamma,b}(X_i)$ and
$\E[\widehat e_{i,b}(\theta_0)\mid X_i]=h_{\alpha,b}(X_i)$.

\paragraph{Weights and averages.}
Put $\omega_b=|\mathcal P_b|/\{n(n-1)\}$, so that
$\sum_b\omega_b=1$, and write
\[
\mathbb U_bf=\frac1{|\mathcal P_b|}
\sum_{(i,j)\in\mathcal P_b}f(W_i,W_j),
\qquad
\widehat\Psi_{b,\tau}(w_i,w_j,\theta)
=\tfrac12\{\widehat\phi_b(w_i,\theta)+\widehat\phi_b(w_j,\theta)\}
-\widehat k_{b,\tau}(w_i,w_j,\theta).
\]
We write $\widehat\Psi_{ij,\tau}(\theta)
=\widehat\Psi_{b(i,j),\tau}(W_i,W_j,\theta)$.
Since both orientations of each pair are included,
\eqref{eq:auto-equation} reads
$\widehat{\mathcal M}_\tau(\theta)
=\sum_b\omega_b\mathbb U_b\widehat\Psi_{b,\tau}(\cdot;\theta)$.

\paragraph{Operators.}
For symmetric $M$, $T_Mf=p'M\E[pf]$ is self-adjoint on
$L_2(P_{0,X})$. The fitted operator of candidate $\tau$ in block
$b$ is $\widehat T_{b,\tau}=T_{\widehat M_{b,\tau}}$, so that
$\widehat T_{b,\tau}f(x)=\E[\widehat K_{b,\tau}(x,X)f(X)]$. Its
comparison operator $T_{b,\tau}$ replaces $\widehat G_b$ in
$\widehat M_{b,\tau}$ by $G_0$, holding the retained eigenspace
or the selected regressors fixed
(Appendix~\ref{app:auto-operators}). For self-adjoint
operators, $A\le B$ means
$\langle f,Af\rangle\le\langle f,Bf\rangle$ for every $f$;
this is the positive-operator order. An operator with
$0\le T\le I$ is a positive contraction:
$0\le\langle f,Tf\rangle\le\|f\|^2$ for every $f$.

\subsection{Remaining curvature}
\label{app:auto-curvature}

This subsection first proves Proposition~\ref{prop:pointwise-cost}
and Corollary~\ref{cor:nonlinear-regression-cost}. It then proves
Proposition~\ref{prop:auto-curvature}, specializes it to the PLM,
and records the additional terms that arise when $F$ is nonlinear.

\begin{proof}[Proof of Proposition~\ref{prop:pointwise-cost}]
Put $r_0=\dot m_0-\alpha_0+\tfrac12a_0V_Z$.
For a nuisance score $s$, differentiation gives the
induced directions
\[
 h_{\gamma,s}=\E[V_Zs\mid X],\qquad
 h_{\alpha,s}=\E[(\dot m_0-\alpha_0)s\mid X]
                  +a_0h_{\gamma,s}.
\]
Polarizing the quadratic-bias identity \eqref{eq:auto-curvature}
therefore represents curvature by the bounded
self-adjoint operator on $L_2^0(P_0)$
\[
 \mathsf B s
 =-\tfrac12\{V_Z\E[r_0s\mid X]+r_0\E[V_Zs\mid X]\}.
\]
By Theorem~\ref{thm:cost}, finite cost is equivalent to
$\Pi_{\mathcal T_\eta}\mathsf B\Pi_{\mathcal T_\eta}$
being Hilbert--Schmidt. Since $\mathcal T_\eta$ has
codimension one in $\mathcal T=L_2^0(P_0)$, this
compression differs from $\mathsf B$ by finite rank.
Hence finite cost is equivalent to $\mathsf B$ being
Hilbert--Schmidt.

Let $v_0=\E[V_Z^2\mid X]$. Direct calculation gives
\[
 \E[(\mathsf B V_Z)^2\mid X]
 =\tfrac14\{v_0^2\E[r_0^2\mid X]
              +3v_0\E[V_Zr_0\mid X]^2\}.
\]
If $r_0\ne0$, nonatomicity supplies infinitely many
disjoint sets $A_j$ of positive covariate probability
on which $\E[r_0^2\mid X]$ is uniformly bounded away
from zero. The functions
$g_j=V_Z\ind\{X\in A_j\}/\|V_Z\ind\{X\in A_j\}\|$ are orthonormal, while
the displayed identity and the bounds on $v_0$ give
$\inf_j\|\mathsf B g_j\|>0$, contradicting the
Hilbert--Schmidt property.
If $r_0=0$, then $\mathsf B=0$, so the curvature
vanishes and $c_{\phi_0}=0$.
\end{proof}

\begin{proof}[Proof of Corollary~\ref{cor:nonlinear-regression-cost}]
Substitute $\dot m_0=\alpha_0=f'(\gamma_0)$ and
$a_0=f''(\gamma_0)$ into
Proposition~\ref{prop:pointwise-cost}, using
$\E[V_Z^2\mid X]>0$.
\end{proof}

\begin{proof}[Proof of Proposition~\ref{prop:auto-curvature}]
Fix $\theta=\theta_0$ and a direction
$h=(h_\gamma,h_\alpha)$, and define the $d$-vectors
\[
a_\gamma=\E[ph_\gamma],\qquad
a_\alpha=\E[ph_\alpha],\qquad
c_\gamma=F''(\gamma_0)[p,h_\gamma],
\]
where $c_\gamma$ has $j$th entry
$F''(\gamma_0)[p_j,h_\gamma]$.

\emph{Step 1: means of the residual vectors.}
Along $\gamma_t=\gamma_0+th_\gamma$ and
$\alpha_t=\alpha_0+th_\alpha$, let $u(t)$ and $v(t)$ be the
residual vectors evaluated at $(\gamma_t,\alpha_t,\theta_0)$, with $\eta_t=(\gamma_t,\alpha_t)$. The
projection equation \eqref{eq:auto-projection}, applied to
each $p_j\in\Gamma$, gives $\E[u(t)]=-ta_\gamma$.
Differentiating under the expectation and using
$F'(\gamma_0)[p_j]=\E[\alpha_0p_j]$,
\[
\E[v(t)]=F'(\gamma_t)[p]-\E[p\alpha_t]
=t(c_\gamma-a_\alpha)+o(t).
\]
In particular, $\E[u_0]=\E[v_0]=0$, so the kernel
\eqref{eq:auto-general-kernel} is degenerate at the truth.

\emph{Step 2: the curvature supplied by the correction.}
The two arguments of the kernel are independent, so
\[
\begin{aligned}
\E[k_{ij}(\eta_t,\theta_0;M)]
&=-\tfrac12\left\{2\E[u(t)]'M\E[v(t)]
+\E[u(t)]'MH_0M\E[u(t)]\right\}\\
&=t^2\left\{a_\gamma'M(c_\gamma-a_\alpha)
-\tfrac12a_\gamma'MH_0Ma_\gamma\right\}+o(t^2).
\end{aligned}
\]
Under the admissibility conditions,
Lemma~\ref{lem:pathwise} identifies the coefficient of
$t^2$ with $(\mathcal Rk_0(M))(h,h)$. Since
$T_Mh_\gamma=p'Ma_\gamma$ and $M$ is symmetric,
\[
a_\gamma'Mc_\gamma=F''(\gamma_0)[T_Mh_\gamma,h_\gamma],
\quad
a_\gamma'Ma_\alpha=\langle h_\alpha,T_Mh_\gamma\rangle,
\quad
a_\gamma'MH_0Ma_\gamma
=F''(\gamma_0)[T_Mh_\gamma,T_Mh_\gamma].
\]

\emph{Step 3: subtraction.}
Subtracting from \eqref{eq:auto-curvature},
\begin{align*}
(B_{\phi_0}-\mathcal Rk_0(M))(h,h)
={}&\tfrac12F''(\gamma_0)[h_\gamma,h_\gamma]
 -F''(\gamma_0)[T_Mh_\gamma,h_\gamma]
 +\tfrac12F''(\gamma_0)[T_Mh_\gamma,T_Mh_\gamma]\\
 &-\langle h_\alpha,(I-T_M)h_\gamma\rangle\\
={}&\tfrac12F''(\gamma_0)[\widetilde h_{\gamma,M},\widetilde h_{\gamma,M}]
 -\langle h_\alpha,\widetilde h_{\gamma,M}\rangle,
\end{align*}
where the second equality uses the symmetry of
$F''(\gamma_0)$. If $M=G_0^{-1}$, $T_M$ is the orthogonal
projection onto $\operatorname{span}(p)$, so $\widetilde h_{\gamma,M}=0$
whenever $h_\gamma\in\operatorname{span}(p)$.
\end{proof}

\paragraph{The PLM.}
Here $F''=0$ and $h_\alpha=-h_e$, so the
proposition gives \eqref{eq:plm-remaining-bias} in the form
\[
(B_{\phi_0}-\mathcal Rk_0(M))(h,h)
=\langle h_e,(I-T_M)h_\gamma\rangle.
\]
With $h_e=p'\beta_e+r_e$ and $h_\gamma=p'\beta_\gamma+r_\gamma$,
where $r_e,r_\gamma$ are orthogonal to $\operatorname{span}(p)$,
$\langle h_e,h_\gamma\rangle=\beta_e'G_0\beta_\gamma+\E[r_er_\gamma]$ and
$\langle h_e,T_Mh_\gamma\rangle=\beta_e'G_0MG_0\beta_\gamma$. Because the PLM bias is 
exactly quadratic, the identity is an exact expectation
identity for fixed fitted regressions and a fixed fitted
matrix.

\paragraph{Nonlinear moments.}
A feasible kernel replaces $u_0$, $v_0$, and $H_0$ by fitted
counterparts; $H_0$ can be estimated by averaging second derivatives of
$m$ over the training sample, and the representer by regularized Riesz
regression. For nonlinear $F$, \eqref{eq:auto-remaining} describes the
quadratic term only. A Lipschitz Hessian and a bounded operator norm
for $T_M$ give a Taylor remainder of order $\|h_\gamma\|^3$, and
estimating $H_0$ adds $\frac12a_M'(\widehat H-H_0)a_M$, $a_M=M\E[ph_\gamma]$,
to the bias of the fitted moment. Both terms vanish in the PLM.

\subsection{Ridge and selected-projection operators}
\label{app:auto-operators}

Proposition~\ref{prop:auto-selection} compares candidates
through population operators. This subsection shows that
these operators are positive contractions and that each
family is ordered.

Fix a retained coefficient space with orthonormal basis
matrix $U$, and suppose $G_{U,0}=U'G_0U$ is positive definite.
For $\zeta\ge0$, put
\[
M_U(\zeta)=U(G_{U,0}+\zeta I)^{-1}U',
\qquad
T_U(\zeta)=T_{M_U(\zeta)}.
\]
Ridge uses $U=\widehat U_{+,b}$: since
$\widehat U_{+,b}'\widehat G_b\widehat U_{+,b}
=\widehat\Lambda_{+,b}$, the matrix $M_U(\zeta)$ is
\eqref{eq:auto-ridge} with $\widehat G_b$ replaced by $G_0$.
Selected projection uses $\zeta=0$ and $U=E_S$, the matrix
that inserts the coordinates of $S$ into the full
dictionary: $M_{E_S}(0)=E_S(E_S'G_0E_S)^{-1}E_S'$ is the
population version of $\widehat M_{b,\tau}$ in
Section~\ref{sec:auto-estimate}. The comparison operators
$T_{b,\tau}$ are these operators for the space retained or
selected on $I_b^c$.

\begin{lemma}[Population ridge and projection operators]
\label{lem:auto-operators}
Let $\nu_j$ be the eigenvalues of $G_{U,0}$.
\begin{enumerate}[label=(\alph*),leftmargin=*]
\item The nonzero eigenvalues of $T_U(\zeta)$ are
$\nu_j/(\nu_j+\zeta)$. Hence $0\le T_U(\zeta)\le I$.
\item $T_U(0)$ is the orthogonal projection onto
$\{p'Uc:c\}$.
\item For fixed $U$, $T_U(\zeta)$ increases in the
positive-operator order as $\zeta$ decreases. Orthogonal
projections onto nested selected spaces are ordered in the
same way.
\end{enumerate}
\end{lemma}

\begin{proof}
Let $A_Uc=p'Uc$ map coefficient vectors into $L_2(P_{0,X})$.
Its adjoint is $A_U^*f=U'\E[pf]$, and $A_U^*A_U=G_{U,0}$. Hence
$Q_U=A_UG_{U,0}^{-1/2}$ is an isometry, and
\[
T_U(\zeta)=A_U(G_{U,0}+\zeta I)^{-1}A_U^*
=Q_U\{G_{U,0}^{1/2}(G_{U,0}+\zeta I)^{-1}G_{U,0}^{1/2}\}Q_U^*.
\]
The middle matrix has eigenvalues
$\nu_j/(\nu_j+\zeta)\in[0,1]$, which proves (a). At
$\zeta=0$ it is the identity, so $T_U(0)=Q_UQ_U^*$ is the
orthogonal projection onto the range of $A_U$, which
proves (b). The eigenvalues in (a) increase as $\zeta$
decreases, with eigenvectors that do not depend on $\zeta$,
and projections onto nested spaces are ordered by
inclusion. This proves (c).
\end{proof}

The retained space may be selected from the training data;
the lemma holds for each realized space. Replacing $G_0$ by
an empirical Gram matrix need not preserve the contraction
property in $L_2(P_{0,X})$; Lemma~\ref{lem:auto-gram} bounds that
replacement error.

The ridge kernel is a Riesz regression: for fixed $x$, the minimizer
of $c'\widehat G_bc-2p(x)'c+\zeta c'c$ over the retained eigenspace of
$\widehat G_b$ is $\widehat M_b(\zeta)p(x)$, so
$\widehat K_b(\cdot,x;\zeta)$ is the ridge-regularized representer of
evaluation at $x$ on $\operatorname{span}(p)$.

\subsection{The cross-fitted estimator}
\label{app:auto-crossfit}

For a within-set block, $\mathcal P_b=\{(i,j)\in I_b^2:i\ne j\}$.
For a rectangular block, write $I_b=I_{b,1}\cup I_{b,2}$ for disjoint
side sets and let $\mathcal P_b=(I_{b,1}\times I_{b,2})\cup
(I_{b,2}\times I_{b,1})$. The sets $\mathcal P_b$ partition all distinct
ordered pairs, and block $b$ is trained on $I_b^c$. The number of
blocks is fixed; the cardinalities of each participating side set
(or $I_b$ for a within-set block) and of $I_b^c$ are bounded below by
a positive constant times $n$. Thus $|\mathcal P_b|\asymp n^2$,
and both terms of every pair use its block's fit. The proofs use this
geometry and these size bounds. In the fifteen-block implementation,
training sizes are $4n/5$ for within-group blocks and $3n/5$ for
between-group blocks. Conditioning on the training information means
conditioning on $\mathcal H_b$.

In the PLM, since $\widehat K_{b,\tau}$ is symmetric and both
orientations of each pair are included, the two terms of
\eqref{eq:plm-estimated-kernel} contribute equally, and
\[
\widehat{\mathcal M}_\tau(\theta)
=\frac1{n(n-1)}\sum_b\sum_{(i,j)\in\mathcal P_b}
\widehat e_{i,b}(\theta)\bigl\{\widehat V_{i,b}
-\widehat K_{ij,\tau}\widehat V_{j,b}\bigr\}
=\widehat A_\tau-\theta\widehat J_\tau,
\]
which gives \eqref{eq:auto-plm-equation}.

\subsection{Automatic selection}
\label{app:auto-select}

Section~\ref{sec:auto-selection} motivated ADA with a
common training sample. In \eqref{eq:auto-equation}, each
pair block has its own training sample, so the objects of
Section~\ref{sec:auto-selection} are averaged over blocks
with the weights $\omega_b$. For candidate $\tau$, let
$T_{b,\tau}$ be its comparison operator in block $b$ and
$\widetilde h_{\gamma,b,\tau}=(I-T_{b,\tau})h_{\gamma,b}$. For
$a\in\{\gamma,\alpha\}$, define
\begin{equation}
\begin{aligned}
b_\tau&=\sum_b\omega_b\left\{
\tfrac12F''(\gamma_0)[\widetilde h_{\gamma,b,\tau},\widetilde h_{\gamma,b,\tau}]
-\langle h_{\alpha,b},\widetilde h_{\gamma,b,\tau}\rangle\right\},
&\mathcal Q_{a,\tau}
&=\sum_b\omega_b\langle h_{a,b},T_{b,\tau}h_{a,b}\rangle,\\
d_{a,\tau}&=\sum_b\omega_b\|h_{a,b}\|^2-\mathcal Q_{a,\tau},
&\delta_a&=\max_\tau|\widehat Q_{a,\tau}-\mathcal Q_{a,\tau}|.
\end{aligned}
\label{eq:app-auto-block-objects}
\end{equation}
Here $b_\tau$ is the block-weighted remaining curvature
\eqref{eq:auto-remaining}, $\mathcal Q_{a,\tau}$ is the
captured error that $\widehat Q_{a,\tau}$ estimates,
$d_{a,\tau}$ is the uncaptured error, and $\delta_a$ is the
estimation error of the selection statistics. With a
common training sample, these reduce to the objects of
Section~\ref{sec:auto-selection}. For a projection,
$d_{a,\tau}$ equals the squared approximation error; for
ridge, it bounds that error from above. The maximum in
$\delta_a$ is over the full candidate library, so
$\delta_a$ also accounts for the empirical Gram matrix
and, for $a=\alpha$, the preliminary estimate
$\widetilde\theta$.

\begin{lemma}[Bias control]
\label{lem:auto-bias}
If $0\le T_{b,\tau}\le I$ for every block, then
\[
|b_\tau|\le\tfrac12C_Fd_{\gamma,\tau}
+\sqrt{d_{\alpha,\tau}d_{\gamma,\tau}}.
\]
With a common training sample, this is
\eqref{eq:auto-bias-control}.
\end{lemma}

\begin{proof}
Let $T_{b,\tau}^\perp=I-T_{b,\tau}$. Then
$0\le T_{b,\tau}^\perp\le I$ and
$(T_{b,\tau}^\perp)^2\le T_{b,\tau}^\perp$, so
$\|T_{b,\tau}^\perp h_{\gamma,b}\|^2
\le\langle h_{\gamma,b},T_{b,\tau}^\perp h_{\gamma,b}\rangle$.
Since $F''(\gamma_0)$ has norm at most $C_F$, the curvature
term of block $b$ is at most
$\frac12C_F\langle h_{\gamma,b},T_{b,\tau}^\perp h_{\gamma,b}\rangle$
in absolute value. Cauchy--Schwarz applied with
$(T_{b,\tau}^\perp)^{1/2}$ gives
\[
|\langle h_{\alpha,b},T_{b,\tau}^\perp h_{\gamma,b}\rangle|
\le\{\langle h_{\alpha,b},T_{b,\tau}^\perp h_{\alpha,b}\rangle
\langle h_{\gamma,b},T_{b,\tau}^\perp h_{\gamma,b}\rangle\}^{1/2}.
\]
Average over blocks with weights $\omega_b$ and apply
Cauchy--Schwarz once more, now to the weights. Since
$d_{a,\tau}=\sum_b\omega_b
\langle h_{a,b},T_{b,\tau}^\perp h_{a,b}\rangle$, this gives
the claim.
\end{proof}

\paragraph{Hessian adjustment for nonlinear $F$.}
For nonlinear $F$, the mean of $v$ also contains
$F''(\gamma_0)[p,h_\gamma]$ (Step~1 in the proof of
Proposition~\ref{prop:auto-curvature}). This term can be removed to
first order by adding $\widehat\chi_{j,b}(X_i)\{Z_i-\widehat\gamma_b(X_i)\}$ to
$(\widehat v_{i,b})_j$ in \eqref{eq:auto-captured}, where $\widehat\chi_{j,b}$ is a
Riesz-regression estimate of $\chi_{j,0}$, the representer of
$F''(\gamma_0)[p_j,\cdot]$; when $F$ is affine, $\chi_{j,0}=0$.

\begin{proposition}[An oracle bound for the selected correction]
\label{prop:auto-selection}
Suppose $q_\alpha,q_\gamma>0$, and put
$\varpi=(q_\gamma/q_\alpha)^{1/2}$. Let $C_F$ bound the
operator norm of $F''(\gamma_0)$, and let $b_\tau$, $d_{a,\tau}$,
and $\delta_a$ be as in \eqref{eq:app-auto-block-objects},
computed with positive-contraction comparison operators
$T_{b,\tau}$. Then
\begin{equation}
|b_{\widehat \tau}|
\le(1+C_F\varpi)
\left[
\frac12\min_{\tau\in\mathcal J_{\mathrm{sel}}}
\{\varpi d_{\alpha,\tau}+\varpi^{-1}d_{\gamma,\tau}\}
+\varpi\delta_\alpha+\varpi^{-1}\delta_\gamma
\right].
\label{eq:auto-oracle}
\end{equation}
For affine $F$, including the PLM, $C_F=0$. If each
population path is nested in the positive-operator order,
the minimum can also be taken over all rule-C-eligible
members of the two paths.
\end{proposition}

The right side has two parts. The first is the bound
\eqref{eq:auto-bias-control}, in weighted arithmetic-mean
form, for the best retained candidate. The second is the
price of estimating the captured errors. The bound compares
ADA's choice with the best retained candidate and does not
require ADA to select a particular family with probability
approaching one. It concerns a bound on remaining bias, not
mean squared error.

\begin{proof}
\emph{Step 1: bias bound.}
By Lemma~\ref{lem:auto-bias},
\begin{equation}
|b_\tau|\le\tfrac12C_Fd_{\gamma,\tau}
+\sqrt{d_{\alpha,\tau}d_{\gamma,\tau}}.
\label{eq:app-auto-bias-bound}
\end{equation}

\emph{Step 2: ADA maximizes an estimated weighted captured
error.} Define
\[
\mathcal C_\tau=\varpi\mathcal Q_{\alpha,\tau}
+\varpi^{-1}\mathcal Q_{\gamma,\tau},
\qquad
\widehat{\mathcal C}_\tau=\varpi\widehat Q_{\alpha,\tau}
+\varpi^{-1}\widehat Q_{\gamma,\tau}.
\]
Multiplying the criterion in \eqref{eq:auto-selection} by
$\sqrt{q_\alpha q_\gamma}$ shows that $\widehat \tau$
maximizes $\widehat{\mathcal C}_\tau$ over
$\mathcal J_{\mathrm{sel}}$. With
$\bar\delta=\varpi\delta_\alpha+\varpi^{-1}\delta_\gamma$,
$|\widehat{\mathcal C}_\tau-\mathcal C_\tau|\le\bar\delta$ for
every candidate. Hence, for every
$\tau\in\mathcal J_{\mathrm{sel}}$,
\[
\mathcal C_{\widehat \tau}
\ge\widehat{\mathcal C}_{\widehat \tau}-\bar\delta
\ge\widehat{\mathcal C}_\tau-\bar\delta
\ge\mathcal C_\tau-2\bar\delta.
\]

\emph{Step 3: from captured to uncaptured error.}
Put $\mathcal E_\tau=\varpi d_{\alpha,\tau}+\varpi^{-1}d_{\gamma,\tau}$.
The total errors $\sum_b\omega_b\|h_{a,b}\|^2$ do not depend
on the candidate, so $\mathcal E_\tau$ is a constant minus
$\mathcal C_\tau$, and Step~2 gives
\begin{equation}
\mathcal E_{\widehat \tau}
\le\min_{\tau\in\mathcal J_{\mathrm{sel}}}\mathcal E_\tau
+2\bar\delta.
\label{eq:app-auto-energy}
\end{equation}

\emph{Step 4: back to the bias.}
The arithmetic--geometric mean inequality gives
$\sqrt{d_{\alpha,\tau}d_{\gamma,\tau}}\le\mathcal E_\tau/2$, and
$d_{\gamma,\tau}\le\varpi\mathcal E_\tau$. Thus
\eqref{eq:app-auto-bias-bound} gives
$|b_\tau|\le\frac12(1+C_F\varpi)\mathcal E_\tau$, and
\eqref{eq:app-auto-energy} proves \eqref{eq:auto-oracle}.

\emph{Step 5: nested paths.}
If each population path is nested in the positive-operator
order, the largest rule-C-eligible index in a family has no
larger $d_{\alpha,\tau}$ or $d_{\gamma,\tau}$ than any other
eligible member of that family. The minimum over the
retained members therefore equals the minimum over all
eligible members. This step does not require the estimated
variance to be monotone along the path.
\end{proof}

The inequality holds for every realized candidate set and
every positive pair of normalizers. The variance budget and
the selection statistics may therefore use the same
observations. Its statistical content depends on
$\delta_\alpha$ and $\delta_\gamma$, and on a budget that
admits a sufficiently accurate correction;
Theorem~\ref{thm:auto-inference}(b) states the
requirement used for inference. The simulations use $c=2$.
This constant specifies a variance tolerance; the theory
does not identify it as a universal optimum.

\subsection{Inference with negligible correction variance}
\label{app:auto-negligible}

This subsection proves Theorem~\ref{thm:auto-inference} and
Corollary~\ref{cor:plm}. After a variance formula for block averages,
it records the residual identities for moments affine in the nuisance
and the behavior of estimated Gram matrices, then expands the
estimating equation and its variance estimator, and finally proves the
results.

\begin{lemma}[Variance of block averages]
\label{lem:auto-blocks}
Condition on $\mathcal H_b$, and let $f$ be a symmetric kernel that is
measurable with respect to it, with Hoeffding decomposition
$f=\bar f+f^{[1]}(w_i)+f^{[1]}(w_j)+f^{[2]}(w_i,w_j)$, where $\bar f=\E f$,
$f^{[1]}(w)=\E[f(w,W)]-\bar f$, and $f^{[2]}$ is degenerate.
\begin{enumerate}[label=(\alph*),leftmargin=*]
\item If $\mathcal P_b$ contains the distinct ordered pairs of a set of
$n_b$ observations, then
$\Var(\mathbb U_bf\mid\mathcal H_b)=4\|f^{[1]}\|^2/n_b
+2\|f^{[2]}\|^2/\{n_b(n_b-1)\}$.
\item If $\mathcal P_b$ contains the pairs, in both orientations,
between disjoint sets $\mathcal S_1$ and $\mathcal S_2$ of $n_{b,1}$
and $n_{b,2}$ observations, then
$\Var(\mathbb U_bf\mid\mathcal H_b)=(n_{b,1}^{-1}+n_{b,2}^{-1})\|f^{[1]}\|^2
+\|f^{[2]}\|^2/(n_{b,1}n_{b,2})$.
\end{enumerate}
\end{lemma}

\begin{proof}
In (a), $\mathbb U_bf-\bar f=2n_b^{-1}\sum_if^{[1]}(W_i)+\mathbb U_bf^{[2]}$. In
(b), symmetry of $f$ and the inclusion of both orientations give
\[
\mathbb U_bf-\bar f
=\frac1{n_{b,1}}\sum_{i\in\mathcal S_1}f^{[1]}(W_i)
+\frac1{n_{b,2}}\sum_{j\in\mathcal S_2}f^{[1]}(W_j)
+\frac1{n_{b,1}n_{b,2}}
\sum_{i\in\mathcal S_1,\,j\in\mathcal S_2}f^{[2]}(W_i,W_j).
\]
In both cases the terms are uncorrelated, and degeneracy makes
distinct pair terms uncorrelated, including pairs that share one
observation.
\end{proof}

\subsubsection{Residual identities and Gram matrices}
\label{app:auto-identities}

Throughout this subsection, Assumption~\ref{ass:auto-model} holds, so
$F$ is affine and $H_0=0$. Fix a block $b$ and a candidate $\tau$, and write
$M=\widehat M_{b,\tau}$, $K_{ij}=\widehat K_{ij,\tau}$,
$\widehat T=\widehat T_{b,\tau}$, $p_i=p(X_i)$, and $h_\gamma$,
$h_\alpha$, $h_{\dot\alpha}$ for the block-$b$ errors. Let
\[
v_0(W)=D_\gamma m(W,\gamma_0,\theta_0)[p]-p(X)\alpha_0(X),
\qquad
\dot v_0(W)=\partial_\theta D_\gamma m(W,\gamma_0,\theta)[p]
-p(X)\,\partial_\theta\alpha_0(X,\theta),
\]
where $v_{0,i}=v_0(W_i)$ and $\dot v_{0,i}=\dot v_0(W_i)$. These do not depend on $\theta$ because $m$ and $\alpha_0(\cdot,\theta)$ are
affine in $\theta$, and let
$\upsilon_b(W)=D_\gamma m(W,\gamma_0,\theta_0)[h_\gamma]-\alpha_0(X)h_\gamma(X)$.
Since $m$ is affine in $\gamma$, for $i\in I_b$,
\begin{equation}
\begin{gathered}
\widehat u_{i,b}=p_i\{V_{Z,i}-h_\gamma(X_i)\},
\qquad
\widehat v_{i,b}(\theta_0)=v_{0,i}-p_ih_\alpha(X_i),\\
\widehat\phi_{i,b}(\theta_0)
=\phi_0(W_i)+\upsilon_b(W_i)+h_\alpha(X_i)V_{Z,i}
-h_\alpha(X_i)h_\gamma(X_i),
\end{gathered}
\label{eq:app-auto-linear-residuals}
\end{equation}
and $\partial_\theta\widehat v_{i,b}=\dot v_{0,i}-p_ih_{\dot\alpha}(X_i)$. By
the definition of the representers, $\E[v_0]=\E[\dot v_0]=0$ and
$\E[\upsilon_b\mid\mathcal H_b]=0$. Assumption~\ref{ass:auto-model}(ii),
applied at two values of $\theta$, gives
\begin{equation}
\E[v_0v_0']\le CG_0,\qquad \E[\dot v_0\dot v_0']\le CG_0,
\qquad \E[\upsilon_b^2\mid\mathcal H_b]\le C\|h_\gamma\|^2.
\label{eq:app-auto-msc}
\end{equation}
In the PLM, $V_Z=V$, $v_0=-p\varepsilon$, $\dot v_0=pV$,
$\upsilon_b=-\varepsilon h_\gamma$, and $h_{\dot\alpha}=h_\gamma$.

We use the leverage and the effective dimension
\[
\widehat{\mathcal L}_{b,\tau}=\sup_x\E[\widehat K_{b,\tau}(X,x)^2]
=\sup_xp(x)'MG_0Mp(x),
\qquad
\widehat{\mathrm{df}}_{b,\tau}=\E[\widehat K_{b,\tau}(X_i,X_j)^2]=\tr(G_0MG_0M),
\]
with $\widehat{\mathrm{df}}_{b,\tau}\le\widehat{\mathcal L}_{b,\tau}$. By
\eqref{eq:app-auto-msc}, for fixed $x$ and $w\in\{v_0,\dot v_0\}$,
\begin{equation}
\E[\{p(x)'Mw\}^2\mid\mathcal H_b]
\le C\,\E[\widehat K_{b,\tau}(x,X)^2]\le C\widehat{\mathcal L}_{b,\tau}.
\label{eq:app-auto-leverage}
\end{equation}

\begin{lemma}[Gram matrices]
\label{lem:auto-gram}
Under Assumption~\ref{ass:auto-first-step}(ii)--(iii), let
$\varrho_b=\|G_0^{-1/2}(\widehat G_b-G_0)G_0^{-1/2}\|_{\op}$.
\begin{enumerate}[label=(\alph*),leftmargin=*]
\item $\max_b\varrho_b=O_p\big(\xi_d\sqrt{\log(2d)/n}\big)=o_p(1)$.
\item On the event $\{\max_b\varrho_b\le1/2\}$, which has probability
approaching one, the numerical threshold in \eqref{eq:auto-ridge}
does not bind when it is below $\lambda_{\min}(G_0)/2$, and for every
candidate $\|\widehat T_{b,\tau}\|_{\op}\le2$,
$\sup_x\widehat K_{b,\tau}(x,x)\le C\xi_d^2$,
$\widehat{\mathcal L}_{b,\tau}\le C\xi_d^2$, and
$\widehat{\mathrm{df}}_{b,\tau}\le Cd\le C\xi_d^2$.
\item On the same event, for all $f,g\in L_2(P_{0,X})$,
$|\langle f,(\widehat T_{b,\tau}-T_{b,\tau})g\rangle|\le2\varrho_b\|f\|\,\|g\|$.
\end{enumerate}
\end{lemma}

\begin{proof}
(a) Conditional on $\mathcal H_n$, the matrix
$G_0^{-1/2}\widehat G_bG_0^{-1/2}$ is an average of
$|I_b^c|\asymp n$ iid positive semidefinite matrices
$G_0^{-1/2}p(X_i)p(X_i)'G_0^{-1/2}$ with mean $I$ and norm
at most $\xi_d^2/\lambda_{\min}(G_0)$. The matrix Bernstein inequality
\citep[Theorem~1.6]{Tropp2012} gives
$\varrho_b=O_p(\sqrt{\xi_d^2\log(2d)/n}+\xi_d^2\log(2d)/n)$, and the second
term is of smaller order. There are finitely many blocks.

(b) On the event, $\widehat G_b\ge\frac12G_0$, so every eigenvalue
of $\widehat G_b$ is at least $\lambda_{\min}(G_0)/2$. For ridge, let the columns of $U$ be an orthonormal basis
of the retained eigenspace and put
$A=U'\widehat G_bU+\zeta I$. Since
$A\ge\frac12U'G_0U$,
\[
\|G_0^{1/2}\widehat M_{b,\tau}G_0^{1/2}\|_{\op}
=\|A^{-1/2}U'G_0UA^{-1/2}\|_{\op}\le2.
\]
The bound is immediate if the retained space is empty. For
selected projection, principal submatrices inherit the bound,
$\widehat G_{S,b}\ge\frac12G_{S,0}$, and
$\|G_0^{1/2}E_S\widehat G_{S,b}^{-1}E_S'G_0^{1/2}\|_{\op}
=\|\widehat G_{S,b}^{-1/2}G_{S,0}\widehat G_{S,b}^{-1/2}\|_{\op}\le2$. Hence
$\|\widehat T_{b,\tau}\|_{\op}\le2$ and
$\|\widehat M_{b,\tau}\|_{\op}\le2/\lambda_{\min}(G_0)$, so
$\sup_xp(x)'\widehat M_{b,\tau}p(x)\le C\xi_d^2$. Next,
$\widehat{\mathcal L}_{b,\tau}\le\|\widehat T_{b,\tau}\|_{\op}
\sup_xp(x)'\widehat M_{b,\tau}p(x)$ and
$\widehat{\mathrm{df}}_{b,\tau}\le\|\widehat T_{b,\tau}\|_{\op}
\tr(G_0^{1/2}\widehat M_{b,\tau}G_0^{1/2})\le4d$. Finally,
$d\lambda_{\min}(G_0)\le\tr G_0=\E\|p(X)\|_2^2\le\xi_d^2$.

(c) Write $\Delta_b=G_0^{-1/2}(\widehat G_b-G_0)G_0^{-1/2}$ and let $M$ be
the comparison matrix. Then
\[
\langle f,(\widehat T-T)g\rangle
=(G_0^{-1/2}\E[pf])'\,G_0^{1/2}(\widehat M-M)G_0^{1/2}\,
(G_0^{-1/2}\E[pg]),
\]
where $\|G_0^{-1/2}\E[pf]\|_2\le\|f\|$. For ridge and for selected
projection, respectively,
\begin{align*}
G_0^{1/2}(\widehat M-M)G_0^{1/2}
&=-(G_0^{1/2}\widehat MG_0^{1/2})\,\Delta_b\,(G_0^{1/2}MG_0^{1/2}),\\
G_0^{1/2}(\widehat M-M)G_0^{1/2}
&=-(G_0^{1/2}E_S\widehat G_{S,b}^{-1}E_S'G_0^{1/2})\,
\Delta_b\,(G_0^{1/2}E_SG_{S,0}^{-1}E_S'G_0^{1/2}),
\end{align*}
and the last factor in each line has norm at most one. In both cases
the norm is at most $2\varrho_b$.
\end{proof}

\subsubsection{Expansion and variance}

\begin{lemma}[Expansion of the estimating equation]
\label{lem:auto-expansion}
Suppose Assumption~\ref{ass:auto-model} and
Assumption~\ref{ass:auto-first-step}(i) hold, and that for candidate $\tau$,
$\max_b\|\widehat T_{b,\tau}\|_{\op}=O_p(1)$ and
$\max_b\widehat{\mathcal L}_{b,\tau}=O_p(n)$. Let
\[
k_{0,b,\tau}(W_i,W_j)
=-\tfrac12\{V_{Z,i}\,p_i'\widehat M_{b,\tau}v_{0,j}
+V_{Z,j}\,p_j'\widehat M_{b,\tau}v_{0,i}\}
\]
be the correction at the true nuisance functions, and put
\[
U^0_\tau=\sum_b\omega_b\,\mathbb U_b\,k_{0,b,\tau},
\qquad
\bar b_\tau=-\sum_b\omega_b\langle h_{\alpha,b},
(I-\widehat T_{b,\tau})h_{\gamma,b}\rangle.
\]
Then
$\widehat{\mathcal M}_\tau$ is affine in $\theta$, so that
$\widehat\theta_\tau-\theta_0=\widehat{\mathcal M}_\tau(\theta_0)/\widehat J_\tau$, and
\begin{enumerate}[label=(\alph*),leftmargin=*]
\item $\widehat{\mathcal M}_\tau(\theta_0)
=\Pn\phi_0-U^0_\tau+\bar b_\tau+o_p(n^{-1/2})$;
\item $k_{0,b,\tau}$ is degenerate given $\mathcal H_b$,
$\E[k_{0,b,\tau}^2\mid\mathcal H_b]\le C\widehat{\mathrm{df}}_{b,\tau}$, and
$\Var(\mathbb U_bk_{0,b,\tau}\mid\mathcal H_b)\le C\widehat{\mathrm{df}}_{b,\tau}/n^2$;
\item $\widehat J_\tau=J_0+o_p(1)$.
\end{enumerate}
With all matrices equal to zero, (a) and (c) imply that
the preliminary estimate without correction satisfies
\[
\begin{aligned}
\widetilde\theta-\theta_0
&=J_0^{-1}\left\{\Pn\phi_0
-\sum_b\omega_b
 \langle h_{\alpha,b},h_{\gamma,b}\rangle\right\}
+o_p(n^{-1/2}+r_n^2)\\
&=O_p(n^{-1/2}+r_n^2).
\end{aligned}
\]
\end{lemma}

\begin{proof}
The functions $m$, $\widehat\alpha_b(\cdot,\theta)$, and
$\widehat v_{i,b}(\theta)$ are affine in $\theta$, and hence so is
$\widehat{\mathcal M}_\tau$.

(a) Fix $b$, condition on $I_b^c$, and write
$K_{ij}$ and $M$ as above. By \eqref{eq:app-auto-linear-residuals},
$\widehat\Psi_{ij,\tau}(\theta_0)
=\Psi_{0,ij}+\Delta^{(1)}_{ij}+\Delta^{(2)}_{ij}$, where
$\Psi_{0,ij}=\frac12\{\phi_0(W_i)+\phi_0(W_j)\}-k_{0,b,\tau}(W_i,W_j)$
and
\begin{align*}
\Delta^{(1)}_{ij}={}&\tfrac12\{\upsilon_b(W_i)+h_\alpha(X_i)V_{Z,i}
+\upsilon_b(W_j)+h_\alpha(X_j)V_{Z,j}\}\\
&-\tfrac12\{V_{Z,i}K_{ij}h_\alpha(X_j)+V_{Z,j}K_{ij}h_\alpha(X_i)
+h_\gamma(X_i)p_i'Mv_{0,j}+h_\gamma(X_j)p_j'Mv_{0,i}\},\\
\Delta^{(2)}_{ij}={}&-\tfrac12\{h_\alpha(X_i)h_\gamma(X_i)
+h_\alpha(X_j)h_\gamma(X_j)\}
+\tfrac12K_{ij}\{h_\alpha(X_i)h_\gamma(X_j)
+h_\alpha(X_j)h_\gamma(X_i)\}.
\end{align*}
The oracle part averages exactly to the stated terms:
$\sum_b\omega_b\mathbb U_b\Psi_0=\Pn\phi_0-U^0_\tau$.

Every term of $\Delta^{(1)}$ is linear in one of $\upsilon_b$, $V_Z$,
or $v_0$, each of which has conditional mean zero given $\mathcal H_b$
($V_Z$ also given $X$), so $\E[\Delta^{(1)}\mid\mathcal H_b]=0$. Its
first projection is
\[
f^{[1]}(w)=\tfrac12\bigl\{\upsilon_b(w)+V_Z\,((I-\widehat T)h_\alpha)(x)
-\E[h_\gamma p']Mv_0(w)\bigr\},
\]
and, by \eqref{eq:app-auto-msc} and
\[
\E[(\E[h_\gamma p']Mv_0)^2]\le C\,\E[h_\gamma p']MG_0M\E[ph_\gamma]
=C\|\widehat Th_\gamma\|^2,
\]
$\|f^{[1]}\|^2\le C(1+\|\widehat T\|_{\op})^2(\|h_\alpha\|^2+\|h_\gamma\|^2)$.
For the degenerate part, \eqref{eq:app-auto-leverage} gives
\[
\E[V_{Z,1}^2K_{12}^2h_\alpha(X_2)^2]\le C\widehat{\mathcal L}_{b,\tau}
\|h_\alpha\|^2,
\qquad
\E[h_\gamma(X_1)^2(p_1'Mv_{0,2})^2]\le C\widehat{\mathcal L}_{b,\tau}\|h_\gamma\|^2,
\]
so
$\|f^{[2]}\|^2\le C(1+\widehat{\mathcal L}_{b,\tau})(\|h_\alpha\|^2+\|h_\gamma\|^2)$.
Since each participating set and training sample has size proportional to $n$,
Lemma~\ref{lem:auto-blocks} gives
\[
\Var(\mathbb U_b\Delta^{(1)}\mid\mathcal H_b)
\le\frac Cn(\|h_\alpha\|^2+\|h_\gamma\|^2)
\left(1+\|\widehat T\|_{\op}^2+\frac{\widehat{\mathcal L}_{b,\tau}}n\right)
=o_p(n^{-1}).
\]
For $\Delta^{(2)}$, $\E[\Delta^{(2)}\mid\mathcal H_b]
=-\langle h_\alpha,h_\gamma\rangle+\langle h_\alpha,\widehat Th_\gamma\rangle
=-\langle h_\alpha,(I-\widehat T)h_\gamma\rangle$, whose weighted sum
over blocks is $\bar b_\tau$. Its centered first projection is
$-\frac12\{h_\alpha h_\gamma-h_\alpha\widehat Th_\gamma
-h_\gamma\widehat Th_\alpha\}$ minus its mean, with squared norm at
most $C(\|h_\alpha\|_\infty^2+\|h_\gamma\|_\infty^2)(1+\|\widehat T\|_{\op})^2
(\|h_\alpha\|^2+\|h_\gamma\|^2)$, and its degenerate part has squared
norm at most $C\|h_\alpha\|_\infty^2\widehat{\mathcal L}_{b,\tau}\|h_\gamma\|^2$.
By Lemma~\ref{lem:auto-blocks} and the sup-norm bounds, the centered
average is $o_p(n^{-1/2})$. Conditional Chebyshev, applied to each of
the finitely many blocks and candidates, proves (a).

(b) Degeneracy follows from $\E[V_Z\mid X]=0$ and $\E[v_0]=0$. By
$\E[V_Z^2\mid X]\le C$ and \eqref{eq:app-auto-msc},
$\E[k_{0,b,\tau}^2\mid\mathcal H_b]\le C\E[V_{Z,1}^2(p_1'Mv_{0,2})^2]
\le C\E[p_1'MG_0Mp_1]=C\widehat{\mathrm{df}}_{b,\tau}$; apply
Lemma~\ref{lem:auto-blocks}.

(c) The derivative of $\widehat{\mathcal M}_\tau$ does not depend on
$\theta$. Its first-order part is the weighted average of
\[
\partial_\theta\widehat\phi_{i,b}(\theta)
=\partial_\theta m(W_i,\gamma_0,\theta)
+\partial_\theta D_\gamma m(W_i,\gamma_0,\theta)[h_\gamma]
+\{\partial_\theta\alpha_0(X_i,\theta)+h_{\dot\alpha}(X_i)\}
\{V_{Z,i}-h_\gamma(X_i)\}.
\]
Because $\partial_\theta\alpha_0(\cdot,\theta)$ represents
$a\mapsto\E\{\partial_\theta D_\gamma m(W,\gamma_0,\theta)[a]\}$, its conditional
mean is $-J_0-\langle h_{\dot\alpha},h_\gamma\rangle$. The first term has
a finite second moment, and the remaining terms differ from
$\partial_\theta\alpha_0(X_i,\theta)V_{Z,i}$, which has mean zero, by
terms with $L_2$ norm $O(\|h_{\dot\alpha}\|+\|h_\gamma\|)=o_p(1)$; the law
of large numbers within each block gives convergence to $-J_0$. The
pair part is
$\frac12\{\widehat u_{i,b}'M(\dot v_{0,j}-p_jh_{\dot\alpha}(X_j))
+\widehat u_{j,b}'M(\dot v_{0,i}-p_ih_{\dot\alpha}(X_i))\}$, which has the
structure of $-\widehat k_{b,\tau}$ with $(v_0,h_\alpha)$ replaced by
$(\dot v_0,h_{\dot\alpha})$. Its conditional mean is
$\langle h_{\dot\alpha},\widehat Th_\gamma\rangle=O_p(r_n^2)$, and the
arguments of (a) and (b) show that it fluctuates by
$O_p(n^{-1/2}+\widehat{\mathrm{df}}_{b,\tau}^{1/2}/n)=O_p(n^{-1/2})$.

Without correction, (a) holds with $U^0_\tau=0$ and $\widehat T=0$, and
$|\langle h_{\alpha,b},h_{\gamma,b}\rangle|\le r_n^2$.
By (c), the estimated inverse sensitivity differs from
$J_0^{-1}$ by $o_p(1)$. Multiplying this difference by
the uncorrected numerator, which is
$O_p(n^{-1/2}+r_n^2)$, gives the stated remainder.

\end{proof}

\begin{lemma}[Variance consistency]
\label{lem:auto-variance}
Under the conditions of Lemma~\ref{lem:auto-expansion}, suppose also
that $\max_b\widehat{\mathrm{df}}_{b,\tau}=o_p(n)$ and
$\widehat\theta_\tau\to_p\theta_0$. Then $n\widehat\sigma_\tau^{\star2}$ and
$n\widehat\sigma_\tau^2$ converge in probability to $\Omega_1/J_0^2$.
\end{lemma}

\begin{proof}
\emph{First-order part of the rows.}
Since $\widehat\phi_b$ is affine in $\theta$, Jensen's
inequality over the partners of $i$ gives
\[
\begin{aligned}
\frac1n\sum_i(\bar\phi_{i,\tau}-\phi_0(W_i))^2
\le{}&
2\sum_b\omega_b\,\mathbb U_b
\bigl[\{\widehat\phi_{i,b}(\theta_0)-\phi_0(W_i)\}^2\bigr]\\
&+2(\widehat\theta_\tau-\theta_0)^2
\sum_b\omega_b\,\mathbb U_b
\bigl[\{\partial_\theta\widehat\phi_{i,b}(\theta)\}^2\bigr].
\end{aligned}
\]
Here each block average treats its integrand as a
function of the ordered pair $(W_i,W_j)$.
By \eqref{eq:app-auto-linear-residuals},
\eqref{eq:app-auto-msc}, the conditional variance bound
for $V_Z$, and the first-step estimator bounds, the first block-weighted
sum is $o_p(1)$ by conditional Markov applied separately
to each block. The derivative-square sum is $O_p(1)$
by the same argument, using the second-moment bound
for $\partial_\theta m$ and the derivative bounds in
\eqref{eq:app-auto-msc}. Since
$\widehat\theta_\tau-\theta_0=o_p(1)$, the right side is
$o_p(1)$. The same separation of the scalar estimation
error from empirical derivative-square averages applies
to the pair terms below.
\emph{Pair part of the rows.} Its oracle part is
$(n-1)^{-1}\sum_{j\ne i}
\{V_{Z,i}p_i'Mv_{0,j}+V_{Z,j}p_j'Mv_{0,i}\}$, with
$M$ the matrix of the block of $(i,j)$. Define
$\mathcal N_{b,i}=\{j:(i,j)\in\mathcal P_b\}$.
Conditional on $\mathcal H_b$ and $X_i$, the sum over
$j\in\mathcal N_{b,i}$ of $p_i'Mv_{0,j}$ has mean zero and
second moment at most
$C|\mathcal N_{b,i}|\,
\E[\widehat K_{b,\tau}(X_i,X)^2\mid\mathcal H_b,X_i]$
by \eqref{eq:app-auto-leverage}.
Since $|\mathcal N_{b,i}|\le n-1$, integration over $X_i$
and blockwise conditional Markov give an empirical mean
square of the oracle row part bounded by
$O_p(\max_b\widehat{\mathrm{df}}_{b,\tau}/n)=o_p(1)$. The
terms involving first-step estimation errors are bounded in the same way: for instance,
$(n-1)^{-1}\sum_j V_{Z,i}K_{ij}h_\alpha(X_j)$ has mean square at most
$C\|\widehat T\|_{\op}^2\|h_\alpha\|^2+C\widehat{\mathcal L}_{b,\tau}\|h_\alpha\|^2/n$,
and $(n-1)^{-1}\sum_j h_\gamma(X_i)p_i'Mv_{0,j}$ at most
$C\widehat{\mathcal L}_{b,\tau}\|h_\gamma\|^2/n$. The terms involving
$\widehat\theta_\tau-\theta_0$ replace $(v_0,h_\alpha)$ by
$(\dot v_0,h_{\dot\alpha})$ and are $o_p(1)$ for the same reason.

\emph{Conclusion.} The mean squares of $\widehat\psi_{i,\tau}-\phi_0(W_i)$ and
$\bar\phi_{i,\tau}-\phi_0(W_i)$ over $i$ are $o_p(1)$. Since
$n^{-1}\sum_i\phi_0(W_i)^2\to_p\Omega_1$ and
$n^{-1}\sum_i\phi_0(W_i)\to_p0$, Cauchy--Schwarz gives
$n^{-1}\sum_i(\widehat\psi_{i,\tau}-\bar\psi_\tau)^2\to_p\Omega_1$ and
$n^{-1}\sum_i\bar\phi_{i,\tau}^2\to_p\Omega_1$. Given $\mathcal H_b$ and
$i\ne j$, \eqref{eq:app-auto-linear-residuals} and
\eqref{eq:app-auto-leverage} give
$\E[\widehat k_{ij,\tau}(\theta_0)^2\mid\mathcal H_b]
\le C(1+\|h_\alpha\|_\infty^2)(1+\|h_\gamma\|_\infty^2)\widehat{\mathrm{df}}_{b,\tau}$,
so $P_\tau/n=O_p(\max_b\widehat{\mathrm{df}}_{b,\tau}/n)=o_p(1)$, including the effect of
$\widehat\theta_\tau$. Together with $\widehat J_\tau\to_pJ_0$ from
Lemma~\ref{lem:auto-expansion}(c), both $n\widehat\sigma_\tau^{\star2}$ and
$n\widehat\sigma_\tau^2$ converge to $\Omega_1/J_0^2$.
\end{proof}

\subsubsection{Proof of Theorem~\ref{thm:auto-inference}}

\begin{proof}[Proof of Theorem~\ref{thm:auto-inference}(a)]
By Lemma~\ref{lem:auto-gram}, with probability approaching one,
$\|\widehat T_{b,\tau}\|_{\op}\le2$, $\widehat{\mathcal L}_{b,\tau}\le C\xi_d^2=o(n)$,
and $\widehat{\mathrm{df}}_{b,\tau}\le Cd=o(n)$, because $\xi_d^2\log(2d)/n\to0$
implies $\xi_d^2/n\to0$. Lemma~\ref{lem:auto-expansion} therefore
applies, and part (b) of that lemma gives
$\sqrt nU^0_\tau=O_p\{(\max_b\widehat{\mathrm{df}}_{b,\tau}/n)^{1/2}\}=o_p(1)$.

\emph{The bias.} Write
$\bar b_\tau=b_\tau-\sum_b\omega_b\langle h_{\alpha,b},
(T_{b,\tau}-\widehat T_{b,\tau})h_{\gamma,b}\rangle$ with
$b_\tau=-\sum_b\omega_b\langle h_{\alpha,b},(I-T_{b,\tau})h_{\gamma,b}\rangle$.
By Lemma~\ref{lem:auto-operators}, $I-T_{b,\tau}$ is a positive
contraction, and Cauchy--Schwarz with $(I-T_{b,\tau})^{1/2}$ gives
\[
\sqrt n|b_\tau|\le\sqrt n\max_b
\|(I-T_{b,\tau})^{1/2}h_{\alpha,b}\|\,
\|(I-T_{b,\tau})^{1/2}h_{\gamma,b}\|\to_p0
\]
under (R1). By Lemma~\ref{lem:auto-gram}(a) and (c), the second term
is at most
\[
2\max_b\varrho_b\,\max_b\|h_{\alpha,b}\|\,\|h_{\gamma,b}\|
=O_p\bigl(\xi_d\sqrt{\log(2d)/n}\bigr)\max_b
\|h_{\alpha,b}\|\,\|h_{\gamma,b}\|=o_p(n^{-1/2})
\]
under (R2). With a known Gram matrix, $\widehat T_{b,\tau}=T_{b,\tau}$ and
this term vanishes, so (R2) is not needed.

\emph{Conclusion.} Hence
$\sqrt n\,\widehat{\mathcal M}_\tau(\theta_0)=\sqrt n\,\Pn\phi_0+o_p(1)$,
and Lemma~\ref{lem:auto-expansion}(c) gives
$\sqrt n(\widehat\theta_\tau-\theta_0)=J_0^{-1}\sqrt n\,\Pn\phi_0+o_p(1)$,
which also gives $\widehat\theta_\tau\to_p\theta_0$. The central limit
theorem gives the limit $N(0,\Omega_1/J_0^2)$, and
Lemma~\ref{lem:auto-variance} gives the consistency of $n\widehat\sigma_\tau^{\star2}$
and $n\widehat\sigma_\tau^2$.
\end{proof}

\begin{lemma}[Selection statistics]
\label{lem:auto-selection-stat}
Under Assumptions~\ref{ass:auto-model} and~\ref{ass:auto-first-step}, if
$\widetilde\theta-\theta_0=O_p(n^{-1/2}+r_n^2)$, then, uniformly in $\tau$
and for $a\in\{\gamma,\alpha\}$,
\[
\delta_a=O_p\Big(\frac{\sqrt d}{n}+\frac{r_n}{\sqrt n}
+\xi_d\sqrt{\frac{\log(2d)}{n}}\,r_n^2
+\big(n^{-1/2}+r_n^2\big)\big(r_n^2+\tfrac{\sqrt d}{n}
+\tfrac{r_n}{\sqrt n}\big)\Big),
\]
where $\delta_a$ is defined in \eqref{eq:app-auto-block-objects}.
\end{lemma}

\begin{proof}
Consider $\widehat Q_{\gamma,\tau}$. By
\eqref{eq:app-auto-linear-residuals}, its block-$b$ summand is
$\{V_{Z,i}-h_\gamma(X_i)\}K_{ij}\{V_{Z,j}-h_\gamma(X_j)\}$, with
conditional mean $\langle h_\gamma,\widehat Th_\gamma\rangle$. By
Lemma~\ref{lem:auto-blocks}, the oracle term
$V_{Z,i}K_{ij}V_{Z,j}$ has conditional standard deviation at most
$C\widehat{\mathrm{df}}_{b,\tau}^{1/2}/n$; the cross terms have first projections
$-V_Z(\widehat Th_\gamma)(X)$ and degenerate parts bounded as in
Lemma~\ref{lem:auto-expansion}, so their standard deviation is
$O_p(r_n/\sqrt n)$; and the last term fluctuates by
$O_p(r_n/\sqrt n)$. Lemma~\ref{lem:auto-gram} then gives
$|\widehat Q_{\gamma,\tau}-\mathcal Q_{\gamma,\tau}|
=O_p(\sqrt d/n+r_n/\sqrt n)+2\max_b\varrho_br_n^2$. At $\theta_0$, the
summand of $\widehat Q_{\alpha,\tau}$ is
$\{v_{0,i}-p_ih_\alpha(X_i)\}'M\{v_{0,j}-p_jh_\alpha(X_j)\}$, with conditional
mean $\langle h_\alpha,\widehat Th_\alpha\rangle$; its oracle part
$v_{0,i}'Mv_{0,j}$ is degenerate with $\E[(v_{0,i}'Mv_{0,j})^2]\le C\widehat{\mathrm{df}}_{b,\tau}$
by \eqref{eq:app-auto-msc}, and the cross terms are bounded as before,
so the same bound holds. Since
$\widehat v_{i,b}(\widetilde\theta)=\widehat v_{i,b}(\theta_0)
+(\widetilde\theta-\theta_0)\{\dot v_{0,i}-p_ih_{\dot\alpha}(X_i)\}$,
\[
\widehat Q_{\alpha,\tau}(\widetilde\theta)
-\widehat Q_{\alpha,\tau}(\theta_0)
=2(\widetilde\theta-\theta_0)\widehat Q_{\alpha\dot\alpha,\tau}
+(\widetilde\theta-\theta_0)^2\widehat Q_{\dot\alpha\dot\alpha,\tau},
\]
where the cross statistic $\widehat Q_{\alpha\dot\alpha,\tau}$ has
conditional mean $\langle h_\alpha,\widehat Th_{\dot\alpha}\rangle
=O_p(r_n^2)$ and $\widehat Q_{\dot\alpha\dot\alpha,\tau}$ has conditional mean
$\langle h_{\dot\alpha},\widehat Th_{\dot\alpha}\rangle=O_p(r_n^2)$, both with
the same fluctuations. The bound follows.
\end{proof}

\begin{proof}[Proof of Theorem~\ref{thm:auto-inference}(b)]
Write $\|h_a\|_\omega^2=\sum_b\omega_b\|h_{a,b}\|^2$.

\emph{Step 1: every candidate.} On the event of
Lemma~\ref{lem:auto-gram}, the conditions of
Lemmas~\ref{lem:auto-expansion} and~\ref{lem:auto-variance} hold for
every candidate, with $d/n\to0$. For every $\tau$,
$|\bar b_\tau|\le Cr_n^2\to_p0$, so $\widehat\theta_\tau\to_p\theta_0$, and
$\max_\tau|n\widehat\sigma_\tau^{\star2}-\Omega_1/J_0^2|=o_p(1)$. Moreover,
$\max_\tau\sqrt n|U^0_\tau|=o_p(1)$ and
$\max_\tau\sqrt n|\bar b_\tau-b_\tau|\le C\sqrt n\max_b\varrho_br_n^2=o_p(1)$ by
(iii).

\emph{Step 2: the budget.} By Step 1,
$\max_\tau|\widehat\sigma_\tau^{\star2}/\widehat\sigma_{f,\mathrm{base}}^{\star2}-1|=o_p(1)$, where $f$ is the
family of $\tau$. Since $c>1$, every candidate satisfies
\eqref{eq:auto-budget} with probability approaching one, and
$\mathcal J_{\mathrm{sel}}$ contains the last member of each family.
By nesting, the last member $\tau_f$ of the family of $\tau^\ast$ satisfies
$d_{a,\tau_f}\le d_{a,\tau^\ast}$ and $\mathcal Q_{a,\tau_f}\ge\mathcal Q_{a,\tau^\ast}$.

\emph{Step 3: the selected bias.} By (i), $\|h_a\|_\omega\ge r_n/C$ with
probability approaching one. By Lemma~\ref{lem:auto-selection-stat}
and (iii), $\sqrt n\max_a\delta_a=o_p(1)$, since every term in the
bound times $\sqrt n$ vanishes: $\sqrt{d/n}\to0$, $r_n\to_p0$,
$\sqrt n\,\xi_d\sqrt{\log(2d)/n}\,r_n^2\to_p0$, and
$\sqrt n(n^{-1/2}+r_n^2)(r_n^2+\sqrt d/n+r_n/\sqrt n)=o_p(1)$ because
$\sqrt n\,r_n^4\to_p0$. Let $E_n=\{\delta_a\le\frac12\underline c\,
\|h_a\|_\omega^2,\ a=\gamma,\alpha\}$.

On $E_n$, $q_a\ge\widehat Q_{a,\tau_f}\ge\mathcal Q_{a,\tau_f}-\delta_a
\ge\frac12\underline c\,\|h_a\|_\omega^2>0$ and
$q_a\le\max_\tau\mathcal Q_{a,\tau}+\delta_a\le(1+\underline c)\|h_a\|_\omega^2$,
because $\mathcal Q_{a,\tau}\le\|h_a\|_\omega^2$. By (i), the ratio
$\varpi=(q_\gamma/q_\alpha)^{1/2}$ in
Proposition~\ref{prop:auto-selection} and its inverse are bounded, and
that proposition, with $C_F=0$, gives
$|b_{\widehat \tau}|\le C(d_{\alpha,\tau^\ast}+d_{\gamma,\tau^\ast}
+\delta_\alpha+\delta_\gamma)$. Hence
$\sqrt n|b_{\widehat \tau}|1\{E_n\}=o_p(1)$ by (iii).

On $E_n^c$, some $\delta_a$ exceeds
$\frac12\underline c\|h_a\|_\omega^2\ge c'r_n^2$, so
$\sqrt n\,r_n^2\le C\sqrt n\max_a\delta_a$. Since $|b_\tau|\le r_n^2$ for
every $\tau$, $\sqrt n|b_{\widehat \tau}|1\{E_n^c\}\le C\sqrt n\max_a\delta_a
=o_p(1)$.

\emph{Step 4: conclusion.} By Steps 1 and 3 and
Lemma~\ref{lem:auto-expansion}, uniformly over the finite library,
$\widehat{\mathcal M}_{\widehat \tau}(\theta_0)=\Pn\phi_0+o_p(n^{-1/2})$,
$\widehat J_{\widehat \tau}\to_pJ_0$, and $n\widehat\sigma_{\widehat \tau}^{\star2}$ and
$n\widehat\sigma_{\widehat \tau}^2$ converge to $\Omega_1/J_0^2$. The rest
of the proof is that of part (a).
\end{proof}

\begin{proof}[Proof of Corollary~\ref{cor:plm}]
(a) The moment $m(W,\gamma,\theta)=(Y-\theta D)\{D-\gamma(X)\}$ is
affine in $\gamma$ and in $\theta$, with
$D_\gamma m(W,\gamma_0,\theta)[a]=-(Y-\theta D)a(X)$. Since
$\E[Y-\theta D\mid X]=\mu_0-\theta\gamma_0$, the representer is
$\alpha_0(\cdot,\theta)=\theta\gamma_0-\mu_0$, with
$\partial_\theta\alpha_0(\cdot,\theta)=\gamma_0$; both are bounded because
$\gamma_0$ and $c_0=\mu_0-\theta_0\gamma_0$ are. Writing
$Y-\theta D=c_0+\varepsilon+(\theta_0-\theta)D$ and
$\E[D^2\mid X]=\gamma_0^2+\E[V^2\mid X]$ shows that
$\E[(Y-\theta D)^2\mid X]$ is bounded, which gives mean-square
continuity, and $V_Z=V$ has bounded conditional variance. At the
truth, $\phi_0=(Y-\theta_0D-c_0)V=\varepsilon V$, $J_0=\E[DV]=\E[V^2]>0$,
$0<\Omega_1=\E[\varepsilon^2V^2]\le C\E[V^2]$, and
$\partial_\theta m(W,\gamma_0,\theta)=-DV$ has a finite second moment
because $\E[V^4\mid X]$ is bounded. With
$\widehat\alpha_b(\cdot,\theta)=\theta\widehat\gamma_b-\widehat\mu_b$,
$h_{\alpha,b}=\theta_0h_{\gamma,b}-(\widehat\mu_b-\mu_0)$ and
$h_{\dot\alpha,b}=h_{\gamma,b}$, so Assumption~\ref{ass:auto-first-step}(i)
follows from the conditions on $\widehat\mu_b$ and $\widehat\gamma_b$.

(b) Scale each B-spline basis by the square root of its
dimension. On quasi-uniform partitions, the eigenvalues
of $G_0$ are then bounded away from zero and infinity,
$\xi_d\le C\sqrt d$, and every H\"older-$s$ function $f$
satisfies
$\inf_\beta\|f-p'\beta\|_\infty\le Cd^{-s/q}$
\citep{Newey1997,BCCK2015}.

Fix a training block and let $N=|I_b^c|\asymp n$.
For $f=\mu_0$ or $\gamma_0$, let $Z_{f,i}=Y_i$ or $D_i$,
respectively, and write
$\epsilon_{f,i}=Z_{f,i}-f(X_i)$. These errors have zero
conditional means and uniformly bounded conditional
variances under the conditions of the corollary.
There is an approximant $f_*\in\operatorname{span}(p_{\rm fs})$
with $\|f-f_*\|_\infty=O(d_{\rm fs}^{-s/q})$; the strict inequality
defining $C_{\rm fs}$ makes $f_*$ feasible for all sufficiently
large $n$. Put
\[
 \|a\|_N^2=\frac1N\sum_{i\in I_b^c}a(X_i)^2,
 \qquad
 \widehat G_{\rm fs}=\frac1N\sum_{i\in I_b^c}
                         p_{\rm fs}(X_i)p_{\rm fs}(X_i)'.
\]
The constrained least-squares basic inequality gives
\[
 \|\widehat f-f_*\|_N
 \le
 2\left\|\widehat G_{\rm fs}^{-1/2}
       \frac1N\sum_{i\in I_b^c}
                  p_{\rm fs}(X_i)\epsilon_{f,i}\right\|_2
 +2\|f-f_*\|_N.
\]
The conditional second moment of the vector norm on the
right is bounded by $Cd_{\rm fs}/N$. Since
$d_{\rm fs}\log(2d_{\rm fs})/n\to0$, empirical and population norms are
equivalent on the first-step estimation space with probability tending
to one. Consequently,
\[
 \|\widehat f-f\|
 =O_p\!\left(\sqrt{d_{\rm fs}/n}+d_{\rm fs}^{-s/q}\right),
 \qquad
 \|\widehat f-f\|_\infty=O_p(1).
\]
The second bound follows directly from the constraint.
Thus the first-step estimators satisfy the boundedness and consistency
requirements in part (a), and
$r_n=O_p\{(n/\log n)^{-s/(2s+q)}\}$.

The correction is the projection candidate, so $T_{b,\tau}=\Pi_d$, the
projection onto the correction space. Because the first-step estimation space is
contained in the correction space,
$(I-\Pi_d)h_{\gamma,b}=-(I-\Pi_d)\gamma_0$ and
$(I-\Pi_d)h_{\alpha,b}=(I-\Pi_d)(\mu_0-\theta_0\gamma_0)
=(I-\Pi_d)c_0$. Both are $O(d^{-s/q})$, so (R1) holds if
$\sqrt n\,d^{-2s/q}\to0$. Condition (R2) holds if
$\sqrt{d\log(2d)}\,(n/\log n)^{-2s/(2s+q)}\to0$, and
Assumption~\ref{ass:auto-first-step}(ii) if $d\log(2d)/n\to0$. Take
$d\asymp n^{\kappa}$. These requirements, together with $d\ge d_{\rm fs}$,
hold if
\[
\max\left\{\frac{q}{4s},\frac{q}{2s+q}\right\}
<\kappa<\min\left\{1,\frac{4s}{2s+q}\right\},
\]
because the inequalities are strict and absorb the logarithms. Since
$s>q/4$, $q/(4s)<1$ and $q/(2s+q)<\min\{1,4s/(2s+q)\}$. The remaining
inequality, $q/(4s)<4s/(2s+q)$, is $16s^2-2qs-q^2>0$, that is,
$s>(1+\sqrt{17})q/16$. For the comparison,
$\sqrt n\,r_n^2\to0$ at these rates requires $2s/(2s+q)>1/2$, that is,
$s>q/2$.
\end{proof}

\paragraph{The variance budget in this regime.}
Variance consistency implies, for each fixed $c>1$,
$\max_\tau|\widehat\sigma_\tau^{\star2}/\widehat\sigma_{f,\mathrm{base}}^{\star2}-1|=o_p(1)$, where $f$ is the
family of $\tau$. Rule C therefore admits the entire finite library with
probability tending to one and retains the last member of each
family. The budget has a finite-sample role here, and the grid cap
determines asymptotic complexity: Assumption~\ref{ass:auto-first-step}(ii)
bounds the dictionary at $d=o(n)$.

\subsection{Inference with nonnegligible correction variance}
\label{app:nonneg}

This subsection proves Theorem~\ref{thm:nonneg-inference}
for overlapping training samples. With $\mathcal H_n$ as
defined in Section~\ref{sec:auto-estimate}, put $\mathcal G_n=\mathcal H_n\vee\sigma(X_1,\ldots,X_n)$.
Since $\mathcal H_n$ is independent of the main sample,
the pairs $(\varepsilon_i,V_i)$ are independent across $i$
conditionally on $\mathcal G_n$, with
\[
\E[\varepsilon_i\mid\mathcal G_n]
=\E[V_i\mid\mathcal G_n]
=\E[\varepsilon_iV_i\mid\mathcal G_n]=0.
\]
Put $\sigma_{\varepsilon,i}^2=\E[\varepsilon_i^2\mid X_i]$,
$\sigma_{V,i}^2=\E[V_i^2\mid X_i]$, and
$\kappa_{22,i}=\E[\varepsilon_i^2V_i^2\mid X_i]$.
The assumptions bound the conditional fourth moments of
$\varepsilon_i$, $V_i$, and $\varepsilon_iV_i$ uniformly.
For any $n\times n$ matrix $A$,
$\|A\|_F^2\le n\|A\|_{\op}^2$ and
$\sum_j A_{ij}^2\le\|A\|_{\op}^2$.

Use the reference matrices from \eqref{eq:nonneg-stability}
to define
\[
N_\tau^\circ=\sum_i\varepsilon_iV_i
-\varepsilon'\mathbb B_\tau^\circ V,\qquad
Z_{\gamma,\tau}^\circ=n^{-1/2}V'\mathbb B_\tau^\circ V,\qquad
Z_{\alpha,\tau}^\circ=n^{-1/2}\varepsilon'\mathbb B_\tau^\circ\varepsilon,
\]
and
\[
\Omega_{2,\tau,n}^\circ=
\frac1n\sum_{i\ne j}((\mathbb B_\tau^\circ)_{ij})^2
\sigma_{\varepsilon,i}^2\sigma_{V,j}^2.
\]
Conditional index matching gives
\begin{equation}
\Var(N_\tau^\circ\mid\mathcal G_n)
=n(\Omega_{1,n}+\Omega_{2,\tau,n}^\circ),\qquad
\Cov(N_\tau^\circ,Z_{a,\tau'}^\circ\mid\mathcal G_n)=0
\label{eq:nonneg-moments}
\end{equation}
for all $\tau,\tau'$ and $a\in\{\gamma,\alpha\}$.
The reversed pair matching vanishes because
$\E[\varepsilon_iV_i\mid X_i]=0$; all other unmatched
indices leave a factor with conditional mean zero.

\begin{lemma}[Linear--quadratic forms]
\label{lem:nonneg-clt}
Let $c\in\R^n$ and $n\times n$ matrices $A_1,A_2,A_3$ be
$\mathcal G_n$-measurable, with zero diagonals, $A_2$ and $A_3$
symmetric, $\max_i|c_i|=O_p(1)$, and $\max_k\|A_k\|_{\op}=O_p(1)$. Put
\[
L_n=\sum_ic_i\varepsilon_iV_i+\varepsilon'A_1V
+\varepsilon'A_2\varepsilon+V'A_3V,
\qquad s_n^2=\Var(L_n\mid\mathcal G_n).
\]
If $s_n^2\ge\underline c\,n$ with probability approaching one for a
constant $\underline c>0$, then
$\sup_x|\PP(L_n/s_n\le x\mid\mathcal G_n)-\Phi(x)|\to_p0$.
\end{lemma}

\begin{proof}
Work on the event where the bounds hold, conditionally on
$\mathcal G_n$. Let $\mathcal F_i$ be generated by $\mathcal G_n$ and
$(\varepsilon_j,V_j)_{j\le i}$, and put
\[
S_i=\sum_{j<i}(A_{1,ij}V_j+2A_{2,ij}\varepsilon_j),
\qquad
T_i=\sum_{j<i}(A_{1,ji}\varepsilon_j+2A_{3,ij}V_j).
\]
Then $L_n=\sum_i\xi_i$ with
$\xi_i=c_i\varepsilon_iV_i+\varepsilon_iS_i+V_iT_i$, a martingale
difference sequence for $(\mathcal F_i)$. By the martingale central
limit theorem \citep[Corollary~3.1]{HallHeyde1980}, it suffices that
$s_n^{-4}\sum_i\E[\xi_i^4\mid\mathcal G_n]\to0$ and
$s_n^{-2}\sum_i\E[\xi_i^2\mid\mathcal F_{i-1}]\to_p1$.

Since $S_i$ and $T_i$ are linear in variables independent of
$(\varepsilon_i,V_i)$ with bounded fourth moments,
$\E[S_i^4+T_i^4\mid\mathcal G_n]\le C\max_k\|A_k\|_{\op}^4$, so
$\sum_i\E[\xi_i^4\mid\mathcal G_n]\le Cn=o(s_n^4)$. Next,
\[
\E[\xi_i^2\mid\mathcal F_{i-1}]
=c_i^2\kappa_{22,i}+\sigma_{\varepsilon,i}^2S_i^2+\sigma_{V,i}^2T_i^2
+2c_i\kappa_{21,i}S_i+2c_i\kappa_{12,i}T_i,
\]
with $\kappa_{21,i}=\E[\varepsilon_i^2V_i\mid X_i]$ and
$\kappa_{12,i}=\E[\varepsilon_iV_i^2\mid X_i]$ bounded; the term in
$S_iT_i$ has coefficient $\E[\varepsilon_iV_i\mid X_i]=0$. The sum
of these conditional variances has mean $s_n^2$, and it remains to
show that its conditional variance is $o(n^2)$. Let $\tilde A_k$ be
the strictly lower triangular part of $A_k$, also including $A_1'$
when bounding $T_i$. The truncation bound of
\citet{KwapienPelczynski1970} gives $\|\tilde A_k\|_{\op}\le C\log(n+1)\|A_k\|_{\op}$.
The linear terms, such as $\sum_ic_i\kappa_{21,i}S_i$, have conditional
variance at most $C\max_k\|\tilde A_k\|_{\op}^2\sum_ic_i^2=O(n\log^2n)$.
The quadratic terms, such as $\sum_i\sigma_{\varepsilon,i}^2S_i^2$, are quadratic
forms $\zeta'\mathcal M\zeta$ in the independent vectors
$\zeta_j=(\varepsilon_j,V_j)$ with
$\|\mathcal M\|_F\le C\max_k\{\|\tilde A_k\|_{\op}\,\|A_k\|_F\}
=O(\sqrt n\log n)$, and such forms have conditional variance at most
$C\|\mathcal M\|_F^2=O(n\log^2n)$. Both are $o(n^2)$.
\end{proof}

\begin{lemma}[Replacement]
\label{lem:nonneg-replacement}
Suppose the moment and first-step estimator conditions of
Theorem~\ref{thm:nonneg-inference} hold and candidate $\tau$
satisfies \eqref{eq:nonneg-stability}. Then:
\begin{enumerate}[label=(\alph*),leftmargin=*]
\item $n\widehat{\mathcal M}_\tau(\theta_0)
=N_\tau^\circ+n\bar b_{\tau,n}+o_p(\sqrt n)$ and
$\widehat J_\tau\to_pJ_0$;
\item $n^{-1}\sum_i(\widehat\psi_{i,\tau}-\bar\psi_\tau)^2
=\Omega_{1,n}+2\Omega_{2,\tau,n}+o_p(1)$,
$P_\tau/n=\Omega_{2,\tau,n}+o_p(1)$, and
$n^{-1}\sum_i\bar\phi_{i,\tau}^2=\Omega_{1,n}+o_p(1)$;
\item if also $\sqrt n\,r_n^4\to_p0$ and
$\widetilde\theta-\theta_0=O_p(n^{-1/2}+r_n^2)$, then
$\sqrt n(\widehat Q_{a,\tau}-\bar Q_{a,\tau,n})
=Z_{a,\tau}^\circ+o_p(1)$, where
\[
\bar Q_{a,\tau,n}=\frac1n\sum_b
h_{a,b}'\mathbb B_\tau^{(b)}h_{a,b},
\qquad a\in\{\gamma,\alpha\}.
\]
\end{enumerate}
Here a superscript $(b)$ retains entries corresponding to
$\mathcal P_b$, and $h_{a,b}$ denotes the vector
$(h_{a,b}(X_i))_{i\le n}$.
The conclusions hold uniformly over a bounded number of candidates.
\end{lemma}

\begin{proof}
Let $\mathcal F_b$ be generated by $\mathcal G_n$ and the
observations in $I_b^c$, and let a superscript $(b)$
retain only entries corresponding to $\mathcal P_b$.
For symmetric zero-diagonal $A$, a within-set restriction
is a principal submatrix. A rectangular restriction has the
form $\left(\begin{smallmatrix}0&C\\C'&0\end{smallmatrix}\right)$
with $C=A_{I_{b,1},I_{b,2}}$. Hence
$\|A^{(b)}\|_{\op}\le\|A\|_{\op}$ in both cases.
Set $\Delta_\tau=\mathbb B_\tau-\mathbb B_\tau^\circ$.
Conditional on $\mathcal F_b$, $\Delta_\tau^{(b)}$ is fixed,
has zero diagonal, and involves independent evaluation observations.
The conditional means of
\[
\varepsilon'\Delta_\tau^{(b)}V,\qquad
V'\Delta_\tau^{(b)}V,\qquad
\varepsilon'\Delta_\tau^{(b)}\varepsilon
\]
are zero and their conditional variances are bounded by
$C\|\Delta_\tau^{(b)}\|_F^2$. Conditional Chebyshev and
\eqref{eq:nonneg-stability}, applied separately to the finitely
many blocks, therefore make all three full-sample forms
$o_p(\sqrt n)$. Moreover,
\[
|\Omega_{2,\tau,n}-\Omega_{2,\tau,n}^\circ|
\le\frac Cn\|\Delta_\tau\|_F
(\|\mathbb B_\tau\|_F+\|\mathbb B_\tau^\circ\|_F)=o_p(1).
\]

(a) Let $I_b$ be the participating observations of block $b$
and put $\|h\|_b^2=\sum_{i\in I_b}h_b(X_i)^2$.
For this step, condition on the smaller training field
$\mathcal H_b=\mathcal H_n\vee\sigma(W_i:i\in I_b^c)$.
Independence of the evaluation observations gives
\[
\E[\|h_{a,b}\|_b^2\mid\mathcal H_b]
=|I_b|\|h_{a,b}\|^2.
\]
Conditional Markov applied to each empirical-to-population
squared-norm ratio, defined as zero when its denominator is
zero, yields
$\max_{b,a}\|h_{a,b}\|_b^2=O_p(nr_n^2)=o_p(n)$.
The following residual calculations condition on $\mathcal F_b$.

At $\theta_0$, $\widehat e_{i,b}=\varepsilon_i+h_{\alpha,b}(X_i)$
and $\widehat V_{i,b}=V_i-h_{\gamma,b}(X_i)$.
Expanding the estimating equation gives
$\sum_i\varepsilon_iV_i-\varepsilon'\mathbb B_\tau V
+n\bar b_{\tau,n}+\sum_b\Lambda_b$, where, writing $m_{b,i}$
for the number of partners of $i$ in $\mathcal P_b$,
\[
\Lambda_b=\sum_{i\in I_b}
\left[V_i\left\{\frac{m_{b,i}}{n-1}h_{\alpha,b}(X_i)
-(\mathbb B_\tau^{(b)}h_{\alpha,b})_i\right\}
-\varepsilon_i\left\{\frac{m_{b,i}}{n-1}h_{\gamma,b}(X_i)
-(\mathbb B_\tau^{(b)}h_{\gamma,b})_i\right\}\right].
\]
Its conditional mean is zero and its conditional variance is
at most $C(1+\|\mathbb B_\tau\|_{\op}^2)
(\|h_{\alpha,b}\|_b^2+\|h_{\gamma,b}\|_b^2)=o_p(n)$.
The preceding matrix replacement proves the moment expansion.
Expanding the derivative in the same way gives
\[
\widehat J_\tau=\frac1n\sum_iV_i^2
-\frac1nV'\mathbb B_\tau^\circ V+o_p(1)\to_pJ_0.
\]
Indeed, the quadratic form is $O_p(\sqrt n)$, and the
terms involving first-step estimation errors are bounded by their empirical norms and the
operator norms. Also $|\bar b_{\tau,n}|=O_p(r_n^2)=o_p(1)$,
so $\widehat\theta_\tau\to_p\theta_0$.

(b) For any symmetric zero-diagonal $A$, define
\[
a_i^0(A)=\varepsilon_iV_i-\varepsilon_i(AV)_i
-V_i(A\varepsilon)_i,\qquad
P^0(A)=\frac12\sum_{i\ne j}A_{ij}^2
(\varepsilon_iV_j+\varepsilon_jV_i)^2.
\]
For $A=\mathbb B_\tau^\circ$, conditional index matching gives
\[
\E\left[\sum_i a_i^0(A)^2\mid\mathcal G_n\right]
=n(\Omega_{1,n}+2\Omega_{2,\tau,n}^\circ),\qquad
\E[P^0(A)\mid\mathcal G_n]=n\Omega_{2,\tau,n}^\circ.
\]
The corresponding conditional variances are bounded by
$Cn(1+\|A\|_{\op}^4)$ and $Cn\|A\|_{\op}^4$.
To obtain these bounds, expand the squared rows and match
indices across their centered products. The first-order
term contributes at most $Cn$. Terms linear in $A$ have
variance at most $Cn\|A\|_{\op}^2$: centering their
single-index factors leaves degenerate pair terms and linear
forms with coefficient vectors $A'\kappa$, where
$\|\kappa\|_2^2\le Cn$. Terms quadratic in $A$ have variance
bounded by a constant times
\[
\|A\|_F^2\|A\|_{\op}^2+
\sum_i\left(\sum_j A_{ij}^2\right)^2
\le2n\|A\|_{\op}^4.
\]
The cross terms are bounded by Cauchy--Schwarz.
Also $n^{-1}\sum_i a_i^0(A)=O_p(n^{-1/2})$.

For the actual matrices, blockwise conditioning gives
\[
\E\left[\sum_i\{
\varepsilon_i^2(\Delta_\tau^{(b)}V)_i^2+
V_i^2(\Delta_\tau^{(b)}\varepsilon)_i^2\}
\mid\mathcal F_b\right]
\le C\|\Delta_\tau^{(b)}\|_F^2.
\]
Thus $n^{-1}\sum_i\{a_i^0(\mathbb B_\tau)
-a_i^0(\mathbb B_\tau^\circ)\}^2=o_p(1)$.
Similarly $P^0(\Delta_\tau)=o_p(n)$, and
\[
|P^0(\mathbb B_\tau)-P^0(\mathbb B_\tau^\circ)|
\le P^0(\Delta_\tau)
+2\{P^0(\Delta_\tau)P^0(\mathbb B_\tau^\circ)\}^{1/2}
=o_p(n).
\]
First-step replacement follows from the same conditioning. For
example, the conditional sums of squares of
$h_{\alpha,b}(X_i)(\mathbb B_\tau^{(b)}V)_i$ and
$\varepsilon_i(\mathbb B_\tau^{(b)}h_{\gamma,b})_i$ are bounded
by $C\|\mathbb B_\tau\|_{\op}^2\|h_{\alpha,b}\|_b^2$ and
$C\|\mathbb B_\tau\|_{\op}^2\|h_{\gamma,b}\|_b^2$.
Terms containing two first-step estimation errors also use the sup-norm bounds.
The same estimates applied to weighted pair squares give
$P_\tau-P^0(\mathbb B_\tau)=o_p(n)$.
Terms containing $\widehat\theta_\tau-\theta_0$ factor this
scalar outside empirical derivative-square averages, which
are $O_p(1)$ by the fourth-moment bounds.
Consequently the fitted and oracle rows differ by $o_p(1)$
in empirical mean square. The same argument gives
$n^{-1}\sum_i(\bar\phi_{i,\tau}-\varepsilon_iV_i)^2=o_p(1)$.
Cauchy--Schwarz and the law of large numbers prove (b).

(c) The blockwise identities are
\[
\widehat Q_{\gamma,\tau}
=\frac1n\sum_b(V-h_{\gamma,b})'
\mathbb B_\tau^{(b)}(V-h_{\gamma,b}),\qquad
\widehat Q_{\alpha,\tau}(\theta_0)
=\frac1n\sum_b(\varepsilon+h_{\alpha,b})'
\mathbb B_\tau^{(b)}(\varepsilon+h_{\alpha,b}).
\]
Subtracting $\bar Q_{a,\tau,n}$ leaves the oracle quadratic
form and linear terms. Each linear term, scaled by $n^{-1/2}$,
has conditional variance at most
$Cn^{-1}\|\mathbb B_\tau\|_{\op}^2\|h_{a,b}\|_b^2=o_p(1)$.
The initial matrix replacement gives $Z_{a,\tau}^\circ$.
Write $t=\widetilde\theta-\theta_0$ and
\[
C_\tau=\frac1n\sum_b\widehat e_b(\theta_0)'
\mathbb B_\tau^{(b)}\widehat V_b
=O_p(n^{-1/2}+r_n^2),\qquad
D_\tau=\frac1n\sum_b\widehat V_b'
\mathbb B_\tau^{(b)}\widehat V_b=O_p(1).
\]
These bounds follow by the same expansion and blockwise
conditioning. Affinity gives
$\widehat Q_{\alpha,\tau}(\widetilde\theta)
-\widehat Q_{\alpha,\tau}(\theta_0)=-2tC_\tau+t^2D_\tau
=O_p(n^{-1}+r_n^4)=o_p(n^{-1/2})$.
Part (a) with zero correction matrices proves the stated
rate for the preliminary estimator without correction.
\end{proof}

For selection, let
\[
s_n=\max\{n^{-1/2},\max_{a,\tau}|\bar Q_{a,\tau,n}|\},\quad
\varkappa_n=(\bar Q_{a,\tau,n}/s_n)_{a,\tau},\quad
w_n=n^{-1/2}/s_n,\quad
\Sigma_n=\Var(Z^\circ\mid\mathcal G_n),
\]
where $Z^\circ=(Z_{a,\tau}^\circ)_{a,\tau}$, and set
$v_{\tau,n}=(\Omega_{1,n}+\Omega_{2,\tau,n})/J_0^2$.

\begin{assumption}[No limiting ties]
\label{ass:nonneg-ties}
(i) $(\varkappa_n,w_n,\Sigma_n,(v_{\tau,n})_\tau)$ converges
in probability to a nonrandom limit
$(\varkappa,w,\Sigma,(v_\tau)_\tau)$;
(ii) for every candidate $\tau$ of family $f$,
$v_\tau\ne c\,v_{f,\mathrm{base}}$, where $c$ is the budget constant;
and (iii) for each permitted priority ordering $\pi$, the
map $g_\pi$ applying \eqref{eq:auto-selection}, with its
omission rule and that priority ordering, to the last member
of each family satisfying $v_\tau<c\,v_{f,\mathrm{base}}$ is continuous at
$\varkappa+w\mathcal Z$ almost surely, where
$\mathcal Z\sim N(0,\Sigma)$.
\end{assumption}

Condition (iii) excludes limiting ties between criteria and
sign changes of the normalizers $q_a$ that make the selection
map discontinuous. With a fixed priority ordering, it permits
random limiting selection when $w>0$ and gives asymptotically
nonrandom selection when $w=0$. With overlapping training,
the signals and estimated variances depend on the same
responses as the estimator; the nonrandom limits in (i),
together with the joint Gaussian limit, deliver the required
asymptotic independence.

\begin{proof}[Proof of Theorem~\ref{thm:nonneg-inference}]
For (a), put $s_{n,\tau}^{\circ2}
=n(\Omega_{1,n}+\Omega_{2,\tau,n}^\circ)$.
Since $\Omega_{1,n}\to_p\E[\varepsilon^2V^2]>0$,
$s_{n,\tau}^{\circ2}\ge\underline c n$ with probability
approaching one. Lemma~\ref{lem:nonneg-clt}, with
$c_i=1$, $A_1=-\mathbb B_\tau^\circ$, and $A_2=A_3=0$,
gives $N_\tau^\circ/s_{n,\tau}^\circ\dto N(0,1)$.
Lemma~\ref{lem:nonneg-replacement}, the negligible bias,
and $\widehat J_\tau\to_pJ_0$ give the studentized limit and
both variance ratios, using
$\Omega_{2,\tau,n}-\Omega_{2,\tau,n}^\circ=o_p(1)$.

For (b), the variance estimates and
Assumption~\ref{ass:nonneg-ties}(i)--(ii) imply that rule C
retains the limiting family representatives with probability
approaching one. Let $\Pi_n$ be the fixed or independently
drawn priority ordering, included in $\mathcal H_n$.
Homogeneity and the replacement lemma give
\[
\widehat Q/s_n=\varkappa_n+w_nZ^\circ+o_p(1),\qquad
\widehat \tau=g_{\Pi_n}(\widehat Q/s_n)
\quad\text{with probability approaching one}.
\]
For fixed $\tau$, set $T_\tau^\circ=N_\tau^\circ/s_{n,\tau}^\circ$.
Apply Lemma~\ref{lem:nonneg-clt} to $\sqrt n$ times each
linear combination of $(T_\tau^\circ,Z^\circ)$ whose limiting
variance is positive. Its coefficients and matrix operator
norms are $O_p(1)$; its conditional variance divided by $n$
converges by \eqref{eq:nonneg-moments} and the assumption.
Combinations with zero limiting variance converge to zero
in probability by conditional Chebyshev. Thus
\[
(T_\tau^\circ,Z^\circ)\dto(T,\mathcal Z),\qquad
T\sim N(0,1),\quad T\text{ independent of }\mathcal Z.
\]
The conditional argument also gives, for every bounded
continuous $f$,
\[
\E[f(T_\tau^\circ,Z^\circ)\mid\mathcal G_n]
\to_p\E[f(T,\mathcal Z)].
\]
Boundedness makes this convergence hold in $L_1$.
Along any subsequence on which the distribution of the finite
priority ordering converges,
$(T_\tau^\circ,Z^\circ,\Pi_n)\dto(T,\mathcal Z,\Pi)$,
with $(T,\mathcal Z)$ independent of $\Pi$.
The nonrandom limits of $\varkappa_n,w_n$ and continuity
of each $g_\pi$ give
$(T_\tau^\circ,\widehat \tau)\dto
(T,g_\Pi(\varkappa+w\mathcal Z))$.
Summing over candidates yields
$\PP(T_{\widehat \tau}^\circ\le x)\to\Phi(x)$ along every
such subsequence and hence along the full sequence.
Finally, first evaluate the uniform moment expansion at
$\widehat \tau$. The selected-bias condition yields
\[
n\widehat J_{\widehat \tau}
(\widehat\theta_{\widehat \tau}-\theta_0)
=N_{\widehat \tau}^\circ+o_p(\sqrt n).
\]
Uniform variance consistency then gives
$(\widehat\theta_{\widehat \tau}-\theta_0)
/\widehat\sigma_{\widehat \tau}^\star
=T_{\widehat \tau}^\circ+o_p(1)$ and the two variance ratios.
\end{proof}
\section{Proofs for Section~\ref{sec:applications}}
\label{app:applications}

\subsection{Binary choice}
\label{app:binary}

\begin{proof}[Proof of Proposition~\ref{prop:binaryclass}]
The unrestricted distribution of $X$ implies that every first-order
orthogonal $\phi\in L_2^0(P_0)$ has $\E[\phi\mid X]=0$.
Orthogonality to $R_Bh(X)$ further requires $\E[\phi R_B\mid X]=0$.
For a fixed $x$ there are four outcome states, all having positive
probability. The conditional mean-zero space has dimension three;
orthogonality to the nonzero $R_B$ leaves dimension two. Both $C_0(W)=C(W,\theta_0)$
and $D_B$ satisfy these restrictions. Conditional independence gives
\[
 \E[D_B^2\mid X]=w_0,\qquad \E[D_BR_B\mid X]=0,
 \qquad \E[D_B\varepsilon_2\mid X]=0.
\]
The conditional odds identity gives $\E[C_0\mid X]=0$, and differentiation
of this identity in the baseline index gives $\E[C_0R_B\mid X]=0$.
Summing over the four states gives
\begin{equation}
 \E[C_0^2\mid X]=J_0,\quad \E[C_0D_B\mid X]=\kappa_0 w_0,\quad
 \E[C_0\varepsilon_2\mid X]=J_0,\quad J_0-\kappa_0^2w_0=J_0^2/e_{B,0}.
 \label{eq:fourstates}
\end{equation}
For example the discordant states have probabilities
$A=(1-p_{1,0})p_{2,0}$ and $B=p_{1,0}(1-p_{2,0})$, with $A/(A+B)=\Lambda(\theta_0)$
and $AB=w_0$. These identities give $J_0=AB/(A+B)$ and verify
\eqref{eq:fourstates} by direct substitution.
The covariance matrix of $(C_0,D_B)$ has determinant $w_0J_0^2/e_{B,0}>0$.
Overlap bounds its eigenvalues above and away from zero. Thus $(C_0,D_B)$
is a conditional basis, and square integrability of $\phi$ is equivalent
to $a,b\in L_2(P_{0,X})$. This proves \eqref{eq:binaryclass}, its uniqueness,
the variance formula, and target sensitivity $\E[aJ_0]$.

The mixed second likelihood derivative along baseline directions is
\[
 u_{h,\widetilde h}=\{R_B^2-V_0\}h(X)\widetilde h(X).
\]
Since $\E[C_0(R_B^2-V_0)\mid X]=0$ and
$\E[D_B(R_B^2-V_0)\mid X]=2w_0$, the curvature is \eqref{eq:binarycurv}.
For a bounded marginal density direction $z(X)$ the mixed likelihood
derivative is $R_Bh(X)z(X)$, whose covariance with $\phi$ is zero.
Pure marginal-density derivatives depend only on $X$ and also have zero
covariance with $\phi$. Finally, projecting $\ell=\varepsilon_2$ onto
the nuisance space gives $(v_{2,0}/V_0)R_B$; its orthogonal residual has
conditional variance $v_{1,0}v_{2,0}/V_0=e_{B,0}$. This proves the efficient-score formula in Proposition~\ref{prop:binaryclass}.
\end{proof}

\begin{proof}[Proof of Theorem~\ref{thm:binary}: correctability and extension]
Apply Proposition~\ref{prop:support} with scalar residual $R_B$, conditional
variance $\Sigma=V_0$, and curvature coefficient $bw_0$. The whitened coefficient
is $bw_0/V_0=be_{B,0}$, yielding the square-summability condition and
\eqref{eq:binarykernel}. Because each factor of this kernel contains
$R_B(W_i)$, its contraction with a pure marginal score $z(X_i)$ is zero.
It therefore satisfies the remaining zero-curvature restrictions as well.

For attainability, fix the coefficient functions $a,b$ and define at a
nearby law
\[
 \phi(W,\lambda)=a(X)C(W,\theta)+b(X)
   \{Y_1-\Lambda(\gamma(X))\}\{Y_2-\Lambda(\gamma(X)+\theta)\}.
\]
Its conditional mean is zero. Its first baseline derivative has mean zero
because it is a linear combination of centered residuals. It does not
depend on the marginal density as a nuisance argument. Thus validity and
uniform first-order orthogonality hold. Define $k(\cdot,\cdot,\lambda)$ by
\eqref{eq:binarykernel}, using the probabilities, variances, residuals,
and atom masses at that law. It is locally degenerate. Under bounded
baseline perturbations and bounded marginal likelihood ratios bounded
away from zero, overlap and the square-summability condition give
square-integrable derivative envelopes for the kernel and moment.
Differentiation is justified and Lemma~\ref{lem:pathwise} establishes
second-order cancellation. Such paths generate the full tangent space.
\end{proof}

\begin{proof}[Proof of Theorem~\ref{thm:binary}: joint optimization]
By Proposition~\ref{prop:binaryclass} and the preceding correctability argument, the
scaled minimum variance is the minimum of
\begin{equation}
 \E[a^2J_0+2ab\kappa_0 w_0+b^2w_0]
 +\frac{2}{n-1}\sum_{x\in\mathcal X_a}b(x)^2e_{B,0}(x)^2,
 \qquad \E[aJ_0]=1,
 \label{eq:binaryobjective}
\end{equation}
with $b=0$ on the nonatomic part and the atomic sum finite.
At an atom of mass $\pi_{x,0}$, set
$\lambda_{n,x}=\frac{2}{n-1}e_{B,0}(x)^2/\pi_{x,0}$. For fixed $a(x)$, completing the
square in its contribution to \eqref{eq:binaryobjective} gives
\[
 b^*(x)=-\frac{\kappa_0 a(x)w_0(x)}{w_0(x)+\lambda_{n,x}}
       =-\kappa_0\rho_{n,0}(x)a(x).
\]
The minimized integrand is $a(x)^2H_{n,0}(x)$. On the nonatomic part $b=0$
and $H_{n,0}=J_0$. Therefore the remaining problem is
\[
 \min_{\E[aJ_0]=1}\E[a^2H_{n,0}].
\]
Cauchy--Schwarz gives $1\le\E[a^2H_{n,0}]\E[J_0^2/H_{n,0}]$, with equality
precisely at $a=J_0/(H_{n,0}Z_{n,0})$. This proves the formulas and uniqueness.

It remains to verify that this minimizer is feasible on infinitely many
atoms. Overlap and \eqref{eq:fourstates} imply
$0<c_\star\le J_0^2/e_{B,0}\le H_{n,0}\le J_0\le C_\star$, and hence
$c_\star\le Z_{n,0}\le C_\star$ and bounded $a_n^*,b_n^*$. Also
$\rho_{n,0}(x)\le\min\{1,C_\star(n-1)\pi_{x,0}\}$. It follows that
\[
 \sum_x b_n^*(x)^2e_{B,0}(x)^2
 \le C_\star\sum_x\min\{1,(n-1)^2\pi_{x,0}^2\}
 \le C_\star(n-1)\sum_x\pi_{x,0}<\infty.
\]
The last inequality uses $\min\{1,t^2\}\le t$ for $t\ge0$.
Attainability follows from the correctability part of Theorem~\ref{thm:binary}.
The inequalities $J_0^2/e_{B,0}\le H_{n,0}\le J_0$ imply
$\E[J_0]\le Z_{n,0}\le\E[e_{B,0}]$. Finally $\rho_{n,0}$ increases with $n$, so
$H_{n,0}$ decreases and $Z_{n,0}$ increases.
\end{proof}

\begin{proof}[Proof of the limit in Theorem~\ref{thm:binary}]
At every atom $\rho_{n,0}(x)\to1$ and $J_0(x)^2/H_{n,0}(x)\to e_{B,0}(x)$.
Off the atoms $\rho_{n,0}=0$ and $J_0^2/H_{n,0}=J_0$. The bound
$J_0^2/H_{n,0}\le e_{B,0}$, with bounded $e_{B,0}$, gives \eqref{eq:binarylimit} by
dominated convergence. The efficient moment has the representation
\[
 a_{\eff,0}(x)=\frac{e_{B,0}(x)}{J_0(x)\E[e_{B,0}]},\qquad
 b_{\eff,0}(x)=-\kappa_0 a_{\eff,0}(x).
\]
If $\theta_0\ne0$, $|b_{\eff,0}(x)e_{B,0}(x)|$ is bounded away from zero
under overlap. Its squared sum over infinitely many atoms diverges, and
it is nonzero on any nonatomic component. The correctability part of Theorem~\ref{thm:binary}
therefore excludes an exact finite-variance correction of this fixed
efficient moment. When $\kappa_0\ne0$, \eqref{eq:fourstates} implies
$e_{B,0}>J_0$ pointwise, yielding the strict loss on a positive-mass nonatomic
component. When $\kappa_0=0$, $e_{B,0}=J_0$, $b_n^*=0$, and there is no gap.
\end{proof}

\subsection{The DiD class and exact correction}
\label{app:didexact}

\paragraph{Regularity.}
Under Assumption~\ref{ass:did}, parametrize the true law by the
covariate law $H_0$, the odds $a_0$, and the treated and control
conditional outcome laws. Bounded perturbations that preserve the
exclusions stay in the model for small perturbation sizes: a tilt
$1+tz(X)$ of $H_0$ with $\E_{H_0}z=0$; a change $a_0+tu(U)$ of the odds; and
tilts $1+th_t(Y,X)$ and $1+th_c(Y,X)$ of the outcome laws with
$\E[h_t\mid D=1,X]=\E[h_c\mid D=0,X]=0$ and $\E[\varepsilon h_c\mid
D=0,X]$ a function of $V$. They do not change the support of $Y$, and
the bounds on $a_0$ and $\sigma_0^2$ remain strict. Finitely many such
perturbations give likelihood ratios that are bounded, bounded away
from zero, and smooth in the perturbation sizes in $L_\infty$, which
gives conditions (i)--(ii) of Appendix~\ref{app:regularity_paths}; the
moment families below are smooth functions of the perturbed
parameters, which gives (iii). The paths in the proofs below are of
this form.

We first record the efficient coefficients used below.
Let
\[
 \mu_{1,0}(X)=\E[Y\mid D=1,X],\qquad
 \Delta_0(X)=\mu_{1,0}(X)-\mu_0(V),\qquad
 b_0(U)=\E_{Q_0}[\Delta_0\mid U],
\]
and define
\[
 H_{\sigma,0}(V)=\E_{Q_0}[\sigma_0^{-2}(X)\mid V],
 \qquad
 \omega_0(V)=\frac{\E_{Q_0}[a_0(U)\mid V]}{H_{\sigma,0}(V)}.
\]
Here $\sigma_0^{-2}(X)=1/\sigma_0^2(X)$.
The efficient member of \eqref{eq:didclass} has
coefficients
\begin{equation}
 \rho_{E,0}=b_0-\Delta_0,\qquad
 \beta_{E,0}=a_0-\omega_0/\sigma_0^2.
 \label{eq:dideffadj}
\end{equation}
Both satisfy the required conditional-centering
restrictions. Writing $\phi_{E,0}=\phi_0(W;\rho_{E,0},\beta_{E,0})$
and $V_{E,0}=\E[\phi_{E,0}^2]$, the variance decomposition is
\begin{equation}
 \E[\phi_0(W;\rho,\beta)^2]
 =V_{E,0}+\frac{c_0}{\pi_0^2}\E_{Q_0}\left[
 e_0(\rho-\rho_{E,0})^2+
 \sigma_0^2(\beta-\beta_{E,0})^2
 \right].
 \label{eq:didvardec}
\end{equation}
The following proof establishes these formulas
together with the characterization of the full
first-order class.
\begin{proof}[Proof of Proposition~\ref{prop:didclass}]
Let $H_0=\mathcal L(X)$, so $dH_0=c_0(1+a_0)dQ_0$. Parameterize the law by $H_0$, $e_0(U)$, and the two conditional outcome laws. Its tangent space is the orthogonal sum of the following subspaces
of $L_2^0(P_0)$ (the displayed score functions are square-integrable):
\begin{align*}
 \cT_X&=\{h(X):h\in L_2^0(H_0)\},\\
 \cT_a&=\{(D-e_0)f(U):f\in L_2(Q_{U,0})\},\\
 \cT_{\rm t}&=\{Dh_t(Y,X):\E[h_t\mid D=1,X]=0\},\\
 \cT_\mu&=\{(1-D)h_c(Y,X):\E[h_c\mid D=0,X]=0,\
          \E[\varepsilon h_c\mid D=0,X]\in L_2(Q_{V,0})\}.
\end{align*}
For the last space, write a score as $\varepsilon f(V)/\sigma_0^2(X)$ plus a component conditionally orthogonal to $1,\varepsilon$. Bounded likelihood tilts generate these spaces. The variance and overlap bounds ensure closedness and permit approximation by bounded directions, including for finite conditional outcome support.

For a mean-zero function, orthogonality to $\cT_{\rm t}$ forces its treated value to be constant in $Y$ given $X$. Orthogonality to the mean-preserving part of $\cT_\mu$ forces its control value to be affine in $\varepsilon$. Orthogonality to $\cT_X$ makes its conditional mean given $X$ zero. Thus every element of $\cA$ has the form $(D-e_0)\widetilde\rho(X)+(1-D)\varepsilon\widetilde\beta(X)$. Its two components are orthogonal with squared norms $c_0\E_{Q_0}[e_0\widetilde\rho^2]$ and $c_0\E_{Q_0}[\sigma_0^2\widetilde\beta^2]$. Pairing with the assignment and control-mean scores gives $\E_{Q_0}[\widetilde\rho\mid U]=\E_{Q_0}[\widetilde\beta\mid V]=0$. Rescale these coefficients by $\pi_0$ to obtain \eqref{eq:didclass}.

The bias identity \eqref{eq:did-early-bias}, or direct differentiation of the target, verifies that $\phi_{D,0}$ is an influence function: its covariance with every model score equals the derivative of the target along that score. All other influence functions differ by an element of $\cA$. To project $\phi_{D,0}$ onto $\cT$, decompose it as
\[
 \pi_0\phi_{D,0}=D(Y-\mu_{1,0})-(1-D)a_0\varepsilon
       +e_0(\Delta_0-\theta_0)+(D-e_0)(\Delta_0-\theta_0).
\]
The treated residual and covariate term already lie in their respective tangent blocks. The assignment projection replaces $\Delta_0$ by $b_0(U)$. The control projection replaces $a_0\varepsilon$ by $\omega_0(V)\varepsilon/\sigma_0^2(X)$, since its normal equation is $\E_{Q_0}[a_0-\omega_0/\sigma_0^2\mid V]=0$. Hence
\begin{equation}
 \pi_0\phi_{E,0}=D(Y-\mu_{1,0})-(1-D)\omega_0\varepsilon/\sigma_0^2
               +e_0(\Delta_0-\theta_0)+(D-e_0)(b_0-\theta_0).
 \label{eq:dideffexplicit}
\end{equation}
The discarded components give \eqref{eq:dideffadj} and \eqref{eq:didvardec}. The nonzero control component guarantees $V_{E,0}>0$.

For local extension, fix $\rho,\beta$ satisfying the restrictions at $P_0$. At a nearby law $P'$ define
\begin{align}
 C_\rho(P';W)&=\frac{D-e'(U)}{\pi'}\{\rho(X)-\E_{Q'}[\rho\mid U]\},\nonumber\\
 C_\beta(P';W)&=\frac{(1-D)(Y-\mu'(V))}{\pi'}
                         \{\beta(X)-\E_{Q'}[\beta\mid V]\}.
 \label{eq:didextensions}
\end{align}
Writing $\cT(P')$ for the full tangent space at $P'$, these are
valid and belong to its orthogonal complement in $L_2^0(P')$, so differentiation of validity makes their expected derivatives zero in every law direction. Add them to $\psi(W,a',\mu',\theta)/\pi'$. Conditional expectations under $Q'$ are ratios of conditional expectations weighted by $dQ'/dQ_0$. Positive $L_\infty$ likelihood ratios make these maps twice differentiable in $L_2$. This proves admissibility. The efficient family has the same regularity, using the variance lower bound.
\end{proof}

\begin{lemma}[A multiplication obstruction]\label{lem:multiplication}
Let $S$ be a function of $X$ with nonatomic law under $Q_0$, and let $r\in L_2(Q_0)$. Suppose $\mathcal E\subset L_2(Q_{S,0})$ and $\mathcal F\subset L_2(Q_0)$ are defined by finitely many bounded moment restrictions. If the form
\[
 (u,z)\mapsto\E_{Q_0}[r(X)u(S)z(X)],\qquad u\in\mathcal E,\ z\in\mathcal F,
\]
initially on bounded functions, has a Hilbert--Schmidt representation, then $r=0$ almost surely.
\end{lemma}
\begin{proof}
Otherwise $w(S)=\E_{Q_0}[r^2\mid S]$ is between constants $0<\delta<L<\infty$ on a set $B$ of positive probability. On the infinite-dimensional space of functions supported on $B$, the map $M_ru=r(X)u(S)$ is bounded and
$\delta\|u\|_{L_2(Q_{S,0})}^2\le\|M_ru\|_{L_2(Q_0)}^2\le L\|u\|_{L_2(Q_{S,0})}^2$. Domain restriction to $\mathcal E$ removes finitely many dimensions, and projection of the image onto $\mathcal F$ changes the map by finite rank. Restoring these finitely many components to a Hilbert--Schmidt map would make $M_r$ Hilbert--Schmidt. But for an orthonormal sequence supported on $B$, $\sum_j\|M_ru_j\|_{L_2(Q_0)}^2\ge\sum_j\delta=\infty$. Bounded test functions are dense in the restricted spaces: truncate and remove the finitely many moment errors using a fixed bounded collection with full-rank moment matrix. Thus the initial bounded form suffices.
\end{proof}

\begin{proof}[Proof of Theorem~\ref{thm:didexact}]
We give the nuisance paths explicitly; they are bounded tilts of the form described above. Curvature depends only on the moment value by Lemma~\ref{lem:pathwise}, so it can be calculated using \eqref{eq:didextensions}.

\emph{Step 1: the adjustment $\rho$.} Choose bounded $z(X),u(U)$ with
\begin{equation}
 \E_{Q_0}z=\E_{Q_0}[a_0(\Delta_0-\theta_0)z]=0,\qquad
 \E_{Q_0}u=\E_{Q_0}[u\Delta_0]=0.
 \label{eq:diddirections}
\end{equation}
Set $dQ_v=(1+vz)dQ_0$, $a_t=a_0+tu$, retain the treated conditional outcome law, and put
\[
 j(t,v)=\frac{\E_{Q_v}[a_t\Delta_0]}{\E_{Q_v}a_t}-\theta_0.
\]
Set the control mean to $\mu_0+j(t,v)$ using its conditional tilt, and its population fraction to $1/\{1+\E_{Q_v}a_t\}$. The target stays $\theta_0$ exactly. At zero, $j=j_t=j_v=0$. The scores are
\[
 S_au=D u(U)/a_0(U),\qquad S_Qz=z(X)-\E_{H_0}z(X),\qquad
 \E_{H_0}z=c_0\E_{Q_0}[a_0z].
\]
These maps are bounded in the relevant $L_2$ norms.

The expectation of the usual normalized DiD family has mixed derivative zero: its bias is $c_0tj(t,v)\E_{Q_0}u/\pi_{t,v}=0$. The $C_\beta$ term also has zero mixed derivative because $j_t=j_v=0$ and its centered coefficient has mean zero at the origin. For the $C_\rho$ term, let $r_v(U)=\E_{Q_v}[\rho\mid U]$. Conditional centering at the true law gives
\[
 \E C_\rho(P_{t,v};W)=\pi_{t,v}^{-1}\E_{H_0}[(e_t-e_0)r_v].
\]
At zero $r_0=0$, $\partial_vr_v=\E_{Q_0}[\rho z\mid U]$, and $\partial_te_t=u/(1+a_0)^2$. Its mixed derivative is therefore
\[
 \frac{c_0}{\pi_0}\E_{Q_0}\left[\frac{u\rho z}{1+a_0}\right].
\]
Compatibility requires the kernel pairing with $S_au,S_Qz$ to equal half this quantity. An $L_2(P_0^2)$ kernel defines a Hilbert--Schmidt operator; composition with the bounded score maps preserves this property. The finitely many restrictions in \eqref{eq:diddirections} do not remove the multiplication obstruction. Lemma~\ref{lem:multiplication} with $S=U$ and $r=\rho/(1+a_0)$ gives $\rho=0$.

\emph{Step 2: the adjustment $\beta$.} Keep $a_0$ fixed, use the same $Q_v$, and choose bounded $f(V)$ with $\E_{Q_0}[a_0f]=0$. Set
\[
 j(t,v)=\frac{\E_{Q_v}[a_0(\Delta_0-tf)]}{\E_{Q_v}a_0}-\theta_0,
 \qquad \mu_{t,v}=\mu_0+tf+j(t,v).
\]
Again $j=j_t=j_v=0$ and the target is fixed. The outcome score is $S_mf=(1-D)\varepsilon f(V)/\sigma_0^2(X)$, a bounded $L_2$ map. The usual moment has zero bias because the odds are correct. For $\beta_v=\beta-\E_{Q_v}[\beta\mid V]$,
\[
 \E C_\beta(P_{t,v};W)=-\frac{c_0}{\pi_{t,v}}\E_{Q_0}[(tf+j)\beta_v],
 \qquad \partial_v\beta_v\big|_0=-\E_{Q_0}[\beta z\mid V].
\]
Its mixed derivative is $c_0\E_{Q_0}[f\beta z]/\pi_0$. The same argument with $S=V$ gives $\beta=0$.

\emph{Step 3: necessity of the density criterion.} Only $\phi_{D,0}$ remains. Choose bounded $u(U),f(V)$ with $\E_{Q_0}u=\E_{Q_0}[u\Delta_0]=\E_{Q_0}[a_0f]=0$. Hold $Q_0$ fixed, set $a_t=a_0+tu$, and change the control mean to
\[
 \mu_{t,r}=\mu_0+rf-tr\E_{Q_0}[uf]/A_0.
\]
The target is exactly constant and the first scores are $S_au,S_mf$. The mixed derivative of the normalized first-order moment is $\E_{Q_0}[uf]/A_0$, so a correction must satisfy
\begin{equation}
 \E[k(W_i,W_j)(S_au)(W_i)(S_mf)(W_j)]
       =\frac1{2A_0}\E_{Q_0}[u(U)f(V)].
 \label{eq:didTnecessary}
\end{equation}
The left side induces a Hilbert--Schmidt map on the restricted spaces. The right side is $T$ restricted and projected onto spaces of finite codimension. Since $T$ is bounded, restoring the missing finite-dimensional components changes it by a finite-rank bounded operator. Thus $T$ itself is Hilbert--Schmidt.

If $(f_j)$ is an orthonormal basis of $L_2(Q_{V,0})$, the Hilbert--Schmidt property makes $R_0(u,v)=\sum_j(Tf_j)(u)f_j(v)$ converge in $L_2(Q_{U,0}\otimes Q_{V,0})$. Testing this representation on indicators gives
\[
 Q_0(U\in A_1,V\in A_2)=\int_{A_1\times A_2}R_0\,dQ_{U,0}\,dQ_{V,0}.
\]
Uniqueness of measures on rectangles identifies $R_0$ with the nonnegative density in \eqref{eq:didexist}. Conversely this density represents $T$ and its squared $L_2$ norm is the Hilbert--Schmidt norm. Standard Borel spaces ensure the needed separability.

\emph{Step 4: sufficiency, including variation of $Q_0$.} The residuals in \eqref{eq:didK0} have zero conditional means given $U$ and $V$, respectively. Therefore $k_0$ is degenerate. Bounded conditional second moments give $\|k_0\|^2\le C_\star\|R_0\|^2<\infty$. At a nearby law $P'$, use its own residuals, $\pi',c'$, and density ratio $R'$. Along the allowed paths,
\[
 R'=R_0\frac{dQ'_{UV}/dQ_{UV,0}}{(dQ'_U/dQ_{U,0})(dQ'_V/dQ_{V,0})}.
\]
This is twice differentiable in $L_2$, with $\|R'-R_0\|_{L_2(Q_{U,0}\otimes Q_{V,0})}$ of first order in path displacement. Write $T(R')$ for integration against the \emph{true} $V$ marginal. Independence of the observations and conditioning yield, at the fixed true target,
\begin{equation}
 \E\{\psi(W,a',\mu',\theta_0)/\pi'\}-\E k(\cdot,\cdot;P')
 =\frac{c_0}{\pi'}\ip{a'-a_0}{\{T-(c_0/c')T(R')\}(\mu'-\mu_0)}_{L_2(Q_{U,0})}.
 \label{eq:didcubic}
\end{equation}
The operator in braces is zero at the truth and changes at first order. The expression is of order at least three along every allowed finite-dimensional nuisance path. This proves joint second-order orthogonality, including the covariate law.

\emph{Step 5: minimum norm and extension.} Recall that $\mathcal Q$ is the closed span of symmetric products of nuisance scores. Every two kernels correcting $\phi_{D,0}$ have the same inner products with these products, by Lemma~\ref{lem:pathwise}. Projection of $k_0$ onto $\mathcal Q$ is therefore the unique minimum-norm correction. The first two steps fixed the first-order value at $\phi_{D,0}$, proving \eqref{eq:didopt}. Double-center the fixed $k_{\min,0}$ under each nearby law as in Proposition~\ref{prop:extension}, and use the ordinary normalized DiD first-order family. Lemma~\ref{lem:pathwise} proves second-order cancellation at the truth. This establishes local oracle attainment without assuming that the quantities defining the optimum are known.
\end{proof}
The projection in Theorem~\ref{thm:didexact} also has a computable population formula. Put
\[
\begin{gathered}
 B_E=(1-D)\varepsilon/\{\sigma_0^2(X)H_{\sigma,0}(V)\},\\
 k_{T,0}(W_i,W_j)=\frac{A_D(W_i)R_0(U_i,V_j)B_E(W_j)+A_D(W_j)R_0(U_j,V_i)B_E(W_i)}{2\pi_0 c_0}.
\end{gathered}
\]
and define $g_{k,0}(w)=\E[k_{T,0}(w,W)\phi_{E,0}(W)]$, $\gamma_{k,0}=\E[g_{k,0}\phi_{E,0}]$. Then
\begin{align}
 k_{\min,0}(w_i,w_j)
 &=k_{T,0}(w_i,w_j)-\frac{\phi_{E,0}(w_i)g_{k,0}(w_j)+g_{k,0}(w_i)\phi_{E,0}(w_j)}{V_{E,0}}
     +\frac{\gamma_{k,0}\phi_{E,0}(w_i)\phi_{E,0}(w_j)}{V_{E,0}^2},\label{eq:didminexplicit}\\
 \|k_{\min,0}\|^2&=\|k_{T,0}\|^2-2\|g_{k,0}\|^2/V_{E,0}+\gamma_{k,0}^2/V_{E,0}^2.\label{eq:didminnorm}
\end{align}
To verify this, $A_Df(U)$ belongs to the assignment tangent space, while the projection of $B_Dg(V)$ onto $\cT$ is $B_Eg(V)$. The difference is orthogonal to all four tangent blocks; for the control-mean block its pairing reduces to $\E_{Q_0}[gf\{1-(\sigma_0^2H_{\sigma,0})^{-1}\}]=0$. Expanding $R_0$ into products in $L_2$ gives $k_{T,0}=(\Pi_{\cT}\otimes\Pi_{\cT})k_0$, where the two projections
act on the two kernel arguments. Within $\cT$, nuisance projection subtracts $\phi_{E,0}\E[\phi_{E,0}\cdot]/V_{E,0}$. Apply it to both arguments of $k_{T,0}$ to obtain \eqref{eq:didminexplicit}; its inner product with $k_{T,0}$ gives \eqref{eq:didminnorm}. Target-preserving scores span $\cT\cap\phi_{E,0}^\perp$: a constant control-mean change removes target variation in any law path, with zero first derivative when the original score has zero pairing with $\phi_{E,0}$.

\begin{proof}[Proof of Corollary~\ref{cor:diddiscrete}]
With finitely many covariate values, $L_2(Q_0)$ is finite dimensional, and
every first-order value $\phi_0(W;\rho,\beta)$ in \eqref{eq:didclass} is
affine in $Y$ given $(D,X)$. Take paths linear in each conditional
outcome law and in the covariate cell probabilities, and hold the
target fixed by the constant shift of the control mean used in the
proof of Theorem~\ref{thm:didexact}. The second likelihood derivative
along such paths is a sum of products of first derivatives of
different components and of terms that depend on $(D,X)$ only through
finitely many cell parameters. Because $\phi$ is affine in $Y$ given
$(D,X)$, $\E[\phi u_{h,g}]$ depends on outcome-law directions only
through $\E[(Y-\mu_{1,0})h_t\mid D=1,X=x]$ and $\E[\varepsilon h_c\mid
D=0,X=x]$ at the finitely many support points $x$, and on the other
directions through finitely many cell parameters. Each of these is a
bounded linear functional of the score. Hence
$B_\phi(h,g)=\sum_{k,l\le m}c_{kl}\langle f_k,s_h\rangle\langle
f_l,s_g\rangle$ for finitely many $f_k\in L_2^0(P_0)$ and a symmetric
matrix $(c_{kl})$, and the kernel
$\sum_{k,l}c_{kl}\{f_k(w_i)f_l(w_j)+f_l(w_i)f_k(w_j)\}/2$ is symmetric,
degenerate, and satisfies $\mathcal Rk=B_\phi$. Thus $c_\phi<\infty$
for every first-order value. In particular $\phi_{E,0}\in\cD$, so
Theorem~\ref{thm:limit} gives $I_{2,n}\uparrow\E[\ell_{\eff}^2]=1/V_{E,0}$,
that is, $nV_{2,n}^*\downarrow V_{E,0}$.
\end{proof}

\subsection{Shared directions}
\label{app:shared}
\begin{proof}[Proof of Theorem~\ref{thm:shared}]
The following likelihood-ratio submodel makes the common curvature explicit:
\begin{equation}
 \frac{dP_{t_\theta,\mathbf t,r}}{dP_0}
 =D\left(1+\frac{\mathbf t'h(C)}{a_0(U)}\right)
 +(1-D)\left[1+\left\{-t_\theta+r'h(C)-\frac{\mathbf t'r}{A_0}\right\}
                                      \frac{\varepsilon}{\sigma_0^2(X)}\right].
 \label{eq:sharedpath}
\end{equation}
Boundedness makes it positive locally. Its expectation is one because $\E_{Q_0}h=0$ and the control perturbation has conditional mean zero. It retains $Q_0$, the treatment fraction, and the treated conditional outcome law. The odds become $a_0+\mathbf t'h$ and the control mean becomes $\mu_0-t_\theta+r'h-\mathbf t'r/A_0$. The restrictions on $h$ give
\[
 \E_{Q_0}[(a_0+\mathbf t'h)(\Delta_0+t_\theta-r'h+\mathbf t'r/A_0)]=A_0\theta_0+A_0t_\theta-\mathbf t'r+\mathbf t'r,
\]
so the target is $\theta_0+t_\theta$. The target score is $\ell_C=-(1-D)\varepsilon/\sigma_0^2$, the nuisance scores are $s_a,s_\mu$, and the mixed likelihood derivative is $u_{t_j,r_l}=\ind\{j=l\}\ell_C/A_0$. Pure second derivatives are zero.

Every normalized first-order value satisfies $\E[\phi\ell_C]=1$, since a score changing the target at unit rate has this covariance with its influence function. Compatibility thus requires
\[
 M_k=\E[k(W_i,W_j)s_a(W_i)s_\mu(W_j)']=I_d/(2A_0).
\]
Whiten the two score vectors to $\bar s_a=G_{a,0}^{-1/2}s_a$, $\bar s_\mu=G_{\mu,0}^{-1/2}s_\mu$. They are centered, orthonormal within blocks, and mutually orthogonal because $D(1-D)=0$. The $d^2$ symmetric kernels
\[
 Q_{ql}(w_i,w_j)=\{\bar s_{a,q}(w_i)\bar s_{\mu,l}(w_j)
                      +\bar s_{a,q}(w_j)\bar s_{\mu,l}(w_i)\}/\sqrt2
\]
are orthonormal. Their pairings with $k$ are $\sqrt2(G_{a,0}^{-1/2}M_kG_{\mu,0}^{-1/2})_{ql}$. Bessel's inequality yields
\[
 \|k\|^2\ge\frac{\|G_{a,0}^{-1/2}G_{\mu,0}^{-1/2}\|_F^2}{2A_0^2}
           =\frac{\tr(G_{a,0}^{-1}G_{\mu,0}^{-1})}{2A_0^2}.
\]
Projection onto these kernels gives
\[
 k_d(w_i,w_j)=\frac{
 s_a(w_i)'G_{a,0}^{-1}G_{\mu,0}^{-1}s_\mu(w_j)
 +s_a(w_j)'G_{a,0}^{-1}G_{\mu,0}^{-1}s_\mu(w_i)}{2A_0}.
\]
It has zero pure-block pairings. Use $\phi_{E,0}$, its admissible family, and a double-centered extension of $k_d$ to attain the finite bound. Every normalized first-order variance is at least $V_{E,0}$. Finally, $G_{a,0}\preceq(c_0/a_-)I_d$, $G_{\mu,0}\preceq(c_0/v_-)I_d$, and
$A_0^2c_0^2=\pi_0^2$, giving the primitive lower bound. Here $A\preceq B$
means that $B-A$ is positive semidefinite.

If $L_2(Q_{C,0})$ is infinite dimensional, imposing orthogonality to $1,\E_{Q_0}[a_0\mid C],\E_{Q_0}[\Delta_0\mid C]$ leaves an infinite-dimensional space. Bounded functions are dense after finite bounded adjustments. Gram--Schmidt supplies arbitrarily long bounded orthonormal systems satisfying the required restrictions. Their kernel-norm lower bounds diverge, excluding one finite-norm kernel that corrects them all.
\end{proof}

\subsection{Feasible DiD corrections}
\label{app:didfeasible}

\begin{proof}[Proof of Proposition~\ref{prop:threebias}]
The doubly robust component $\psi(W,\widehat a,\widehat\mu,\theta_0)$ has expectation
$c_0\E_{Q_0}[\delta_a\delta_\mu]$ by conditioning on $X$. Since the observed
covariate law is $c_0(1+a_0)dQ_0$ and $\E_{Q_0}[f-\widehat r\mid U]=-\delta_r$,
\[
 \E[(D-\widehat e)(f-\widehat r)]
 =c_0\E_{Q_0}\Bigl[\frac{a_0-\widehat a}{1+\widehat a}(f-\widehat r)\Bigr]
 =c_0\E_{Q_0}\Bigl[\frac{\delta_a}{1+\widehat a}\delta_r\Bigr].
\]
Likewise
$\E[(1-D)(Y-\widehat\mu)(v-\widehat t)]=-c_0\E_{Q_0}[\delta_\mu(v-\widehat
t)]=c_0\E_{Q_0}[\delta_\mu\delta_t]$.
\end{proof}

\paragraph{Estimator.}
We use independent training and $n$ evaluation observations, with the
fitted functions of Section~\ref{sec:didapprox}. Write $X=(C,X_V,X_U)$,
$U=(C,X_U)$, and $V=(C,X_V)$, of dimensions $d_C$, $d_U\ge1$, and $d_V\ge1$, and suppose that, under $Q_0$, $X$ has a
density $q_0$ on a unit cube, bounded above and away from zero; $C$ may
be empty. For evaluation observation $i$, put
$A_i=D_i-(1-D_i)\widehat a(U_i)$, $B_i=(1-D_i)\{Y_i-\widehat\mu(V_i)\}$,
$S_i^f=(1-D_i)\{f(X_i)-\widehat r(U_i)\}$, and
$S_i^v=(1-D_i)\{v(X_i)-\widehat t(V_i)\}$, and let $\widehat c$
estimate $c_0$ on the training sample. Partition the supports of $U$ and
$V$ into $K_U$ and $K_V$ regular cells, with cell-projection kernels
$L_U(u,u')=\sum_{b}\ind\{u,u'\in b\}/Q_{U,0}(b)$ and $L_V$ defined
likewise. Partition the support of $C$ into $J$ cells $b$, let
$P_C^Uh=\E_{Q_0}[h(U)\mid X_U,b(C)]$ and $P_C^Vl=\E_{Q_0}[l(V)\mid X_V,b(C)]$,
and approximate $T$ by $T_J=P_C^UT+TP_C^V-P_C^UTP_C^V$, so that
$T-T_J=(I-P_C^U)T(I-P_C^V)$. With $q_{U,0}$, $q_{V,0}$ the marginal densities
and $G_0$, $G_{U,0}$, $G_{V,0}$ their integrals over a common cell, the kernel of
$T_J$ relative to $Q_{U,0}\otimes Q_{V,0}$ at $u=(c_1,z)$, $v=(c_2,x)$ is
\begin{equation}
 R_J(u,v)=\sum_b\ind\{c_1,c_2\in b\}\Bigl\{
 \frac{q_0(c_2,x,z)}{G_{U,0}(b,z)q_{V,0}(c_2,x)}
 +\frac{q_0(c_1,x,z)}{q_{U,0}(c_1,z)G_{V,0}(b,x)}
 -\frac{G_0(b,x,z)}{G_{U,0}(b,z)G_{V,0}(b,x)}\Bigr\},
 \label{eq:RJ}
\end{equation}
as one verifies by integrating a test function of $(c_2,x)$: the three
terms give $P_C^UTf$, $TP_C^Vf$, and $P_C^UTP_C^Vf$. When $C$ is empty,
$R_J(u,v)=q_0(x,z)/\{q_{U,0}(z)q_{V,0}(x)\}$. Fitted kernels
$\widehat R_J,\widehat L_U,\widehat L_V$ replace $q_0$ by a positive
bounded fitted density $\widehat q$. With
$\sym L_{ij}=(L_{ij}+L_{ji})/2$, define
\begin{gather*}
 \widehat k_{a\mu,ij}=\widehat c^{-1}\sym\{A_i\widehat R_J(U_i,V_j)B_j\},
 \qquad
 \widehat k_{\mu t,ij}=\widehat c^{-1}\sym\{B_i\widehat L_V(V_i,V_j)S_j^v\},\\
 \widehat k_{ar,ij}=\widehat c^{-1}\sym\Bigl\{\frac{A_i\widehat L_U(U_i,U_j)
 S_j^f}{1+\widehat a(U_i)}\Bigr\},
\end{gather*}
where $\widehat k_{a\mu,ij}=\widehat k_{a\mu}(W_i,W_j)$,
and likewise for the other two corrections. On $\{\Pn D>0\}$, let $\widehat\theta$ solve
$\Pn\widehat\psi(\cdot,\widehat\theta)-\Un(\widehat k_{a\mu}+\widehat k_{ar}+\widehat k_{\mu t})=0$,
with $\widehat\psi(W,\theta)$ in \eqref{eq:didfitmoment}.
Set $\widehat\theta=0$ when $\Pn D=0$, before applying
the clipping below. All three kernels are
degenerate at the true nuisance values.

Let $r_k=\|\delta_k\|_{L_2(Q_0)}$ for $k\in\{a,\mu,r,t\}$,
$e_q=\|\widehat q-q_0\|_\infty$, $e_c=|\widehat c-c_0|$, and
$w_a=\delta_a/(1+\widehat a)$.

\begin{assumption}[Approximation and estimated operators]
\label{ass:didrates}
The fitted functions and $f,v$ are uniformly bounded, the fitted
density and odds are bounded above and away from zero, and
$\widehat c$ is bounded away from zero and one. With probability
approaching one, the following H\"older seminorms are bounded:
$\delta_a,\delta_\mu$ in $C$, of orders $s_{aC},s_{\mu C}\in(0,1]$,
uniformly in the other coordinates; $w_a,\delta_r$ on $U$, of orders
$s_a,s_r\in(0,1]$; and $\delta_\mu,\delta_t$ on $V$, of orders
$s_\mu,s_t\in(0,1]$. The quantities $r_a,r_\mu,r_r,r_t,e_q,e_c$
converge to zero in probability. Put $D_n=J+K_U+K_V$ and
\begin{equation}
 B_n=J^{-(s_{aC}+s_{\mu C})/d_C}+K_U^{-(s_a+s_r)/d_U}
 +K_V^{-(s_\mu+s_t)/d_V}
 +(e_q+e_c)(r_ar_\mu+r_ar_r+r_\mu r_t),
 \label{eq:didbiasrate}
\end{equation}
omitting the first term when $C$ is empty.
\end{assumption}

\begin{theorem}[Feasible DiD correction and recovery of efficiency]
\label{thm:didfeasible}
Suppose Assumptions~\ref{ass:did} and \ref{ass:didrates} hold, and clip
$\widehat\theta$ to a fixed bounded interval containing $\theta_0$ in
its interior. On training events where the stated bounds hold,
\begin{equation}
 \E[(\widehat\theta-\theta_0)^2\mid\mathrm{training}]
 \le C_\star\{n^{-1}+D_n/n^2+B_n^2\}+C_\star e^{-c_\star n}.
 \label{eq:didrisk}
\end{equation}
If, in addition, $D_n=o(n)$, $\sqrt n\,B_n=o_p(1)$, and $f\to f_0$,
$v\to v_0$ in $L_2(Q_0)$ for fixed bounded $f_0,v_0$, then
$\widehat\theta-\theta_0=\Pn\phi_0(\cdot;\rho_0,\beta_0)+o_p(n^{-1/2})$, with
$\rho_0=f_0-\E_{Q_0}[f_0\mid U]$ and $\beta_0=v_0-\E_{Q_0}[v_0\mid V]$. If
$f_0=-\Delta_0$ and $v_0=a_0-\omega_0/\sigma_0^2$, the influence function is
$\phi_{E,0}$, the asymptotic variance is $V_{E,0}$, and the sample variance of
the fitted unnormalized moment values, divided by $(\Pn D)^2$, is consistent.
The asymptotic linear expansion and consistency of the
variance estimator also hold for cross-fitted averages
over a fixed number of folds, under the corresponding
conditions for each fold. The conditional risk bound
\eqref{eq:didrisk} concerns the independent-training
construction. 
\end{theorem}

\begin{proof}
\emph{Residual identity.} Conditioning gives, for bounded $h(U)$ and
$l(V)$, $\E[Ah]=-c_0\E_{Q_0}[\delta_ah]$, $\E[Bl]=-c_0\E_{Q_0}[\delta_\mu l]$,
$\E[S^fh]=-c_0\E_{Q_0}[\delta_rh]$, and $\E[S^vl]=-c_0\E_{Q_0}[\delta_tl]$. Let
$\widehat T_J,\widehat P_U,\widehat P_V$ integrate the fitted kernels
against the true marginals. Independence within a pair and
\eqref{eq:threebias} give
\begin{equation}
\begin{aligned}
 \E\widehat\psi(W,\theta_0)-\E\widehat k
 =c_0\big[&\ip{\delta_a}{\{T-(c_0/\widehat c)\widehat T_J\}\delta_\mu}_{Q_{U,0}}
 +\ip{w_a}{\{I-(c_0/\widehat c)\widehat P_U\}\delta_r}_{Q_{U,0}}\\
 &+\ip{\delta_\mu}{\{I-(c_0/\widehat c)\widehat P_V\}\delta_t}_{Q_{V,0}}\big],
\end{aligned}
 \label{eq:didresidual}
\end{equation}
where $\widehat k=\widehat k_{a\mu}+\widehat k_{ar}+\widehat k_{\mu t}$.

\emph{Bias.} Since $P_U$ is an orthogonal projection,
$\ip{w_a}{(I-P_U)\delta_r}=\ip{(I-P_U)w_a}{(I-P_U)\delta_r}$, and a
H\"older function of order $s$ has cell-projection error of order
$K^{-s/d}$ on a regular $d$-dimensional partition; this bounds the $U$
and $V$ products. For the common coordinate,
$|\ip{\delta_a}{(T-T_J)\delta_\mu}|=|\ip{(I-P_C^U)\delta_a}
{T(I-P_C^V)\delta_\mu}|\le C_\star J^{-(s_{aC}+s_{\mu C})/d_C}$ by
contraction of $T$. Positivity gives $Q_{U,0}(b)\asymp K_U^{-1}$ and
$|\widehat Q_U(b)-Q_{U,0}(b)|\le e_q|b|$, so $\|\widehat P_U-P_U\|_{\op}\le
Ce_q$ and likewise for $\widehat P_V$; each term of \eqref{eq:RJ} is
of order $J$ on pairs in the same common cell and changes by at most
$C_\star e_qJ$, and the Schur bound gives $\|\widehat
T_J-T_J\|_{\op}\le Ce_q$, uniformly in the resolutions. Including
$c_0/\widehat c$ adds $C_\star e_c$. Cauchy--Schwarz in
\eqref{eq:didresidual} gives $|\E\widehat\psi(W,\theta_0)-\E\widehat k|\le
C_\star B_n$.

\emph{Variance.} Bounded residuals and positivity give
$\|\widehat k_{a\mu}\|^2\le C_\star J$, $\|\widehat k_{ar}\|^2\le C_\star K_U$,
and $\|\widehat k_{\mu t}\|^2\le C_\star K_V$: each cell kernel is of size
$K$ on a set of probability of order $K^{-1}$. With $k^{[1]}$ and $k^{[2]}$ the
first projection and degenerate part of $\widehat k$, the signed-mean
identities and bounded operator norms give $\|k^{[1]}\|\le
C_\star(r_a+r_\mu+r_r+r_t)$ and $\|k^{[2]}\|^2\le C_\star D_n$, so the
estimating equation has conditional variance at most
$C_\star/n+C_\star D_n/n^2$. Its derivative is $-\Pn D$, and
$\{\Pn D<\pi_0/2\}$ has exponentially small probability, which gives
\eqref{eq:didrisk} after clipping.

\emph{Linear expansion and efficiency.} The fitted first-order moment
at $\theta_0$ converges in $L_2(P_0)$ to $\pi_0\phi_0(W;\rho_0,\beta_0)$, so
its centered empirical replacement error is $o_p(n^{-1/2})$; the
kernel projection is negligible by the bound on $\|k^{[1]}\|$, the
degenerate term because $D_n=o(n)$, and the bias because $\sqrt
nB_n=o_p(1)$. Division by $\Pn D\to_p\pi_0$ gives the expansion, and
finitely many folds add finitely many negligible remainders. With
$f_0=-\Delta_0$ and $v_0=a_0-\omega_0/\sigma_0^2$, $\rho_0=b_0-\Delta_0$ and
$\beta_0=\beta_{E,0}$, so Proposition~\ref{prop:didclass} identifies
$\phi_{E,0}$; consistency of the variance estimator follows from
boundedness and $L_2$ consistency.
\end{proof}

\section{Simulation details}
\label{app:plm-sim}

\paragraph{Designs.}
The designs are those of Section~\ref{sec:numerical}. In the
low-dimensional designs, \eqref{eq:plm-sim-errors} gives
$\E[\varepsilon\mid D,X]=0$ and
$\Var(\varepsilon\mid D,X)=\frac12\sigma_{\varepsilon,0}(X)^2[1+\{D-\gamma_0(X)\}^2/
\sigma_{V,0}(X)^2]$. The number of terms in $h_s$ is fixed at eleven across
sample sizes, so the functions are smooth, and $s$ controls the decay
of their coefficients and hence the difficulty of approximating them
with the dictionaries.

\paragraph{First-step estimation and cross-fitting.}
In the low-dimensional designs, one multi-output random forest
estimates $(\mu_0,\gamma_0)$ from $(Y,D)$, with 100 trees, minimum leaf
size five, and all covariates as split candidates, so the two
regressions share the tree partitions. In C1 and C2, separate
heteroskedastic plug-in Lasso regressions for $Y$ and $D$ use an
intercept and training-standardized covariates; with $M$ training
observations and $p$ covariates, the penalty is
$1.1M^{-1/2}\Phi^{-1}\{1-a/(2p)\}$ with $a=0.1/\log\{\max(p,M)\}$, the
loadings are $\{|I_b^c|^{-1}\sum_{i\in I_b^c}Z_{ij}^2r_i^2\}^{1/2}$ with $Z_{ij}$ the standardized regressor and $r_i$ the
current residual, and two loading updates follow an initial fit. The
pair blocks of Section~\ref{sec:numerical} are drawn after the data;
within-group blocks are trained on $4n/5$ observations and
between-group blocks on $3n/5$.

\paragraph{Corrections and selection.}
With $c_j(x)=\sqrt2\cos(\pi jx)$, the ridge dictionary in one
dimension is $(1,c_1/2,\ldots,c_{d-1}/d)'$ with $d<n/4$; Two uses the
tensor products $c_j(x_1)c_l(x_2)/\{(1+j)(1+l)\}$ ordered by
$((1+j)(1+l),j+l,j)$; C1 and C2 use $(1,X')'$. Ridge retains the
eigenvalues of $\widehat G_b$ above $10^{-10}$ times the largest. In C1
and C2, each penalty solves
$\sum_a\widehat\nu_a/(\widehat\nu_a+\zeta_\tau)=d_\tau$ for the eigenvalues
$\widehat\nu_a$ of the first block's training Gram matrix, and the
penalties are reused across blocks. Selected projection orders the
nonconstant features by the training LAR paths for $Y$ and $D$, after
standardization: a feature's score is the larger of its normalized
entry steps on the two paths, features entering neither path come
last, and ties are broken by the sum of absolute marginal correlations
with $Y$ and $D$. The constant enters first, and the selected Gram
matrix is stabilized by $10^{-10}I$. Rule C checks eligibility over the
entire grid. ADA uses the block residuals and, for
$\widetilde\theta$, the first-order DML estimate; a component whose
maximum $q_a$ is nonpositive is omitted, and ties favor selected
projection.

\paragraph{Methods.}
HOIF retains the constant and the first $d_{\mathrm H}-1$ nonconstant dictionary
functions, with the same stabilization. TL uses five folds,
training-standardized features, the first $\min\{250,\lfloor
M/2\rfloor\}$ unscaled cosine or tensor functions in the
low-dimensional designs, with $M=4n/5$, and the covariates in C1 and
C2; its regression penalties use the plug-in loadings above, its
nodewise regressions use plug-in penalties, and TL-CV selects the two
regression penalties by three-fold cross-validation over twenty values.
Its standard error is the usual one computed from its own cross-fitted
Lasso residuals.

\paragraph{Reporting.}
Replication $r$ uses data seed $1000n+r$, and all methods use the same
data within a replication. Bias is the mean estimation error, SD the
standard deviation, RMSE the root mean squared error, and interval
length and standard errors are medians across replications. Coverage
uses the standard normal critical value, with Monte Carlo standard
error $\{\widehat p(1-\widehat p)/500\}^{1/2}$.
Table~\ref{tab:plm-extra} adds designs B and C2 to the simulation results together with the methods eHOIF and TL. Table \ref{tab:plm-bias} reports the bias,
standard deviation, and interval length of every method.

\begin{table}[htbp]
\centering
\footnotesize
\setlength{\tabcolsep}{4pt}
\caption{Coverage of nominal 95\% intervals (RMSE$\times100$), all designs and methods}
\label{tab:plm-extra}
\begin{tabular}{llcccccccc}
\toprule
Design & $n$ & DML & eHOIF & sHOIF & TL & TL-CV & Ridge & SP & ADA\\
\midrule
A      & 500  & .71 (12.2) & .52 (608) & .80 (10.5) & .01 (176) & .55 (250) & .96 (7.9) & .90 (9.6) & .96 (8.0)\\
       & 1000 & .74 (7.9) & .60 (7810) & .79 (7.4) & .01 (21.5) & .82 (7.3) & .95 (5.7) & .93 (6.1) & .95 (5.7)\\
B      & 500  & .73 (12.9) & .52 (421) & .81 (11.3) & .02 (180) & .50 (275) & .96 (8.8) & .90 (10.4) & .95 (8.8)\\
       & 1000 & .76 (8.6) & .63 (8199) & .81 (8.0) & .03 (22.1) & .80 (8.3) & .95 (6.3) & .93 (6.6) & .95 (6.3)\\
Local  & 500  & .57 (17.2) & .47 (801) & .76 (13.2) & .00 (43.9) & .51 (69.4) & .96 (9.4) & .82 (14.0) & .96 (9.5)\\
       & 1000 & .65 (10.5) & .54 (758) & .75 (9.0) & .00 (32.5) & .72 (34.0) & .97 (6.3) & .88 (7.7) & .97 (6.3)\\
Sparse & 500  & .72 (12.6) & .55 (176) & .73 (12.3) & .74 (11.8) & .88 (10.0) & .95 (8.9) & .95 (9.3) & .95 (9.2)\\
       & 1000 & .80 (8.2) & .64 (1495) & .80 (8.2) & .89 (6.4) & .92 (6.2) & .94 (6.2) & .94 (6.3) & .94 (6.2)\\
Two    & 500  & .50 (17.4) & .66 (141) & .56 (16.0) & .04 (30.9) & .61 (24.8) & .96 (10.0) & .77 (13.4) & .96 (10.1)\\
       & 1000 & .48 (12.7) & .65 (25.3) & .52 (11.9) & .00 (26.0) & .67 (10.5) & .94 (7.4) & .80 (9.0) & .94 (7.4)\\
C1     & 500  & .32 (11.1) & .81 (6.7) & .94 (4.4) & .79 (6.2) & .92 (5.1) & .45 (10.3) & .96 (5.8) & .96 (5.9)\\
       & 1000 & .53 (6.5) & .90 (3.5) & .93 (3.2) & .90 (3.7) & .93 (3.5) & .65 (5.9) & .95 (3.7) & .95 (3.8)\\
C2     & 500  & .46 (13.0) & .76 (9.0) & .91 (6.5) & .80 (7.7) & .90 (7.1) & .56 (12.2) & .95 (8.0) & .95 (8.0)\\
       & 1000 & .64 (7.8) & .89 (4.8) & .93 (4.5) & .91 (4.8) & .94 (4.6) & .74 (7.2) & .95 (5.0) & .95 (5.0)\\
Rough  & 500  & .17 (24.1) & .23 (1215) & .25 (22.0) & .00 (54.3) & .33 (563) & .73 (14.9) & .65 (16.5) & .74 (14.7)\\
       & 1000 & .22 (16.2) & .24 (502) & .27 (15.2) & .00 (33.6) & .50 (13.4) & .83 (9.1) & .74 (10.4) & .84 (9.0)\\
\bottomrule
\end{tabular}

\begin{minipage}{.97\textwidth}
\footnotesize
\medskip
Ridge, SP, and ADA use the pair blocks with shared fits; DML is five-fold DML; eHOIF and
sHOIF use $d_{\mathrm H}=24$ and $32$; TL and TL-CV use their own five-fold scheme. 500
replications per cell.
\end{minipage}
\end{table}

\begin{table}[htbp]
\centering
\footnotesize
\caption{Bias, standard deviation, and median interval length ($\times100$)}
\label{tab:plm-bias}
\begin{tabular}{lllrrrrrrrr}
\toprule
Design & $n$ & & DML & eHOIF & sHOIF & TL & TL-CV & Ridge & SP & ADA\\
\midrule
A      & 500  & Bias & 9.5 & 15.7 & 7.4 & 45.5 & $-$8.1 & 1.0 & 4.8 & 1.0\\
       &      & SD & 7.6 & 608 & 7.5 & 170 & 250 & 7.9 & 8.3 & 7.9\\
       &      & Length & 28.1 & 63.8 & 28.1 & 25.5 & 33.9 & 34.1 & 32.5 & 34.1\\
       & 1000 & Bias & 5.8 & 364 & 5.0 & 20.7 & 2.4 & $-$0.1 & 2.3 & $-$0.1\\
       &      & SD & 5.5 & 7810 & 5.5 & 5.9 & 6.9 & 5.7 & 5.6 & 5.7\\
       &      & Length & 19.5 & 130 & 19.5 & 17.0 & 19.3 & 23.4 & 22.6 & 23.4\\
\addlinespace
B      & 500  & Bias & 9.8 & 12.9 & 7.7 & 46.3 & $-$6.4 & 1.1 & 5.1 & 1.1\\
       &      & SD & 8.4 & 421 & 8.3 & 174 & 275 & 8.7 & 9.1 & 8.8\\
       &      & Length & 30.4 & 58.0 & 30.3 & 26.6 & 37.5 & 36.4 & 34.8 & 36.5\\
       & 1000 & Bias & 6.1 & 335 & 5.2 & 21.1 & 2.4 & 0.0 & 2.4 & 0.0\\
       &      & SD & 6.1 & 8200 & 6.0 & 6.5 & 7.9 & 6.3 & 6.1 & 6.3\\
       &      & Length & 21.3 & 126 & 21.2 & 18.1 & 21.0 & 25.2 & 24.3 & 25.2\\
\addlinespace
Local  & 500  & Bias & 14.8 & $-$16.4 & 10.3 & 43.1 & 7.2 & 2.2 & 10.5 & 2.3\\
       &      & SD & 8.8 & 802 & 8.3 & 8.3 & 69.1 & 9.1 & 9.3 & 9.3\\
       &      & Length & 33.1 & 78.5 & 33.4 & 30.3 & 34.9 & 42.5 & 39.1 & 42.5\\
       & 1000 & Bias & 8.6 & 41.1 & 6.7 & 31.8 & 4.7 & $-$0.5 & 4.2 & $-$0.5\\
       &      & SD & 6.2 & 758 & 6.0 & 6.9 & 33.7 & 6.3 & 6.5 & 6.3\\
       &      & Length & 21.9 & 44.4 & 21.9 & 21.6 & 21.5 & 27.6 & 25.8 & 27.6\\
\addlinespace
Sparse & 500  & Bias & 9.5 & 13.9 & 9.2 & 8.0 & $-$0.4 & 1.6 & $-$1.1 & 0.1\\
       &      & SD & 8.3 & 176 & 8.2 & 8.7 & 10.0 & 8.8 & 9.3 & 9.2\\
       &      & Length & 29.5 & 38.9 & 29.4 & 27.7 & 30.0 & 34.7 & 35.2 & 35.3\\
       & 1000 & Bias & 5.7 & 71.3 & 5.6 & 2.2 & 0.4 & 0.3 & 0.7 & 0.3\\
       &      & SD & 6.0 & 1495 & 5.9 & 6.0 & 6.2 & 6.2 & 6.2 & 6.2\\
       &      & Length & 21.1 & 36.8 & 21.1 & 20.5 & 21.7 & 24.4 & 24.1 & 24.4\\
\addlinespace
Two    & 500  & Bias & 15.4 & $-$3.9 & 13.8 & 29.8 & 8.3 & 0.0 & 9.9 & 0.0\\
       &      & SD & 8.2 & 141 & 8.2 & 8.2 & 23.4 & 10.1 & 9.0 & 10.1\\
       &      & Length & 30.8 & 35.4 & 30.7 & 30.1 & 31.5 & 39.3 & 34.9 & 39.3\\
       & 1000 & Bias & 11.2 & 4.7 & 10.3 & 25.3 & 6.9 & $-$2.1 & 6.5 & $-$2.1\\
       &      & SD & 6.0 & 24.8 & 6.0 & 5.9 & 7.9 & 7.1 & 6.2 & 7.1\\
       &      & Length & 21.8 & 22.3 & 21.7 & 21.5 & 21.9 & 27.8 & 24.3 & 27.8\\
\addlinespace
C1     & 500  & Bias & 10.2 & $-$3.4 & 0.7 & 3.9 & 0.7 & 9.5 & $-$2.9 & $-$2.9\\
       &      & SD & 4.2 & 5.7 & 4.4 & 4.8 & 5.0 & 4.2 & 5.1 & 5.1\\
       &      & Length & 16.4 & 17.1 & 16.6 & 16.5 & 17.2 & 17.4 & 22.9 & 22.9\\
       & 1000 & Bias & 5.7 & $-$0.7 & 0.4 & 1.8 & 0.5 & 5.0 & $-$1.1 & $-$1.1\\
       &      & SD & 3.1 & 3.4 & 3.2 & 3.2 & 3.5 & 3.2 & 3.6 & 3.6\\
       &      & Length & 12.0 & 11.9 & 12.0 & 12.0 & 12.3 & 12.4 & 14.7 & 14.7\\
\addlinespace
C2     & 500  & Bias & 11.4 & $-$4.2 & 0.7 & 4.0 & 0.6 & 10.5 & $-$3.4 & $-$3.4\\
       &      & SD & 6.3 & 7.9 & 6.5 & 6.6 & 7.1 & 6.2 & 7.3 & 7.3\\
       &      & Length & 21.3 & 21.2 & 21.5 & 21.4 & 23.6 & 22.5 & 31.3 & 31.3\\
       & 1000 & Bias & 6.5 & $-$0.9 & 0.4 & 1.9 & 0.5 & 5.7 & $-$1.6 & $-$1.5\\
       &      & SD & 4.4 & 4.7 & 4.5 & 4.4 & 4.6 & 4.4 & 4.7 & 4.7\\
       &      & Length & 16.1 & 15.6 & 16.1 & 16.1 & 17.0 & 16.6 & 20.0 & 20.0\\
\addlinespace
Rough  & 500  & Bias & 22.8 & 89.0 & 20.6 & 53.5 & $-$52.7 & 12.3 & 14.0 & 12.0\\
       &      & SD & 8.0 & 1213 & 7.9 & 9.2 & 561 & 8.6 & 8.9 & 8.6\\
       &      & Length & 29.3 & 72.6 & 29.2 & 21.9 & 58.9 & 35.8 & 35.0 & 36.0\\
       & 1000 & Bias & 15.1 & 35.2 & 14.0 & 32.8 & 9.9 & 6.6 & 8.3 & 6.5\\
       &      & SD & 5.8 & 502 & 5.8 & 7.5 & 9.0 & 6.2 & 6.3 & 6.2\\
       &      & Length & 20.9 & 40.0 & 20.9 & 16.8 & 20.5 & 25.3 & 24.6 & 25.3\\
\bottomrule
\end{tabular}

\begin{minipage}{.97\textwidth}
\footnotesize
\medskip
Methods as in Table~\ref{tab:plm-extra}. Length is the median length of the nominal 95\%
interval.
\end{minipage}
\end{table}

\end{document}